\documentclass[trackchanges,floatfix]{aastex631}

\newcommand{\swift}{\textit{Swift}}
\newcommand{\nustar}{NuSTAR}
\newcommand{\g}{$\gamma$-ray}

\definecolor{darkspringgreen}{rgb}{0.09, 0.59, 0.16}

\usepackage{subfigure}
\usepackage{amsmath,multirow, url,hyperref}
\usepackage[normalem]{ulem}
 \usepackage{booktabs}
\shorttitle{VERITAS and Multiwavelength Observations of the Blazar B3 2247+381}
\shortauthors{The VERITAS Collaboration et al.}
\graphicspath{{./}{figures/}}

\begin{document}

\title{VERITAS and multiwavelength observations of the Blazar B3 2247+381 in response to an IceCube neutrino alert}

\author[0000-0002-2028-9230]{A.~Acharyya}\affiliation{CP3-Origins, University of Southern Denmark, Campusvej 55, 5230 Odense M, Denmark}
\author[0000-0002-9021-6192]{C.~B.~Adams}\affiliation{Physics Department, Columbia University, New York, NY 10027, USA}
\author[0000-0002-3886-3739]{P.~Bangale}\affiliation{Department of Physics and Astronomy and the Bartol Research Institute, University of Delaware, Newark, DE 19716, USA}
\author[0000-0002-9675-7328]{J.~T.~Bartkoske}\affiliation{Department of Physics and Astronomy, University of Utah, Salt Lake City, UT 84112, USA}
\author[0000-0003-2098-170X]{W.~Benbow}\affiliation{Center for Astrophysics $|$ Harvard \& Smithsonian, Cambridge, MA 02138, USA}
\author[0000-0001-6391-9661]{J.~H.~Buckley}\affiliation{Department of Physics, Washington University, St. Louis, MO 63130, USA}
\author{J.~L.~Christiansen}\affiliation{Physics Department, California Polytechnic State University, San Luis Obispo, CA 94307, USA}
\author[0000-0003-1716-4119]{A.~Duerr}\affiliation{Department of Physics and Astronomy, University of Utah, Salt Lake City, UT 84112, USA}
\author[0000-0002-1853-863X]{M.~Errando}\affiliation{Department of Physics, Washington University, St. Louis, MO 63130, USA}
\author{M.~Escobar~Godoy}\affiliation{Santa Cruz Institute for Particle Physics and Department of Physics, University of California, Santa Cruz, CA 95064, USA}
\author[0000-0002-5068-7344]{A.~Falcone}\affiliation{Dept. of Astronomy and Astrophysics, 525 Davey Lab, Pennsylvania State University, University Park, PA 16802, USA}
\author[0000-0001-6674-4238]{Q.~Feng}\affiliation{Department of Physics and Astronomy, University of Utah, Salt Lake City, UT 84112, USA}
\author[0000-0002-2944-6060]{J.~Foote}\affiliation{Department of Physics and Astronomy and the Bartol Research Institute, University of Delaware, Newark, DE 19716, USA}
\author[0000-0002-1067-8558]{L.~Fortson}\affiliation{School of Physics and Astronomy, University of Minnesota, Minneapolis, MN 55455, USA}
\author[0000-0003-1614-1273]{A.~Furniss}\affiliation{Santa Cruz Institute for Particle Physics and Department of Physics, University of California, Santa Cruz, CA 95064, USA}
\author{G.~Gallagher}\affiliation{Department of Physics and Astronomy, Ball State University, Muncie, IN 47306, USA}
\author[0000-0002-0109-4737]{W.~Hanlon}\affiliation{Center for Astrophysics $|$ Harvard \& Smithsonian, Cambridge, MA 02138, USA}
\author[0000-0002-8513-5603]{D.~Hanna}\affiliation{Physics Department, McGill University, Montreal, QC H3A 2T8, Canada}
\author[0000-0003-3878-1677]{O.~Hervet}\affiliation{Santa Cruz Institute for Particle Physics and Department of Physics, University of California, Santa Cruz, CA 95064, USA}
\author[0000-0001-6951-2299]{C.~E.~Hinrichs}\affiliation{Center for Astrophysics $|$ Harvard \& Smithsonian, Cambridge, MA 02138, USA} \affiliation{Department of Physics and Astronomy, Dartmouth College, 6127 Wilder Laboratory, Hanover, NH 03755 USA}
\author{J.~Hoang}\affiliation{Santa Cruz Institute for Particle Physics and Department of Physics, University of California, Santa Cruz, CA 95064, USA}
\author[0000-0002-6833-0474]{J.~Holder}\affiliation{Department of Physics and Astronomy and the Bartol Research Institute, University of Delaware, Newark, DE 19716, USA}
\author[0000-0002-1432-7771]{T.~B.~Humensky}\affiliation{Dept. of Physics, University of Maryland, College Park, MD 20742, USA} \affiliation{NASA GSFC, Greenbelt, MD 20771, USA}
\author[0000-0002-1089-1754]{W.~Jin}\affiliation{Department of Physics and Astronomy, University of California, Los Angeles, CA 90095, USA}
\author[0009-0008-2688-0815]{M.~N.~Johnson}\affiliation{Santa Cruz Institute for Particle Physics and Department of Physics, University of California, Santa Cruz, CA 95064, USA}
\author[0000-0002-3638-0637]{P.~Kaaret}\affiliation{Department of Physics and Astronomy, University of Iowa, Van Allen Hall, Iowa City, IA 52242, USA}
\author{M.~Kertzman}\affiliation{Department of Physics and Astronomy, DePauw University, Greencastle, IN 46135-0037, USA}
\author{M.~Kherlakian}\affiliation{Deutsches Elektronen-Synchrotron DESY, Platanenallee 6, D-15738 Zeuthen, Germany}
\author[0000-0003-4785-0101]{D.~Kieda}\affiliation{Department of Physics and Astronomy, University of Utah, Salt Lake City, UT 84112, USA}
\author[0000-0002-4260-9186]{T.~K.~Kleiner}\affiliation{Deutsches Elektronen-Synchrotron DESY, Platanenallee 6, D-15738 Zeuthen, Germany}
\author[0000-0002-4289-7106]{N.~Korzoun}\affiliation{Department of Physics and Astronomy and the Bartol Research Institute, University of Delaware, Newark, DE 19716, USA}
\author{F.~Krennrich}\affiliation{Department of Physics and Astronomy, Iowa State University, Ames, IA 50011, USA}
\author[0000-0002-5167-1221]{S.~Kumar}\affiliation{Dept. of Physics, University of Maryland, College Park, MD 20742, USA} 
\author[0000-0003-4641-4201]{M.~J.~Lang}\affiliation{School of Natural Sciences, University of Galway, University Road, Galway, H91 TK33, Ireland}
\author[0000-0003-3802-1619]{M.~Lundy}\affiliation{Physics Department, McGill University, Montreal, QC H3A 2T8, Canada}
\author{C.~E~McGrath}\affiliation{School of Physics, University College Dublin, Belfield, Dublin 4, Ireland}
\author{E.~Meyer}\affiliation{Department of Physics, University of Maryland, Baltimore County, Baltimore MD 21250, USA}
\author[0000-0001-7106-8502]{M.~J.~Millard}\affiliation{Department of Physics and Astronomy, University of Iowa, Van Allen Hall, Iowa City, IA 52242, USA}
\author{J.~Millis}\affiliation{Department of Physics and Astronomy, Ball State University, Muncie, IN 47306, USA} \affiliation{Department of Physics, Anderson University, 1100 East 5th Street, Anderson, IN 46012}
\author[0000-0001-5937-446X]{C.~L.~Mooney}\affiliation{Department of Physics and Astronomy and the Bartol Research Institute, University of Delaware, Newark, DE 19716, USA}
\author[0000-0002-1499-2667]{P.~Moriarty}\affiliation{School of Natural Sciences, University of Galway, University Road, Galway, H91 TK33, Ireland}
\author[0000-0002-3223-0754]{R.~Mukherjee}\affiliation{Department of Physics and Astronomy, Barnard College, Columbia University, NY 10027, USA}
\author[0000-0002-6121-3443]{W.~Ning}\affiliation{Department of Physics and Astronomy, University of California, Los Angeles, CA 90095, USA}
\author[0000-0002-9296-2981]{S.~O'Brien}\affiliation{Physics Department, McGill University, Montreal, QC H3A 2T8, Canada} \affiliation{Arthur B. McDonald Canadian Astroparticle Physics Research Institute, 64 Bader Lane, Queens University, Kingston, ON Canada, K7L 3N6}
\author[0000-0002-4837-5253]{R.~A.~Ong}\affiliation{Department of Physics and Astronomy, University of California, Los Angeles, CA 90095, USA}
\author[0000-0001-7861-1707]{M.~Pohl}\affiliation{Institute of Physics and Astronomy, University of Potsdam, 14476 Potsdam-Golm, Germany} \affiliation{Deutsches Elektronen-Synchrotron DESY, Platanenallee 6, D-15738 Zeuthen, Germany}
\author[0000-0002-0529-1973]{E.~Pueschel}\affiliation{Fakult{\"a}t f{\"u}r Physik {\&} Astronomie, Ruhr-Universit{\"a}t Bochum, D-44780 Bochum, Germany}
\author[0000-0002-4855-2694]{J.~Quinn}\affiliation{School of Physics, University College Dublin, Belfield, Dublin 4, Ireland}
\author{P.~L.~Rabinowitz}\affiliation{Department of Physics, Washington University, St. Louis, MO 63130, USA}
\author[0000-0002-5351-3323]{K.~Ragan}\affiliation{Physics Department, McGill University, Montreal, QC H3A 2T8, Canada}
\author{P.~T.~Reynolds}\affiliation{Department of Physical Sciences, Munster Technological University, Bishopstown, Cork, T12 P928, Ireland}
\author[0000-0002-7523-7366]{D.~Ribeiro}\affiliation{School of Physics and Astronomy, University of Minnesota, Minneapolis, MN 55455, USA}
\author{E.~Roache}\affiliation{Center for Astrophysics $|$ Harvard \& Smithsonian, Cambridge, MA 02138, USA}
\author[0000-0001-6662-5925]{J.~L.~Ryan}\affiliation{Department of Physics and Astronomy, University of California, Los Angeles, CA 90095, USA}
\author[0000-0003-1387-8915]{I.~Sadeh}\affiliation{Deutsches Elektronen-Synchrotron DESY, Platanenallee 6, D-15738 Zeuthen, Germany}
\author{A.~C.~Sadun}\affiliation{Department of Physics, University of Colorado Denver, Campus Box 157, P.O. Box 173364, Denver CO 80217, USA}
\author[0000-0002-3171-5039]{L.~Saha}\affiliation{Center for Astrophysics $|$ Harvard \& Smithsonian, Cambridge, MA 02138, USA}
\author{M.~Santander}\affiliation{Dept. of Physics and Astronomy, University of Alabama, Tuscaloosa, AL 35487, USA}
\author{G.~H.~Sembroski}\affiliation{Department of Physics and Astronomy, Purdue University, West Lafayette, IN 47907, USA}
\author[0000-0002-9856-989X]{R.~Shang}\affiliation{Department of Physics and Astronomy, Barnard College, Columbia University, NY 10027, USA}
\author[0000-0003-3407-9936]{M.~Splettstoesser}\affiliation{Santa Cruz Institute for Particle Physics and Department of Physics, University of California, Santa Cruz, CA 95064, USA}
\author[0000-0002-9852-2469]{D.~Tak}\affiliation{SNU Astronomy Research Center, Seoul National University, Seoul 08826, Republic of Korea.}
\author{A.~K.~Talluri}\affiliation{School of Physics and Astronomy, University of Minnesota, Minneapolis, MN 55455, USA}
\author{J.~V.~Tucci}\affiliation{Department of Physics, Indiana University-Purdue University Indianapolis, Indianapolis, IN 46202, USA}
\author[0000-0002-8090-6528]{J.~Valverde}\affiliation{Department of Physics, University of Maryland, Baltimore County, Baltimore MD 21250, USA} \affiliation{NASA GSFC, Greenbelt, MD 20771, USA}
\author[0000-0003-2740-9714]{D.~A.~Williams}\affiliation{Santa Cruz Institute for Particle Physics and Department of Physics, University of California, Santa Cruz, CA 95064, USA}
\author[0000-0002-2730-2733]{S.~L.~Wong}\affiliation{Physics Department, McGill University, Montreal, QC H3A 2T8, Canada}
\author[0009-0001-6471-1405]{J.~Woo}\affiliation{Columbia Astrophysics Laboratory, Columbia University, New York, NY 10027, USA}
\collaboration{69}{The VERITAS Collaboration}

\affiliation{III. Physikalisches Institut, RWTH Aachen University, D-52056 Aachen, Germany}
\affiliation{Department of Physics, University of Adelaide, Adelaide, 5005, Australia}
\affiliation{Dept. of Physics and Astronomy, University of Alaska Anchorage, 3211 Providence Dr., Anchorage, AK 99508, USA}
\affiliation{Dept. of Physics, University of Texas at Arlington, 502 Yates St., Science Hall Rm 108, Box 19059, Arlington, TX 76019, USA}
\affiliation{School of Physics and Center for Relativistic Astrophysics, Georgia Institute of Technology, Atlanta, GA 30332, USA}
\affiliation{Dept. of Physics, Southern University, Baton Rouge, LA 70813, USA}
\affiliation{Dept. of Physics, University of California, Berkeley, CA 94720, USA}
\affiliation{Lawrence Berkeley National Laboratory, Berkeley, CA 94720, USA}
\affiliation{Institut f{\"u}r Physik, Humboldt-Universit{\"a}t zu Berlin, D-12489 Berlin, Germany}
\affiliation{Fakult{\"a}t f{\"u}r Physik {\&} Astronomie, Ruhr-Universit{\"a}t Bochum, D-44780 Bochum, Germany}
\affiliation{Universit{\'e} Libre de Bruxelles, Science Faculty CP230, B-1050 Brussels, Belgium}
\affiliation{Vrije Universiteit Brussel (VUB), Dienst ELEM, B-1050 Brussels, Belgium}
\affiliation{Dept. of Physics, Simon Fraser University, Burnaby, BC V5A 1S6, Canada}
\affiliation{Department of Physics and Laboratory for Particle Physics and Cosmology, Harvard University, Cambridge, MA 02138, USA}
\affiliation{Dept. of Physics, Massachusetts Institute of Technology, Cambridge, MA 02139, USA}
\affiliation{Dept. of Physics and The International Center for Hadron Astrophysics, Chiba University, Chiba 263-8522, Japan}
\affiliation{Department of Physics, Loyola University Chicago, Chicago, IL 60660, USA}
\affiliation{Dept. of Physics and Astronomy, University of Canterbury, Private Bag 4800, Christchurch, New Zealand}
\affiliation{Dept. of Physics, University of Maryland, College Park, MD 20742, USA}
\affiliation{Dept. of Astronomy, Ohio State University, Columbus, OH 43210, USA}
\affiliation{Dept. of Physics and Center for Cosmology and Astro-Particle Physics, Ohio State University, Columbus, OH 43210, USA}
\affiliation{Niels Bohr Institute, University of Copenhagen, DK-2100 Copenhagen, Denmark}
\affiliation{Dept. of Physics, TU Dortmund University, D-44221 Dortmund, Germany}
\affiliation{Dept. of Physics and Astronomy, Michigan State University, East Lansing, MI 48824, USA}
\affiliation{Dept. of Physics, University of Alberta, Edmonton, Alberta, T6G 2E1, Canada}
\affiliation{Erlangen Centre for Astroparticle Physics, Friedrich-Alexander-Universit{\"a}t Erlangen-N{\"u}rnberg, D-91058 Erlangen, Germany}
\affiliation{Physik-department, Technische Universit{\"a}t M{\"u}nchen, D-85748 Garching, Germany}
\affiliation{D{\'e}partement de physique nucl{\'e}aire et corpusculaire, Universit{\'e} de Gen{\`e}ve, CH-1211 Gen{\`e}ve, Switzerland}
\affiliation{Dept. of Physics and Astronomy, University of Gent, B-9000 Gent, Belgium}
\affiliation{Dept. of Physics and Astronomy, University of California, Irvine, CA 92697, USA}
\affiliation{Karlsruhe Institute of Technology, Institute for Astroparticle Physics, D-76021 Karlsruhe, Germany}
\affiliation{Karlsruhe Institute of Technology, Institute of Experimental Particle Physics, D-76021 Karlsruhe, Germany}
\affiliation{Dept. of Physics, Engineering Physics, and Astronomy, Queen's University, Kingston, ON K7L 3N6, Canada}
\affiliation{Department of Physics {\&} Astronomy, University of Nevada, Las Vegas, NV 89154, USA}
\affiliation{Nevada Center for Astrophysics, University of Nevada, Las Vegas, NV 89154, USA}
\affiliation{Dept. of Physics and Astronomy, University of Kansas, Lawrence, KS 66045, USA}
\affiliation{Centre for Cosmology, Particle Physics and Phenomenology - CP3, Universit{\'e} catholique de Louvain, Louvain-la-Neuve, Belgium}
\affiliation{Department of Physics, Mercer University, Macon, GA 31207-0001, USA}
\affiliation{Dept. of Astronomy, University of Wisconsin{\textemdash}Madison, Madison, WI 53706, USA}
\affiliation{Dept. of Physics and Wisconsin IceCube Particle Astrophysics Center, University of Wisconsin{\textemdash}Madison, Madison, WI 53706, USA}
\affiliation{Institute of Physics, University of Mainz, Staudinger Weg 7, D-55099 Mainz, Germany}
\affiliation{Department of Physics, Marquette University, Milwaukee, WI 53201, USA}
\affiliation{Institut f{\"u}r Kernphysik, Westf{\"a}lische Wilhelms-Universit{\"a}t M{\"u}nster, D-48149 M{\"u}nster, Germany}
\affiliation{Bartol Research Institute and Dept. of Physics and Astronomy, University of Delaware, Newark, DE 19716, USA}
\affiliation{Dept. of Physics, Yale University, New Haven, CT 06520, USA}
\affiliation{Columbia Astrophysics and Nevis Laboratories, Columbia University, New York, NY 10027, USA}
\affiliation{Dept. of Physics, University of Oxford, Parks Road, Oxford OX1 3PU, United Kingdom}
\affiliation{Dipartimento di Fisica e Astronomia Galileo Galilei, Universit{\`a} Degli Studi di Padova, I-35122 Padova PD, Italy}
\affiliation{Dept. of Physics, Drexel University, 3141 Chestnut Street, Philadelphia, PA 19104, USA}
\affiliation{Physics Department, South Dakota School of Mines and Technology, Rapid City, SD 57701, USA}
\affiliation{Dept. of Physics, University of Wisconsin, River Falls, WI 54022, USA}
\affiliation{Dept. of Physics and Astronomy, University of Rochester, Rochester, NY 14627, USA}
\affiliation{Department of Physics and Astronomy, University of Utah, Salt Lake City, UT 84112, USA}
\affiliation{Dept. of Physics, Chung-Ang University, Seoul 06974, Republic of Korea}
\affiliation{Oskar Klein Centre and Dept. of Physics, Stockholm University, SE-10691 Stockholm, Sweden}
\affiliation{Dept. of Physics and Astronomy, Stony Brook University, Stony Brook, NY 11794-3800, USA}
\affiliation{Dept. of Physics, Sungkyunkwan University, Suwon 16419, Republic of Korea}
\affiliation{Institute of Basic Science, Sungkyunkwan University, Suwon 16419, Republic of Korea}
\affiliation{Institute of Physics, Academia Sinica, Taipei, 11529, Taiwan}
\affiliation{Dept. of Physics and Astronomy, University of Alabama, Tuscaloosa, AL 35487, USA}
\affiliation{Dept. of Astronomy and Astrophysics, Pennsylvania State University, University Park, PA 16802, USA}
\affiliation{Dept. of Physics, Pennsylvania State University, University Park, PA 16802, USA}
\affiliation{Dept. of Physics and Astronomy, Uppsala University, Box 516, SE-75120 Uppsala, Sweden}
\affiliation{Dept. of Physics, University of Wuppertal, D-42119 Wuppertal, Germany}
\affiliation{Deutsches Elektronen-Synchrotron DESY, Platanenallee 6, D-15738 Zeuthen, Germany}

\author[0000-0001-6141-4205]{R. Abbasi}
\affiliation{Department of Physics, Loyola University Chicago, Chicago, IL 60660, USA}

\author[0000-0001-8952-588X]{M. Ackermann}
\affiliation{Deutsches Elektronen-Synchrotron DESY, Platanenallee 6, D-15738 Zeuthen, Germany}

\author{J. Adams}
\affiliation{Dept. of Physics and Astronomy, University of Canterbury, Private Bag 4800, Christchurch, New Zealand}

\author[0000-0002-9714-8866]{S. K. Agarwalla}
\altaffiliation{also at Institute of Physics, Sachivalaya Marg, Sainik School Post, Bhubaneswar 751005, India}
\affiliation{Dept. of Physics and Wisconsin IceCube Particle Astrophysics Center, University of Wisconsin{\textemdash}Madison, Madison, WI 53706, USA}

\author[0000-0003-2252-9514]{J. A. Aguilar}
\affiliation{Universit{\'e} Libre de Bruxelles, Science Faculty CP230, B-1050 Brussels, Belgium}

\author[0000-0003-0709-5631]{M. Ahlers}
\affiliation{Niels Bohr Institute, University of Copenhagen, DK-2100 Copenhagen, Denmark}

\author[0000-0002-9534-9189]{J.M. Alameddine}
\affiliation{Dept. of Physics, TU Dortmund University, D-44221 Dortmund, Germany}

\author{N. M. Amin}
\affiliation{Department of Physics and Astronomy and the Bartol Research Institute, University of Delaware, Newark, DE 19716, USA}

\author[0000-0001-9394-0007]{K. Andeen}
\affiliation{Department of Physics, Marquette University, Milwaukee, WI 53201, USA}

\author[0000-0003-4186-4182]{C. Arg{\"u}elles}
\affiliation{Department of Physics and Laboratory for Particle Physics and Cosmology, Harvard University, Cambridge, MA 02138, USA}

% ...

\date{\today}

\collaboration{427}{IceCube Collaboration}

\author{P.~Drake}\affiliation{Physics Department, Columbia University, New York, NY 10027, USA}
\author{E.~Spira-Savett}\affiliation{Department of Physics and Astronomy, Barnard College, Columbia University, NY 10027, USA}
\author{P.~Lusen }\affiliation{Santa Cruz Institute for Particle Physics and Department of Physics, University of California, Santa Cruz, CA 95064, USA}
\author{K.~Mori}\affiliation{Columbia Astrophysics Laboratory, Columbia University, New York, NY 10027, USA}
\nocollaboration{5}

\correspondingauthor{Atreya Acharyya}
\email{atreya@cp3.sdu.dk}

\correspondingauthor{Samantha Wong}
\email{samantha.wong2@mail.mcgill.ca}

\correspondingauthor{}
\email{analysis@icecube.wisc.edu}

\begin{abstract}
While the sources of the diffuse astrophysical neutrino flux detected by the IceCube Neutrino Observatory are still largely unknown, one of the promising methods used towards understanding this is investigating the potential temporal and spatial correlations between neutrino alerts and the electromagnetic radiation from blazars. We report on the multiwavelength target-of-opportunity observations of the blazar B3~2247+381, taken in response to an IceCube multiplet alert for a cluster of muon neutrino events compatible with the source location between May~20,~2022 and November~10,~2022. B3~2247+381 was not detected with VERITAS during this time period. The source was found to be in a low-flux state in the optical, ultraviolet and gamma-ray bands for the time interval corresponding to the neutrino event, but was detected in the hard X-ray band with \emph{NuSTAR} during this period. We find the multiwavelength spectral energy distribution is well described using a simple one-zone leptonic synchrotron self-Compton radiation model. Moreover, assuming the neutrinos originate from hadronic processes within the jet, the neutrino flux would be accompanied by a photon flux from the cascade emission, and the integrated photon flux required in such a case would significantly exceed the total multiwavelength fluxes and the VERITAS upper limits presented here. The lack of flaring activity observed with VERITAS, combined with the low multiwavelength flux levels, and given the significance of the neutrino excess is at 3$\sigma$ level (uncorrected for trials), makes B3 2247+381 an unlikely source of the IceCube multiplet. We conclude that the neutrino excess is likely a background fluctuation.

\end{abstract}



\section{Introduction} 
\label{sec:intro}

The IceCube Neutrino Observatory \citep{icecube} has opened up a new window to extreme environments in the Universe, where cosmic rays are accelerated to ultra-high energies. Any detection of a TeV -- PeV neutrino emitting source would help to directly answer some century-long open questions on the origin of cosmic rays. The neutrinos observed must be produced in cosmic-ray interactions \citep{2017ARNPS..67...45M}. 
Moreover, while both gamma rays and neutrinos point directly back to their sources, the latter are also not substantially attenuated as they travel through space.
Higher energy gamma rays readily interact with background photons leading to pair-production and attenuation.
The IceCube Observatory has detected a diffuse flux of astrophysical neutrinos \citep{IceCube13}, with the observed isotropic distribution of arrival directions suggesting an extragalactic origin. A number of  extragalactic sources are proposed as candidate high-energy neutrino emitters. These include blazars, a subclass of active galactic nuclei (AGN) with relativistic jets pointing along our line-of-sight (for a review see \cite{2023ecnp.book..483M}).


The IceCube Collaboration has reported evidence for neutrino emission from GeV gamma-ray sources, TXS 0506+056 \citep{eaat1378} and NGC 1068 \citep{2022Sci...378..538I}. While the gamma-ray flux from the active galaxy NGC~1068 is attenuated and not variable, the TeV detected blazar TXS 0506+056 exhibits gamma-ray flaring episodes \citep{2022ApJ...927..197A}. This, in addition to, the difference in AGN type of the sources and their respecitive redshifts, along the two sources representing different neutrino spectra \citep{2022Sci...378..538I}, suggests the existence of at least two populations of extragalactic neutrino sources. The 2017 gamma-ray flare of TXS 0506+056 in spatial and temporal coincidence with the $\sim$  290 TeV neutrino event IceCube-170922A \citep{GCN21916, eaat1378} also highlights the importance of flaring blazars in the search for neutrino sources. 

More recently, IceCube detected a track-like event with an energy 171 TeV on December 8, 2021, which was found to be near the blazar PKS 0735+178 \citep{2021GCN.31191....1I}. An association between the IceCube event and the PKS 0735+178 flare was investigated in multiwavelength studies \citep{2023MNRAS.519.1396S, 2023_Qi}, but no firm connection was established. The above alerts are examples of single high energy ($>$ 100 TeV) neutrino events; other such studies can be found in \cite{2016_Kadler}, \cite{2020ApJ...893..162F}, \cite{2020A&A...640L...4G}, \cite{2022ApJ...932L..25L} and \cite{2023ApJ...954...75A}. These alerts have been publicly distributed by IceCube since 2016. The information about these events is shared using the General Coordinates Network (GCN),\footnote{\url{https://gcn.nasa.gov} (accessed on May 3, 2024)} an open-source platform created by NASA to receive and transmit low-latency alerts about astronomical transient phenomena. These single-event alerts have a typical localization uncertainty of $\sim 1^{\circ}$ and the region of interest (RoI) defined by the neutrino localization uncertainty often contains potential neutrino sources such as AGN or transient sources \citep{2017APh....92...30A, 2023ApJS..269...25A}.



While higher energy neutrinos are believed to have a higher probability of being of astrophysical origin, it should also be noted that the observation of multiple neutrinos from a single location is also a signature for neutrinos of astrophysical origin. In light of this, the IceCube Collaboration has been operating a complimentary approach known as the Gamma-ray Follow-Up (GFU) program\footnote{\url{https://icecube.wisc.edu/science/real-time-alerts/} (accessed on May 3, 2024)}  \citep{2016JInst..1111009I, 2022_gfu} since 2012. The goal of the GFU program is to enable prompt follow-up investigations of known gamma-ray sources for which IceCube has detected a cluster of candidate neutrino events, typically having energies $\sim$ 1 TeV, above a pre-defined significance, using imaging atmospheric Cherenkov telescopes (IACTs). 

The GFU program utilizes a realtime neutrino event selection that selects events that are consistent with muon tracks arising from muon neutrino interactions in or near the IceCube detector volume.   
After significant data reduction, this sample is dominated by events arising from the interaction of cosmic rays in the atmosphere, with atmospheric neutrinos – an irreducible background - dominating the sample from the northern sky, and atmospheric muons dominating in the southern sky.
The GFU searches time windows that span from seconds to as long as 180 days, looking for significant clustering in time and space of neutrinos in this sample.  The alert thresholds are set so that random background fluctuations are suppressed to low levels,  requiring a pre-trials statistical significance of $3 \sigma$, and generate alerts at a frequency of  about 10 per year.

The IceCube GFU program selected a list of 190 sources\footnote{\url{https://user-web.icecube.wisc.edu/~npark/data/VERITAS_shortList_redShift.txt} (accessed on May 3, 2024)} to be monitored for neutrino clusters and observable with VERITAS. These objects include all extragalactic TeV sources detected with IACTs\footnote{\url{http://tevcat2.uchicago.edu/} (accessed on May 3, 2024)}, the Galactic Center, and the Crab Nebula, as well as sources from the 3FGL \citep{Acero_2015} or 3FHL \citep{Ajello_2017} catalogs based upon the following criteria \citep{2022_gfu}:
    \begin{itemize}
    \item Extragalactic sources having a known redshift, z~$\leq$~1.0. This is done because sources at higher redshifts are difficult to detect with current generation IACTs due to absorbtion by extragalactic background light (EBL).
    \item 3FGL sources with variability index $>$ 77.2 and 3FHL sources for which the number of Bayesian blocks from variability analysis~$>$~1. This is done because the majority of detections with current generation IACTs occur during flares.
    \item Sources with a maximum elevation of $> 45^{\circ}$ at the VERITAS site, in order to allow for optimal observations with VERITAS.
    \item Assuming the source produces gamma-ray flares with a 10-fold increase on the average \emph{Fermi}-LAT flux, the extrapolated flux above 100 GeV exceeds the VERITAS $5 \sigma$ sensitivity within 5 hours of observations, allowing for detection in short observing times.
    \end{itemize}

\begin{figure*}
\centering
\includegraphics[width=0.8 \linewidth]{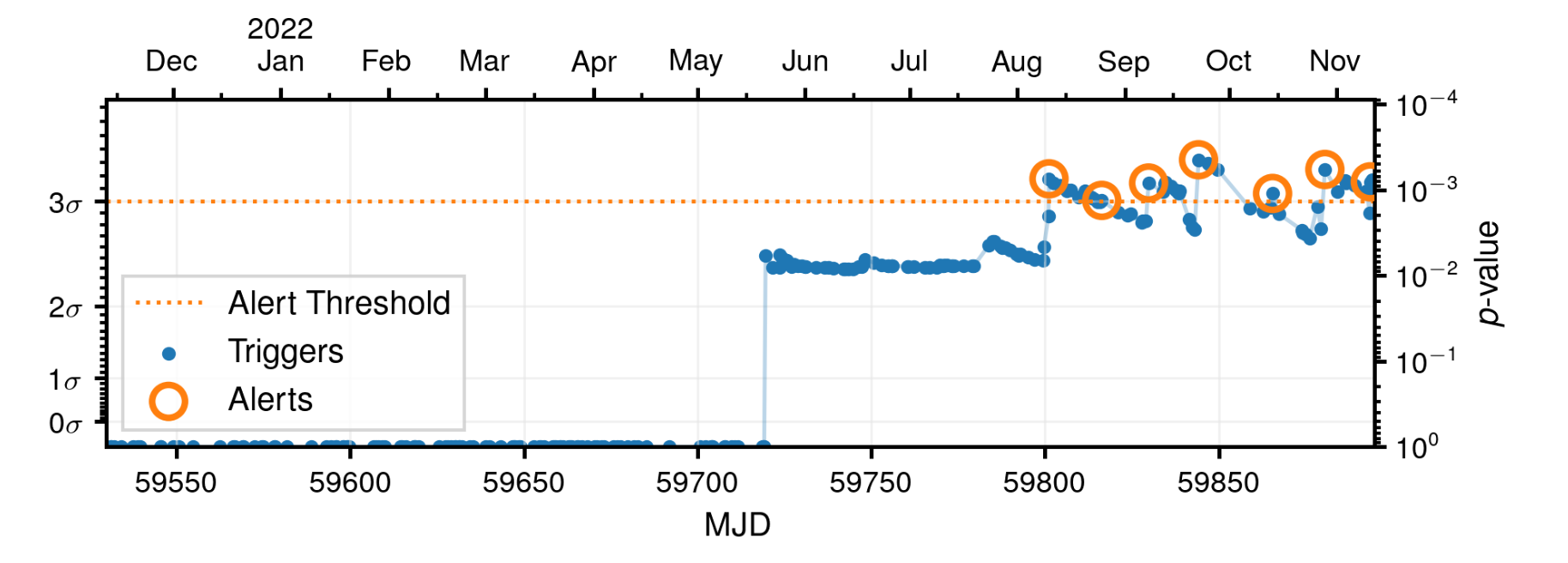}
\caption{The evolution of alert significance at the location of B3 2247+381. In total, 7 alerts were received over a duration of 174 days between May~20, 2022 and November~10, 2022 and the significance of these alerts, taking into account their directions, angular uncertainties and energies, as well as the duration of the time window, was found to be $3.2 \sigma$. It should be noted, this significance value is not fully corrected for trials as the GFU search includes trials corrections for the multiple time windows searched per source but not for the number of catalog entries searched.}
\label{fig:trigger}
\end{figure*}

This study investigates a GFU alert report privately shared with VERITAS by IceCube. It contained information on alerts of a cluster of muon neutrino candidate events from directions compatible with the source B3 2247+381, initially comprising 4 alerts received  at a significance of $> 3 \sigma$ between August~10, 2022 and September~22, 2022. These initial alerts triggered the VERITAS target-of-opportunity (ToO) observations and initiated a multiwavelength campaign including \emph{NuSTAR} observations. In total, there were 7  alerts over a duration of 174 days between May~20, 2022 and November~10, 2022, and are shown in Fig.~\ref{fig:trigger}. 
It should be noted that these 7 alerts were likely not independent and rather one primary alert developing in realtime, and retriggering multiple times as more events came in.
The significance of these alerts, taking into account the event directions, angular uncertainties, event energies, and the time duration of the time window was found to be $3.2 \sigma$. The corresponding false alert rate, quantifying how often the observed significance (or higher) is found at this location in a background-only scenario, is 0.0355 per year. 

The initial event on May 20, 2022 (MJD 59719) corresponded to a muon neutrino candidate event with an energy proxy of $E_{\mu}=~80.6$~TeV. The energy proxy is determined from the energy loss of the muon track and is described in Section 9.1 of \cite{Aartsen_2014}. This initial event pushed the significance of the neutrino cluster above 2.5$\sigma$, which then reached the alert threshold of 3$\sigma$ around August 9, 2022 (MJD 59800), resulting in a GFU alert to be triggered. Subsequent neutrinos with much lower energy proxies, $0.5$ TeV  $<  E_{\mu} < 6.1$ TeV, continued to push the significance of the neutrino multiplets from this location to above 3$\sigma$, triggering six more GFU alerts, even though the cumulative significance of the neutrino multiplet did not continue to grow, as seen in Fig. \ref{fig:trigger}. 

The privately distributed alert stream was formed by clusters of neutrino events in time and space around B3~2247+381. 
It should be noted that the $3.2 \sigma$ significance stated in the IceCube report is not fully corrected for trials as the GFU search includes trials corrections for the multiple time windows searched per source but not for the number of catalog entries searched.
Furthermore, the GFU sample is dominated at this location by neutrino events with energies between 0.5 TeV and 6 TeV, hence these are highly likely to be atmospheric neutrino background.
As the associated source of GFU alerts is already known, the goal of IACT and multiwavelength observations is to investigate possible changes to the state of the source, for example spectral changes or flaring states.
 
The BL Lacertae object B3 2247+381 ($z~=~0.119$; \cite{1998ApJ...494...47F}) was first detected in the very high energy (VHE) regime with the MAGIC telescopes, based on 14.2 hours of good quality gamma-ray data collected between September~30,~2010 and October~30, 2010 \citep{2012A&A...539A.118A}. These observations were triggered by a high optical state and yielded a significance level of 5.6$\sigma$. The observations revealed a relatively soft VHE spectrum with a photon index of $-3.2 \pm 0.6$. Furthermore, no significant short-term flux variability was observed and the spectral energy distribution (SED) was successfully modelled using a one-zone synchrotron self-Compton (SSC) model. Moreover, the observed flux was found to be consistent with an upper limit obtained with MAGIC during a prior observation taken in 2006 during a low optical state, and no connection between the high optical state and the VHE gamma-ray emission could be established. The inclusion of B3 2247+381 in the IceCube GFU list was motivated by this initial detection at VHE energies.


The aim of this paper is to investigate the VERITAS and multiwavelength follow-up to the GFU alert. We study B3 2247+381 in other wavebands to characterize the transition between the low-energy and high-energy components of the broadband SED. Furthermore, we aim to present a proof-of-concept study of a follow-up to an alert within the framework of the GFU program and also investigate if there were any significant changes to the electromagnetic emission state of the source during the time interval corresponding to the alert.

\section{Observations and Data Analysis}
\label{sec:obs_data}
%

\subsection{VERITAS}

The Very Energetic Radiation Imaging Telescope Array System (VERITAS) is an array comprising four 12 m IACTs. It is located at the Fred Lawrence Whipple Observatory (FLWO) in southern Arizona, USA ($30^{\circ}$ 40' N, $110^{\circ}$ 57' W, 1.3 km above sea level; \citealt{vts_paper}).  Each telescope contains a camera comprising 499 photomultiplier tubes, and covering a field of view diameter of 3.5$^{\circ}$. 
When the highest-energy photons enter the atmosphere, they produce an air shower containing particles which  in turn produce a burst of blue light known as Cherenkov radiation. Gamma-ray observatories like VERITAS effectively use the whole atmosphere as their detector, and track the blue visible light produced from the air shower using optical telescopes.

VERITAS is capable of detecting gamma rays having energies in the range from 85~GeV to above 30~TeV, with an energy resolution of $ \Delta E/E~\sim$~15\% (at 1~TeV) and an angular resolution of $\sim$~0.1$^{\circ}$ (68\% containment at 1~TeV). In its current configuration, VERITAS typically detects a source with a flux of 1\% of the steady state gamma-ray flux of the Crab Nebula at a statistical significance of $>5\sigma$ after 25 hours of observation. However, it should be noted that the time required to reach a $5\sigma$ detection is longer for sources with a spectrum softer than the Crab Nebula (photon index $\Gamma > 2.49$) or those observed at large zenith angles ($\theta_{z} > 40^{\circ}$) \citep{Park15}.

Following the IceCube alert, VERITAS started ToO observations of B3~2247+381 on September~22, 2022 (MJD 59844).
VERITAS observed the source for 5 hours between September~23, 2022 and October~3, 2022 (MJD 59845 - MJD 59855) at a mean elevation of $73^{\circ}$ while hindered by poor weather on several nights during this period. 
These observations were performed using a standard ``wobble'' observing mode \citep{FOMIN1994151} with a $0.5^{\circ}$ offset in each of the four cardinal directions in order to simultaneously determine the background event rate. Furthermore, quality cuts were applied to the data set to remove events affected by bad weather. The VERITAS data were analyzed using the \texttt{Eventdisplay} analysis package \citep{Maier17}, and independently confirmed with the \texttt{VEGAS} analysis package \citep{VEGAS}, yielding consistent results.

The VERITAS analysis parameterizes the principal moments of the elliptical shower images before applying a set of cuts to these parameters in order to reject cosmic-ray background events. The cuts are determined from a boosted decision tree algorithm~\citep{2017APh....89....1K}, optimized for soft-spectrum sources ($\Gamma \sim 4$), and previously trained on gamma-ray shower simulations. During this process, we rejected events having fewer than two telescope images. Gamma-ray candidate events which fall within a squared angular distance, $\theta^{2}~\leq$~0.008~deg$^{2}$, between the reconstructed event origin and B3 2247+381 are considered in the source (\textsc{ON}) region. Furthermore, the background is estimated using the reflected region model \citep{2007A&A...466.1219B}, where circular background \textsc{OFF} regions having the same size as the \textsc{ON} region are placed at the same radial distance from the center of the camera. 

B3 2247+381 was not detected with VERITAS during the time interval investigated in this study. An excess of only 11 gamma-ray candidate events was recorded in the source region with $N_{\text{ON}}$ = 141 \textsc{ON} events, $N_{\text{OFF}}$ = 2384 \textsc{OFF} events, a background normalization factor, $\alpha = 0.055$, and $N_{\text{Excess}} = N_{\text{ON}} - \alpha N_{\text{OFF}} = 10.5$, corresponding to a statistical significance of 0.88$\sigma$, calculated following the method of \cite{1983ApJ...272..317L}. The upper limit at 99\% confidence level for the average integral flux above 200 GeV is $3.6 \times 10^{-12}$ cm$^{-2}$ s$^{-1}$, or 1.5\% of the Crab Nebula flux above the same energy threshold. All VERITAS upper limits were computed assuming a power-law spectrum with a photon index of 3 \citep[following][]{Rolke05}.

Furthermore, the analysis of archival observations of B3 2247+281 taken with VERITAS and comprising a total of 27 hours of quality-selected data accumulated from September 2009 to November 2020 found an excess of 65 gamma-ray candidate events in the source region with $N_{\text{ON}}$ = 836 \textsc{ON} events, $N_{\text{OFF}}$ = 14130 \textsc{OFF} events, a background normalization factor, $\alpha = 0.055$, and corresponding to a statistical significance of 2.2$\sigma$ and a flux upper limit of $1.7 \times 10^{-12}$ cm$^{-2}$ s$^{-1}$ above a threshold of 200 GeV.
This can be compared to the time-averaged flux limit during the IceCube alert period of $3.6 \times 10^{-12}$ cm$^{-2}$ s$^{-1}$. The differences between these limits likely originate from the much greater exposure time of the archival data (301 minutes of data during the IceCube alert period compared to 1683 minutes of archival data), which leads to a more constraining limit, rather than any intrinsic gamma-ray variability of the source. 
For comparison, the integral flux of B3 2247+381 above 200 GeV during the 2010 MAGIC detection was (5.0 $\pm$ $0.6_{\text{stat}}$ $\pm$ $1.1_{\text{sys}}$) $\times~10^{-12}$ cm$^{-2}$ s$^{-1}$ \citep{2012A&A...539A.118A}.
A lightcurve of the VERITAS flux above a threshold of 200~GeV and binned in monthly intervals is shown in Fig.~\ref{fig:full_lc}. The VERITAS energy spectral limits, shown in the multiwavelength SED in Fig.~\ref{fig:sed}, were derived using a spectral binning of 5 energy bins per decade in energy.
\subsection{Fermi-LAT}
\label{subsec:LAT}

The \emph{Fermi}-Large Area Telescope (LAT; \citealt{Fermi_LAT}) is a pair conversion telescope capable of detecting gamma-ray photons in the energy range from 20~MeV to above 500 GeV. 
The pair conversion process forms the basis for the operation of the \emph{Fermi}-LAT by providing a unique signature for gamma rays. This distinguishes them from charged cosmic rays and allows a determination of the incident photon directions via the reconstruction of the trajectories of the electron positron pairs. Primarily operating in survey mode, the \emph{Fermi}-LAT scans the entire sky every three hours. In this paper, we initially analyzed \textit{Fermi}-LAT data during the IceCube neutrino alert period, between MJD~59719 and MJD~59893, corresponding to midnight on May~20, 2022 until midnight on November~10, 2022. Throughout the analysis, we use the \textit{Fermi} Science Tools version 2.20\footnote{\url{http://fermi.gsfc.nasa.gov/ssc/data/analysis/software} (accessed on May 3, 2024)}, \texttt{FERMIPY} version 1.2\footnote{\url{http://fermipy.readthedocs.io} (accessed on May 3, 2024)} \citep{wood2017fermipy}, in conjunction with the latest \textit{PASS} 8 instrument response functions (IRFs; \cite{atwood2013pass}).

The \textit{Fermi}-LAT data were processed using a binned maximum likelihood analysis. Photons with energies between 100 MeV and 300 GeV detected within a RoI of radius 10$^{\circ}$ centered on the location of B3 2247+381 were selected for the analysis. We selected only photon events from within a maximum zenith angle of 90$^{\circ}$ in order to reduce contamination from background photons from the Earth's limb, produced from the cosmic-ray interactions with the upper atmosphere. 

A spatial bin size of 0.1$^\circ$ per pixel and two energy bins per decade were used. All sources contained in the 4FGL-DR3 catalog \citep{4fgl_dr3} within 20$^\circ$ of the RoI center, were included in the model with their spectral parameters fixed to their catalog values. This takes into account the gamma-ray emission from sources lying outside the RoI which might contribute photons to the data, especially at low energies, due to the size of the point spread function of the \textit{Fermi}-LAT. Since the time interval considered in this work is beyond that covered in the 4FGL-DR3 catalog, the \textit{gtfindsrc} routine was also applied to search for any additional point sources present in the data and not included in the catalog. 
No significant additional point sources, having a test statistic (TS; \cite{RN7}) $\geq$ 9 (roughly corresponding to a significance of $\sim$ 3$\sigma$) were detected, indicating that all sources in the data had been accounted for. Moreover, the contributions from the isotropic and Galactic diffuse backgrounds were modeled using the most recent templates for isotropic and Galactic diffuse emission, iso\_P8R3\_SOURCE\_V3\_v1.txt and gll\_iem\_v07.fits, respectively.

The normalization factor for both the isotropic and Galactic diffuse emission templates were left free along with the spectral normalization of all modeled sources within the RoI. Moreover, the spectral shape parameters of all modeled sources within 3$^{\circ}$ of B3 2247+381 were left free to vary while those of the remaining sources were fixed to the values reported in the 4FGL-DR3 catalog. 
B3 2247+381 was found to have an integral flux upper limit of $1.43  \times 10^{-9}$ $\text{cm}^{-2}\text{s}^{-1}$ in the energy range 100 MeV -- 300 GeV during the IceCube neutrino alert period. 



We then repeated the same analysis routine over a wider time interval August 4, 2008, the start of the \textit{Fermi}-LAT mission until midnight on June 1, 2023 (MJD 54683 -- MJD 60096). B3 2247+381 was detected at a statistical significance of 23$\sigma$ (TS = 542.3), corresponding to an integral flux of $(2.84~\pm~0.42)~\times 10^{-9}$~$\text{cm}^{-2}\text{s}^{-1}$.
The spectrum was found to be best modeled by a power law (PL):
\begin{equation}
    \centering
   \frac{dN}{dE}=N_{0} \left(\frac{E}{E_0}\right)^{-\Gamma}
	\label{LP_spectrum}
\end{equation}
where $N_{0}$ is the normalization, $E_{0}$ is the pivot energy, and $\Gamma$ is the spectral index. 
The corresponding best-fit spectral parameters are $N_{0} = (2.25 \pm 0.17) \times 10^{-14}$ $\text{cm}^{-2}\text{s}^{-1} \text{MeV}^{-1}$, and $\Gamma~=~1.71 \pm 0.05$. These are, within uncertainties, compatible with the spectral values listed in the 4FGL-DR3 catalog \citep{4fgl_dr3} for this source, explicitly $N_{0} = (2.60 \pm 0.19) \times 10^{-14}$ $\text{cm}^{-2}\text{s}^{-1} \text{MeV}^{-1}$ and $\Gamma = 1.74 \pm 0.05$. 

\begin{figure*}[h]
\centering
\includegraphics[width=\linewidth]{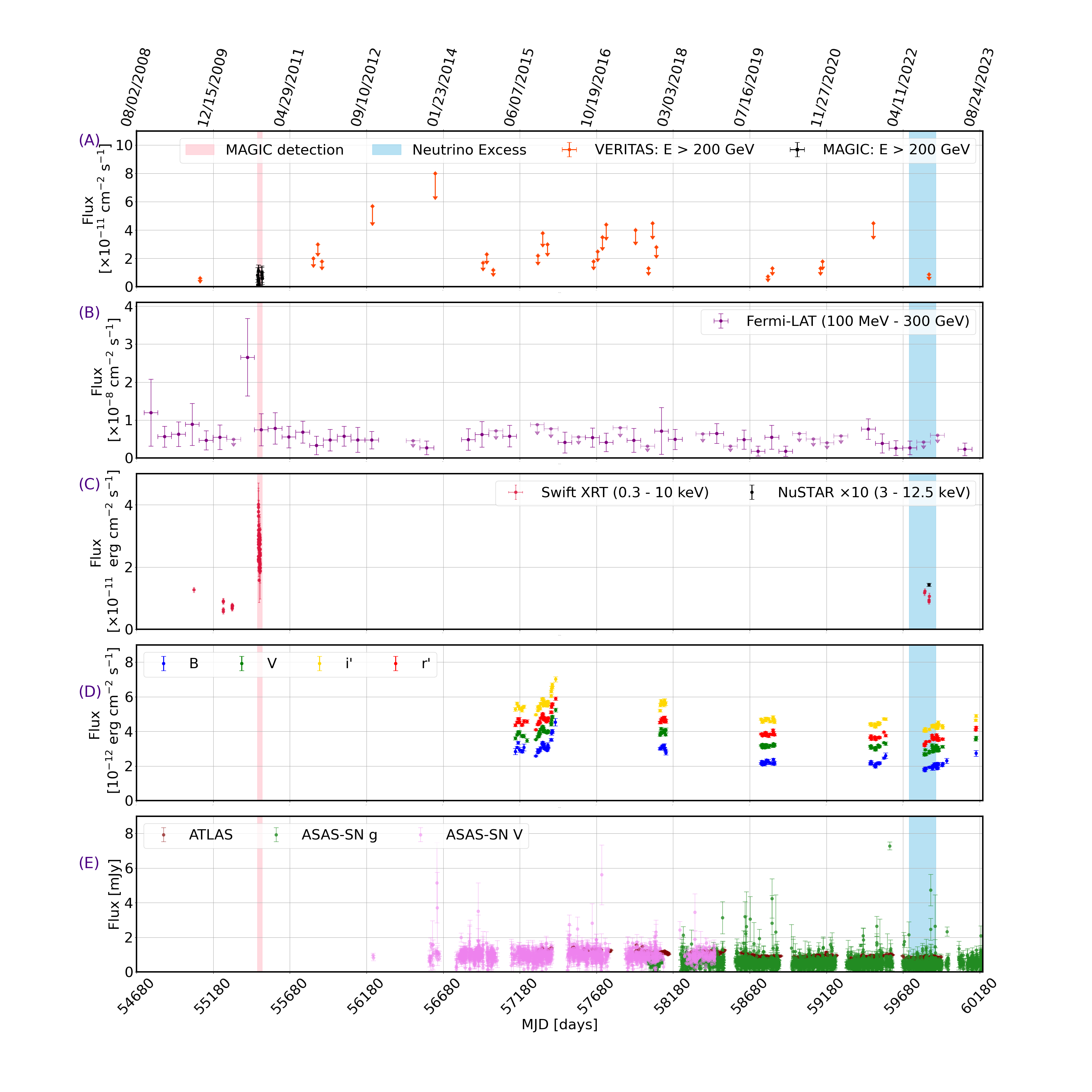}
\caption{Multiwavelength lightcurves of B3 2247+381 between August 4, 2008 until midnight on June 1, 2023 (MJD 54683 -- MJD 60096). The blue narrow band corresponds to the IceCube neutrino excess period and the red narrow band corresponds to the time interval of the MAGIC detection \citep{2012A&A...539A.118A}. \textbf{(A)} Monthly-binned VHE lightcurve for VERITAS observations above an energy threshold of 200 GeV. Also shown for comparison are MAGIC data from the 2010 detection \citep{2012A&A...539A.118A}. \textbf{(B)} 90-day-binned \emph{Fermi}-LAT lightcurve in the 0.1--300 GeV band. Upper limits are shown for individual bins having a significance of lower than $2\sigma$. \textbf{(C)}  \emph{Swift}-XRT and \emph{NuSTAR} fluxes in the 0.3--10 keV band and 3--12.5 keV band, respectively. To aid visual comparison, the \emph{NuSTAR} data have been scaled up by a factor of 10. \textbf{(D)} Optical lightcurve in all filters using data from the 48” optical telescope at the FLWO. Errors plotted are statistical uncertainties only. \textbf{(E)} Optical flux observations: Optical data from the ASAS-SN Sky Patrol in the g and V band  (green and magenta, respectively) and ATLAS R band data (in red).}
\label{fig:full_lc}
\end{figure*}

In order to investigate the temporal variability of the gamma-ray flux of the source during the full time interval, the lightcurve of the integral flux in the energy range 100~MeV--300~GeV, shown in Fig.~\ref{fig:full_lc}, is calculated with 90-day bins and keeping the spectral index fixed to the catalog value. We find no evidence of variability for B3 2247+381 in the long-term \textit{Fermi}-LAT lightcurve (see Section \ref{sec:discuss}) or in the \textit{Fermi}-LAT differential energy spectrum, shown in the multiwavelength SED in Fig.~\ref{fig:sed}, which considers the wider time interval, from MJD~54683 to MJD~60096, for improved statistics. 

\subsection{\nustar{}}
\label{subsec:nustar}

The Nuclear Spectroscopic Telescope Array (\emph{NuSTAR}) is a space-based telescope capable of detecting hard X-ray photons in the energy range between 3--79 keV. \emph{NuSTAR} has an angular resolution of 18” (FWHM) and two focal plane modules (FPMA and FPMB) with a $13' \times 13'$ field of view \citep{2013_NuSTAR}. The \emph{NuSTAR} ToO observations of B3~2247+381 (ObsID = 80802636002, exposure = 40.5 ks) took place on September 26, 2022 (MJD~59848), following the IceCube alert received via MoU. The \emph{NuSTAR} data were processed using \emph{NuSTAR} data analysis software (\texttt{NuSTARDAS}) version~2.1.1 contained within \texttt{HEASOFT} version 6.29.

The source spectra were extracted from a circular region of radius $1'$, chosen to maximize the source significance with respect to the background. The spectra were binned such that there was a 5$\sigma$ significance within each bin. The binned spectra, shown in the multiwavelength SED in Fig.~\ref{fig:sed}, from FPMA and FPMB were simultaneously fitted using \texttt{XSPEC} \citep{xspec} in the energy range between 3 keV and 12 keV, as beyond this the background begins to dominate. B3 2247+381 was strongly detected in the hard X-ray band with \emph{NuSTAR} at a significance of $42\sigma$ during these observations. The \emph{NuSTAR} spectra were well fit by an absorbed power law with the Galactic column density $N_H$ in the direction of B3 2247+381 held fixed at 1.57 $\times 10^{21}$ atoms cm$^{-2}$ \citep{column_density_paper}. The best-fit photon index of the observations was found to be $\Gamma=2.89\pm 0.12$\ and a $\chi^2_{red}= 70.57/63 = 1.12$ was obtained. The time-averaged integral flux of B3 2247+381 measured with \emph{NuSTAR} during the period investigated is $(1.45 \pm 0.05) \times 10^{-12}$~ erg~ cm$^{-2}$~ s$^{-1}$. 
\subsection{\swift{}}
\label{subsec:swift}

The X-Ray Telescope (XRT), onboard the Neil Gehrels \emph{Swift} Observatory, is a grazing-incidence focusing X-ray telescope sensitive to energies from 0.2~keV to 10~keV \citep{2004AIPC..727..637G, 2005_Burrows}. \emph{Swift}-XRT observations of B3 2247+381 were taken for the time interval between May~20, 2022 until midnight on November~10, 2022 (MJD~59719 -- MJD~59893). A total observing time of $\sim$5.5 ks was accumulated during this period in photon-counting (PC) mode. The \emph{Swift}-XRT lightcurve in the energy range  0.3--10 keV was retrieved from the public online tool ``the \swift{}-XRT data products generator"~\footnote{\url{https://www.swift.ac.uk/user_objects/index.php} (accessed on May 3, 2024)} \citep{Evans07, Evans09} and is shown in Fig.~\ref{fig:full_lc}.



The Ultraviolet/Optical Telescope (UVOT), also onboard the Neil Gehrels Swift Observatory, is a photon-counting telescope. It is sensitive to photons having energies ranging roughly between 1.9 eV and 7.3 eV \citep{2005SS_Roming}. \emph{Swift}-UVOT observations are performed in parallel to the \emph{Swift}-XRT measurements in six filters with central wavelengths of $V$ (5468 \AA), $B$ (4392 \AA), $U$ (3465 \AA), $UVW1$ (2600 \AA), $UVM2$ (2246 \AA), and $UVW2$ (1928 \AA). The \emph{Swift}-UVOT data were analyzed with \texttt{HEASOFT} version 6.29. A source region with a radius of $5.0''$ centered on B3~2247+381, and a background region of the same size, away from the blazar and containing no known sources in any band, were used to extract signal and background counts, respectively. The magnitude of the source was then computed using \texttt{uvotsource} and converted to flux using the zero-point for each of the UVOT filters from \citet{Poole08}. 

The measured optical-UV fluxes, obtained from \emph{Swift}-UVOT observations between September~23, 2022 and November~10, 2022 (MJD 59845 - MJD 59893) are shown in the multiwavelength SED in Fig.~\ref{fig:sed}. This includes correcting for interstellar extinction using the approach of \cite{2009_Roming}, assuming a reddening of  $E(B-V) = 0.0167$ \citep{2011_reddenning}. To build a consistent SED, the contribution of the host-galaxy emission has been removed in the UVOT filters, using the R filter and the HIRAC-camera observations of TeV J2250+384 (B3 2247+381) with the Nordic Optical Telescope (NOT) \citep{2003A&A...400...95N}. Assuming the simplest model of \cite{deVaucouleurs_1991}, in conjunction with an elliptical-galaxy template from \texttt{PEGASE.2} \citep{Fioc_1999}, the host estimation in the UVOT bands within a 5" aperture was estimated at V: $2.60 \times10^{-12}$, B: $1.63 \times10^{-12}$, U: $4.92\times10^{-13}$, UVW1: $1.82\times10^{-13}$, UVM2: $1.57\times10^{-13}$, and UVW2: $1.74\times10^{-13}$ erg cm$^{-2}$ s$^{-1}$, respectively.

\subsection{Optical}

The All-Sky Automated Survey for Supernovae (ASAS-SN; \citealt{2014Shappee, 2017Kochanek}) is an automated program aimed at routinely surveying the entire visible sky for bright transient sources with minimal observational bias. It can reach a depth of roughly 17 mag in \emph{V} band and 18.5 mag in \emph{g} band. While ASAS-SN originally comprised two stations, located at the Cerro Tololo International Observatory (CTIO, Chile) and the Haleakala Observatory (Hawaii), respectively, in 2017 three further stations were added, including a second unit at CTIO and one unit each at the McDonald Observatory (Texas) and the South African Astrophysical Observatory. ASAS-SN observations are available in two optical bands: the \emph{V} band,  centered at $\sim$551 nm and corresponding to observations from the two original stations, and the \emph{g} band, centered at $\sim$480 nm and corresponding to observations performed with the three new stations. For the purpose of this study, we obtain B3 2247+381 observations from the ASAS-SN Sky Patrol\footnote{\url{https://asas-sn.osu.edu/} (accessed on May 3, 2024)} for the time interval between August 4, 2008 until midnight on June 1, 2023 (MJD~54683 -- MJD~60096), and shown in Fig.~\ref{fig:full_lc}.

The Asteroid Terrestrial-impact Last Alert System (ATLAS; \cite{ATLAS_main}) is a high cadence all-sky survey system comprising four independent units, one each at Haleakala and Mauna Loa in the Hawaiian islands in the~Northern~Hemisphere and one each at the El Sauce Observatory, Chile and the South African Astronomical Observatory in the Southern Hemisphere. ATLAS is optimized to be efficient for finding potentially dangerous asteroids, as well as in tracking and searching for highly variable and transient sources, such as AGN. Optical observations of B3 2247+381 with ATLAS in the \emph{R} band, centered at $\sim$679~nm and having a typical cadence of one data point per two days, are shown in Fig.~\ref{fig:full_lc}.

Images of B3 2247+381 have also been collected with the 48” optical telescope at the FLWO \footnote{\url{http://www.sao.arizona.edu/FLWO/48/48.html} (accessed on May 3, 2024)} as part of an ongoing study of more than seventy blazars that aims to track and analyze their behavior over time. These data were collected using SDSS r$^\prime$, SDSS i$^\prime$, Harris V, and Harris B filters. Images are processed by an automated pipeline that uses aperture photometry to determine magnitudes. For each image taken at FLWO, a measurement is made of the blazar’s magnitude as well as the magnitude of its check star, which is used to establish and maintain the quality of photometric data. Additional quality is enforced by removing outliers. A total of 517 images were collected over 135 nights, between May 2015 and July 2023. An optical lightcurve of B3 2247+381, showing the resulting 453 good quality measurements in the different bands with the 48” optical telescope, is presented in Fig.~\ref{fig:full_lc}. The data have not been corrected for optical extinction or any potential host galaxy contribution.

\section{Discussion}
\label{sec:discuss}

The SEDs of blazars typically display two broad emission features. While the first feature in the radio to X-ray waveband is commonly attributed to synchrotron emission from relativistic electrons and positrons within the jet, the origin of the second feature in the X-ray to gamma-ray band is less clear. In leptonic models (for example \citealt{1974_Jones, 1994_Sikora}), this emission is attributed to the IC scattering between the energetic leptons in the jet and a field of lower energy photons. These may be the same synchrotron photons produced by the leptons (synchrotron self-Compton [SSC] model) in the case of BL Lacertae objects like B3 2247+381, or photons from an external source (external inverse Compton [EIC] model) in the case of Flat Spectrum Radio Quasars. The sources of external photons include the accretion disk, the broad-line region, and the molecular torus.

The connection between the parent cosmic-ray proton energy and the energies of the neutrinos and photons produced in their interactions is not straightforward nor fully understood. While it is expected that a neutrino would carry approximately $5\%$ of the parent proton energy, Doppler boosting in the jet can lead to the neutrino gaining back energy in the observer frame. This can result in a neutrino with an energy of ~100 TeV in the observer frame being produced by protons with maximum energies in the jet frame as low as ~100 TeV (depending on the Doppler boost factor). It is anyway possible for the maximum proton energy to be much higher, with some models (for example \citealt{2019MNRAS.483L..12C}) deriving maximum proton energies above $10^{17}$ eV.

As the medium is expected to be opaque to the TeV-PeV pionic photons being produced, the electromagnetic cascading of these gamma rays results in some, or all, of their energy being redistributed into lower energy observable bands (keV to MeV), posing an additional challenge to deriving proton energy information from electromagnetic and neutrino observations. Additionally, disentangling the hadronic contribution to the SED is particularly challenging as it is likely subdominant when compared to the main leptonic component (for example \cite{2018_Keivani}), unless the proton energies are well above the PeV range (for example \citealt{2019MNRAS.483L..12C}). The modeling of these different contributions and their connections is an active area of research in the blazar modeling community.

\cite{2019_Gao} demonstrate that a moderate enhancement in the number of cosmic rays during a flare leads to a proportional rise in the neutrino flux, which is accompanied by hard X-rays and TeV gamma rays. Therefore, a multi-messenger approach combining both neutrino and electromagnetic observations is vital for neutrino source identification. Moreover, hard X-ray observations from \emph{NuSTAR}, as well as gamma-ray observations at TeV energies have provided some of the strongest constraints on hadronic emission models in blazars \citep{2023_Qi}. In cases where gamma-rays and neutrinos cannot be linked with a simple model, or VHE radiation is not observed, other models can be used that associate neutrinos with hard X-rays cascading from gamma-ray interactions \citep{icecube_hard_x_ray}. Such “hybrid models” combine a leptonic explanation of the observed emission with a hadronic component, the strength of which is dictated by the X-ray observations \citep{2019_Gao}. Finally, in addition to SEDs, \cite{2019_Gao} also highlight the importance of hard X-ray and gamma-ray variability since the strongest temporal correlations are expected between these two bands with neutrinos.


We characterize the gamma-ray variability of B3 2247+381 by evaluating the goodness-of-fit parameter for a constant value function fit to the lightcurve. As seen in panel (B) of Fig.~\ref{fig:full_lc}, the \textit{Fermi}-LAT lightcurve is very well fit by a constant function at a flux value of ~($3.2 \pm 0.3)~\times 10^{-9}$~$\text{cm}^{-2}\text{s}^{-1}$, with a $\chi^{2}/\text{n.d.f} = 40.8/52 = 0.8$, which suggests the absence of any GeV variability in the source for the time interval investigated. 
Furthermore, a $\chi^{2}/\text{n.d.f} = 152/52 = 2.9$ was obtained when testing for variability using the  \textit{$\text{ts}_{\text{var}}$} routine in \texttt{FERMIPY}, and also accounts for the upper limits in the lightcurve. The alternate hypothesis that a source's lightcurve is not constant, \textit{$\text{ts}_{\text{var}}$}, as defined in equation 4 of \cite{2fgl} is distributed as as a $\chi^2$ distribution with 23 degrees of freedom, where \textit{$\text{ts}_{\text{var}}$} $> 41.6$ corresponds to a variable source with 99\% confidence. For comparison, no variability was observed for the source during the MAGIC detection, where a fit of a constant function to the \textit{Fermi}-LAT flux points yielded a flux value of ~$(3.7 \pm 0.5)~\times 10^{-9}$~$\text{cm}^{-2}\text{s}^{-1}$, with a $\chi^{2}/\text{n.d.f} = 0.7$ \citep{2012A&A...539A.118A}. 

There is evidence of variability on monthly timescales in the long-term X-ray data, as seen in the increased X-ray flux measured by \textit{Swift}-XRT in Figure \ref{fig:full_lc}. This lightcurve is poorly fit by a constant function, corresponding to a $\chi^{2}/\text{n.d.f} = 42.2$. It should be noted that there were no \emph{Swift}-XRT observations of B3 2247+381 for the time period between the 2010 MAGIC detection and the IceCube alert, making it difficult for a detailed study of the long-term variability. Nevertheless, there was an increase in the X-ray flux state of B3~2247+381 corresponding to the MAGIC detection, with an average photon count rate of $0.58~\pm~0.11$~cts/s, compared to $0.23~\pm~0.03$~cts/s for the period corresponding to the neutrino alert. For comparison, no strong X-ray intra-night variability was found by \cite{2012A&A...539A.118A} during the MAGIC detection.

The optical data of B3 2247+381 show stronger evidence of variability, with the ASAS-SN \emph{g} observations in particular showing an increase in flux during the time interval corresponding to the neutrino alert as seen in  Fig.~\ref{fig:full_lc}. All three optical lightcurves are fit by a constant function, with $\chi^{2}/\text{n.d.f}$ of 2.04, 1.94 and 1.61 for the ATLAS, ASAS-SN~\emph{g} and ASAS-SN \emph{V} data respectively. The time-averaged optical flux measured with ATLAS in the \emph{R} band for the entire observation period was $0.95 \pm  0.05$ mJy. For comparison, B3 2247+381 was observed to be in a steady state of $\sim 1.8$ mJy between 2006--2009 in the \emph{R} band by the Tuorla group\footnote{\url{https://users.utu.fi/kani/1m/index.html} (accessed on May 3, 2024)}, using the 35~cm telescope at the KVA observatory on La Palma, Canary Islands, Spain before the optical flare in late summer in 2010, when it reached an average flux of 2.4 mJy \citep{2012A&A...539A.118A}. The FLWO B,V,r$^\prime$ and i$^\prime$ data in Fig.~\ref{fig:full_lc} also support the conclusion that B3 2247+381 was not in a particularly bright optical state at the time of the neutrino alert.

\begin{figure*}[h]
\centering
\includegraphics[width=0.99 \linewidth]{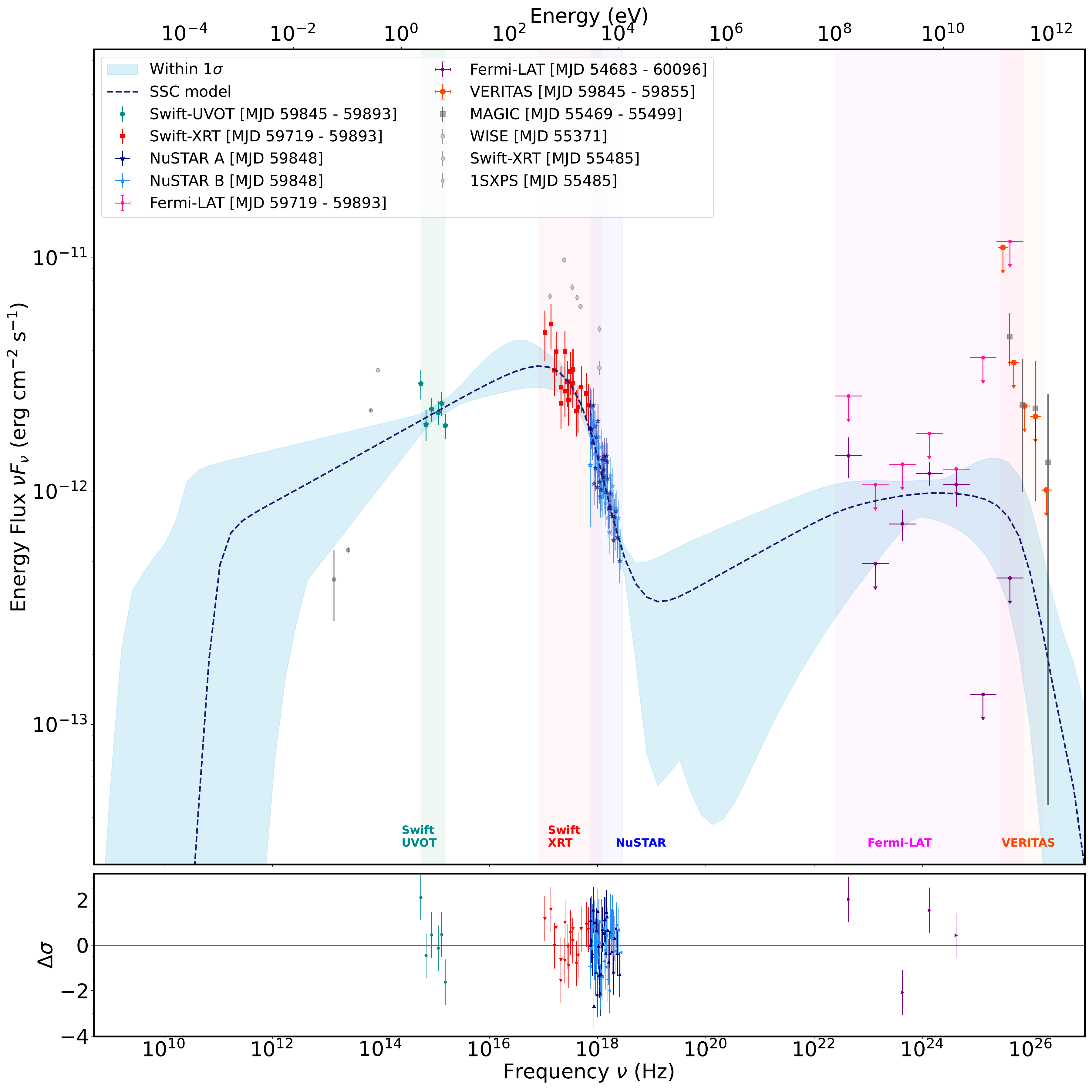}
\caption{The broadband SED of B3~2247+381. The VERITAS spectral limits are time-averaged over the time interval of the ToO observations between September~23, 2022 and October~3, 2022 (MJD 59845 -- 59855). The \emph{Fermi}-LAT spectrum, time-averaged over a wider time interval between August 4, 2008 and June 1, 2023 (MJD 54683 -- 60096), is shown using purple points while those corresponding to the IceCube neutrino alert window between May~20, 2022 and November~10, 2022 (MJD 59719 -- 59893) is shown using pink points. The \emph{NuSTAR} spectra were obtained on September~26, 2022 (MJD 59848). The \emph{Swift}-UVOT and \emph{Swift}-XRT observations correspond to the time interval between September~23, 2022 and November~10, 2022 (MJD 59845 -- 59893).  The grey points represent archival data, provided by the Space Science Data Center--ASI, taken with Wide-field Infrared Survey Explorer (\emph{WISE}) and the \emph{Swift}-XRT. It should be noted that none of these archival data are considered in the fitting, as we consider only contemporaneous datasets. The MAGIC data from the 2010 detection \citep{2012A&A...539A.118A} are shown with gray square markers. A minimum significance of two standard deviations is required for each flux point and upper limits at 99\% confidence level are quoted otherwise. The blue curve and shaded region represents the one-zone synchrotron and SSC radiation model fit and the corresponding $1 \sigma$ confidence interval respectively, obtained using \texttt{Bjet\_MCMC}, with the residuals for each SED point considered shown in the bottom panel, and the parameters shown in Table~\ref{tab:table1}. The \texttt{Bjet\_MCMC} shows the SED including the effect of the extragalactic background light (EBL) using the \cite{2017A&A...603A..34F} model. The VERITAS upper limits have therefore not been corrected for EBL absorption.}
\label{fig:sed}
\end{figure*}

It should be noted that we cannot conclusively claim nor rule out any associations between the neutrino multiplet and the blazar, based on these observations alone. In the case of previous IceCube neutrino alerts with possible blazar counterparts, TXS 0506+056 \citep{2022ApJ...927..197A} and PKS 0735+178 \citep{2023_Qi}, flaring states were detected through multiwavelength observations that were temporally coincident with the neutrino alerts, strengthening the statistical significance of the correlation between the neutrino and electromagnetic emission. Nevertheless, we still aim to shed light on the mechanisms responsible for the emission from B3 2247+381 during the time interval investigated by modeling the multiwavelength SED. The multiwavelength broadband SED of B3 2247+381, shown in Fig.~\ref{fig:sed}, comprises the upper limits obtained with VERITAS, along with observations with \emph{NuSTAR} and optical telescopes. Although these observations were not conducted exactly simultaneously, the lack of observed variability on nightly timescales in any of these wavelengths allows for the combination of these observations into a broadband SED.
The \emph{Fermi}-LAT SED points considered are time-averaged over a wider interval due to low statistics and no apparent variability in the gamma-ray flux from MJD 54683 to MJD 60096.
The synchrotron component is loosely constrained by the relatively flat optical/UV spectra, while the hard X-ray spectra observed with \emph{NuSTAR} is found to occur in the low-energy component of the SED, as opposed to characterizing the transition from low- to high-energy, as seen in the blazar PKS 0735+178 \citep{2023_Qi}, with the transition occurring well above 10~keV.

In order to interpret the multiwavelength behavior of B3 2247+381, we model the broadband observations using \texttt{Bjet\_MCMC}\footnote{\url{https://github.com/Ohervet/Bjet_MCMC} (accessed on May 3, 2024)} \citep{2023_Bjet}, a new public tool that can simply fit the broadband SEDs of blazars. \texttt{Bjet\_MCMC} has previously been used to model several jetted AGN emitting in the VHE band, such as AP Librae \citep{refId0}, HESS J1943+213 \citep{Archer_2018}, 1ES 1215+303 \citep{Valverde_2020}, and also PKS 1222+216 and TON~599 \citep{Adams_2022}. We consider a simple one-zone pure SSC radiation model consisting of a single blob with a broken power law differential electron energy density (EED), a scenario most commonly used to model BL Lacertae objects such as B3 2247+381.We expect a SSC contribution to blazar SEDs, regardless of whether or not the IceCube alert originates from astrophysical neutrinos associated with B3 2247+381. If the neutrino emission indeed originates from B3 2247+381, we would expect an additional hadronic component to the SED, therefore a simple one-zone SSC model would not be able to fully explain the observed SED, as was seen in the case of \cite{2023_Qi}. The SED fit is performed using the Markov-chain Monte-Carlo (MCMC) method. 

The pure SSC model contains by default nine free parameters, including three describing the blob, namely the Doppler factor, $\delta$, the magnetic-field strength, $B$, and the blob radius, $R$, and six parameters which define the EED, namely the particle density factor, $N_{e}$, the two spectral indices of the spectrum, $n_{1}$ and $n_{2}$, the break Lorentz factor of the electron distribution, $\gamma_{\text{break}}$, and the minimum and maximum Lorentz factors of the EED, $\gamma_{\text{min}}$ and $\gamma_{\text{max}}$, respectively. The redshift is fixed $z~=~0.119$ \citep{1998ApJ...494...47F} and a flat $\Lambda$CDM cosmology is assumed \citep{Bennett_2014}. This choice of model also allows for comparison with the parameters obtained during the 2010 MAGIC detection \citep{2012A&A...539A.118A}. We consider a wide range of parameters in order to make as few assumptions as possible, and many parameters are in log scale to allow the walkers to move through multiple orders of magnitude.

\begin{table}
\centering
\caption{The fit parameters and the parameter range corresponding to the $1\sigma$ confidence level of a simple one-zone synchrotron and SSC radiation model fitted to the multiwavength broadband SED of B3 2247+381. Also tabulated is the redshift, $z~=~0.119$, which is a fixed value during model fitting \citep{1998ApJ...494...47F}. The model is shown in Fig.~\ref{fig:sed}. Also shown for comparison are the parameters obtained during the 2010 MAGIC detection \citep{2012A&A...539A.118A}.}
\label{tab:table1}
\begin{tabular}{lccccr} 
\hline
Parameter &Description &Scale &Parameters &$1\sigma$ range & MAGIC  \cr
\hline
 ${\text{B [G]}}$ &magnetic field strength &$\text{log}_{10}$   &-2.0  &[-4, -0.7] &-1.2 \cr
 $N_{e} [\text{cm}^{-3}]$  &particle density factor &$\text{log}_{10}$  &6.0  &[3.5, 7.3] &3.3\cr
 $\gamma_{\text{min}}$ &low-energy cutoff &$\text{log}_{10}$  &0.8  &[0, 2.4] &3.5 \cr
 $\gamma_{\text{max}}$ &high-energy cutoff &$\text{log}_{10}$  &6.9  &[5.1, 7.3] &5.8 \cr
 $\gamma_{\text{break}}$ &energy break &$\text{log}_{10}$  &5.5   &[4.9, 6.6] &4.9\cr
 $n_{1}$ &first index &linear  &2.75  &[2.68, 2.79] &2.0  \cr
 $n_{2}$ &second index &linear  &5.65  &[4.86, 6.55] &4.4 \cr
 $\delta$ &Doppler factor &linear  &73  &[38, 98] &35 \cr
 $ R  [\text{cm}]$ &blob radius &linear &$9.4 \times 10^{15}$ &[$4.4 \times 10^{14}$, $1.9 \times 10^{18}$]   &$8 \times 10^{15}$ \cr
 $z$ &redshift &linear &0.119 &--  &0.119  \cr
 \hline
\end{tabular}
\end{table}

For this model, we considered 100 walkers, 5000 steps, and a burn-in phase of 1000 steps and the MCMC walkers display a good general $\chi^{2}$ convergence. Fig.~\ref{fig:walkers} (see \autoref{sec:appendix}) shows the \texttt{Bjet\_MCMC} $\chi^{2}$ values of walkers at each step for the SED fit for all walkers, smallest $\chi^{2}$ and median $\chi^{2}$ respectively. Fig.~\ref{fig:corner} (see \autoref{sec:appendix}) shows a corner plot of the posterior probability distribution of the free parameters from the SED fit.
The values of the parameters used in the SED fit are reported in Table~\ref{tab:table1}, and also shown for comparison are the parameters used to fit the SED during the MAGIC detection \citep{2012A&A...539A.118A}. The fitted model returned a $\chi^{2}$ = 95.5 with 84 degrees of freedom, corresponding to $\chi^{2}_{\text{red}} = 1.14$. 

Most of the parameters are found to have good agreement with the values obtained in \cite{2012A&A...539A.118A} and typical values seen for BL Lacertae objects \citep{2010MNRAS.401.1570T}. The \texttt{Bjet\_MCMC} SED fit investigates wide ranges for all parameters \citep{2023_Bjet}, and probes outside the parameter range of \cite{2012A&A...539A.118A}. Moreover, the \cite{2012A&A...539A.118A} fitting algorithm acts primarily on the electron normalization, source radius and Doppler factor, with only slight changes to the other parameters. However, B3 2247+381 is relatively weak in the \emph{Fermi}-LAT energy range, and as a result no statistically significant flux variability could be observed in timescales of days to months. It should be noted that large uncertainties obtained for the parameters prevent us from drawing any strong conclusions.  Finally, the TeV non-detection severely limits a detailed study of the IC component of the SED.

As there is no evidence of association between the neutrino flare and the blazar, we remain agnostic about the origin of the gamma-ray emission. The one-zone leptonic scenario seems to reasonably fit the SED, which would justify the absence of an association of an increase in multi-wavelength emission during the time of the neutrino alert.
In this paper, we briefly explore the viability of the neutrino emission, assuming the alerts are associated with B3~2247+381. In a hadronic emission scenario, a significant number of accelerated protons or ions may exist in the jet and neutrino emission may occur through two different channels, of which p$\gamma$  interactions \citep{1995PhR...258..173G} are more likely. Neutrino emission through the $pp$ channel would require very large particle density in the jet and hence a large jet power \citep{2000A&A...354..395P}.

As seen in Fig.~\ref{fig:sed}, the VERITAS upper limits are not constraining and since there is no direct evidence for a cut-off in the GeV gamma-ray spectrum observed by \textit{Fermi}-LAT, we can infer the maximum neutrino flux that provides a VHE gamma-ray non-detection in a hadronic scenario. Assuming no absorption of the gamma rays due to pair production in the broad and narrow line regions, the VERITAS upper limits can be used to make an approximation of the possibility of neutrino production, as the EBL absorption is relatively low at the redshift $z~=~0.119$ \citep{1998ApJ...494...47F}. 
Using the simulated response of IceCube, with GFU event selection to a neutrino flux, the observed event excess (12.7 events with a best fit neutrino spectral index of -2.33 over a 174.0 day time period) would correspond a neutrino flux of
\begin{equation}
E F_\nu (E) \approx 5.8\times 10^{-11} \ \mathrm{erg\,cm^{-2}\,s^{-1}}.
\label{nuflux}
\end{equation}
Furthermore, the target photons that are required for pion production would be observed at $\sim$ 100 keV, above the energies covered by \textit{NuSTAR} observations, but the flux in the SSC model is low, which mandates a high proton source power. 
Given the significance of the neutrino excess at 3$\sigma$ level (uncorrected for trials), and that B3 2247+381 is an unlikely source of the neutrino excess, we conclude that the neutrino excess is likely caused by a fluctuation in the irreducible atmospheric neutrino background.
Furthermore, the cascade emission of charged and neutral pions in the 100 TeV energy band provides a gamma-ray flux constraint at a few hundred GeV, similar to the neutrino flux \citep{2017ApJ...843..109G}. The upper limits obtained with VERITAS are a factor $\sim$ 20 lower and, assuming no gamma-ray absorption above 100 GeV, the expected rate of neutrinos would be around $\sim$ 0.05 per month and any association would likely be a chance coincidence.

\section{Conclusions} 
\label{sec:conclusions}

In this paper, we report on the VERITAS and multiwavelength follow-up observations of B3 2247+381, triggered by a Gamma-ray Follow-Up alert received by private communication from IceCube. The report contained information on alerts for a cluster of muon neutrino candidate events from directions compatible with B3 2247+381 received at a significance of $> 3 \sigma$  between May~20, 2022 and November~10, 2022. While these observations did not yield a detection with VERITAS, B3 2247+381 was, for the first time, significantly detected in the hard X-ray band with \emph{NuSTAR}. 

The broadband SED of B3 2247+381 comprising the upper limits obtained with VERITAS, as well as extensive follow-up observations across the electromagnetic spectrum, in particular the time-averaged \emph{Fermi}-LAT data, and contemporaneous \emph{Swift} and \emph{NuSTAR} observations were investigated. We considered a simple one-zone synchrotron and SSC radiation model consisting of a single blob accelerating electrons with a broken power law electron energy density. A good agreement was seen between the fitted model and the multiwavelength spectral data over the entire energy range considered, and this broadly agrees with the results presented in \cite{2012A&A...539A.118A}.

Furthermore, this investigation also serves as one of the first VERITAS follow-ups to an alert within the framework of the GFU program. It should be noted that while the primary objective of the GFU program is to notify VERITAS and other instruments in the collective observatory network of potential astrophysical neutrino flares from directions compatible with known gamma-ray emitters, the duration of the flares is not predetermined and can span from seconds to the order of months. In this study, we report no significant changes to the gamma-ray emission state of the source despite the significant detection in the hard X-ray regime for the period investigated. Having just a spatial correlation alone makes this source different from previously studied sources with VERITAS, namely TXS 0506+056 \citep{2018ApJ...861L..20A} and PKS 0735+178 \citep{2023_Qi}, both of which coincided both spatially and temporally with a gamma-ray flaring state. 
In looking for spatial coincidence between the neutrino multiplet and gamma-ray sources, we also note that, unlike TXS 0506+056 and PKS 0735+178, which were independently associated with IceCube alerts, B3 2247+381 was a-priori selected for GFU monitoring, which includes already known or likely VHE gamma-ray emitters. The two situations are therefore different.

Moreover, assuming the neutrinos originate from hadronic processes within the jet, the neutrino flux would be accompanied by a photon flux from the cascade emission, and the integrated photon flux required in such a case would significantly exceed the total multiwavelength fluxes and the VERITAS upper limits presented here. Therefore, due to no evidence of flaring activity with VERITAS, combined with the low multiwavelength fluxes, B3~2247+381 is disfavored as the origin of the IceCube multiplet. The neutrino hotspot seen with IceCube may still be of astrophysical origin and further multiwavelength observations of B3~2247+381 and a better understanding of instrument uncertainties might allow for a potential study of multi-zone contributions along the jet and confirm whether or not hadronic contributions and variability can be entirely ruled out.
More generally, future study of  AGN potentially associated with neutrino alerts, in particular those also coincident with flaring periods that have improved statistics, are required to draw further conclusions. For example, the viability of neutrino emission and its possible implications on the multiwavelength SED are investigated for PKS 0735+178 in \cite{2023_Qi}. 

This study also highlights some of the challenges in searching for neutrino-emitting blazars, including the limited localization precision of the IceCube Observatory and the effect of poor weather conditions on IACT observations. Furthermore, the large number of gamma-ray blazars as potential counterparts makes the association between neutrino events and a gamma-ray blazar difficult. As in this study, the electromagnetic emission can often be explained by invoking leptonic models alone without any further need for an hadronic component. Moreover, the jet power and the proton luminosity parameters required in lepto-hadronic models are often too high, and exceed the Eddington limit, for a short period of activity. 

These challenges can only be addressed with further multiwavelength follow-up investigations of flaring blazars in temporal and spatial coincidence with astrophysical neutrino alerts, both single high energy neutrino events and those within the framework of the GFU program. Furthermore, as seen in \cite{IceCube:2023kyd}, such studies are even more effective when we combine observations from multiple IACTs in order to provide a more complete coverage of the entire sky. This also helps to account for cases where the visibility of a source from a single observatory site is adversely affected due to factors such as bad weather, the presence of the Sun or Moon, or technical problems. 

Despite a non-detection at VHE wavelengths, this observation campaign will inform future IceCube GFU follow-up campaigns with VERITAS. We have raised our GFU alert threshold to trigger on neutrino events with a significance $>3.5\sigma$ and will continue to monitor the alerts for a further increase in significance. A rising neutrino alert significance would indicate that the alert location is continuing to emit high energy neutrinos. In contrast, in the case of the alert at the B3 2247+381 location, the initial rise in significance is due to a single event, which then continues to boost the cumulative significance of subsequent events of lower energy and significance, without further rise in significance. Continuing to observe GFU alerts with IACTs, such as VERITAS, provides important opportunities to explore the viability of blazars as neutrino sources. Since we expect hadronic emission mechanisms to produce VHE gamma rays along with IceCube-detected neutrinos, VHE SED points or upper limits are important for constraining whether or not a hadronic component is necessary to model the multiwavelength blazar emission, which would motivate a physical connection with the observed neutrinos. VERITAS receives 1-2 GFU events per observing season, which allows for sufficient observations to be conducted to detect the levels of flaring activity expected from the blazars that have been selected for GFU targets.

Furthermore, future neutrino detectors such as IceCube-Gen2 \citep{2023arXiv230809427I}, KM3NeT (for example \cite{2024arXiv240208363A}), P-ONE \citep{Twagirayezu:2023Sd}, and others will strongly improve the all-sky sensitivity to high energy neutrino events, as their acceptances are complementary to IceCube's.
Finally, neutrino follow-up efforts remain a major focus of the wider multi-messenger community (for example \cite{2022icrc.confE.960T, 2024A&A...687A..59G, XRT_IC220303A, 2023arXiv230906164A}) and will hopefully result in directly constraining the century-old puzzle of the origin of cosmic rays in the future.



\begin{acknowledgments}
\section*{Acknowledgments}

This research is supported by grants from the U.S. Department of Energy Office of Science, the U.S. National Science Foundation and the Smithsonian Institution, by NSERC in Canada, and by the Helmholtz Association in Germany. This research used resources provided by the Open Science Grid, which is supported by the National Science Foundation and the U.S. Department of Energy's Office of Science, and resources of the National Energy Research Scientific Computing Center (NERSC), a U.S. Department of Energy Office of Science User Facility operated under Contract No. DE-AC02-05CH11231. We acknowledge the excellent work of the technical support staff at the Fred Lawrence Whipple Observatory and at the collaborating institutions in the construction and operation of the instrument.

The authors gratefully acknowledge the support from the following agencies and institutions:
USA {\textendash} U.S. National Science Foundation-Office of Polar Programs,
U.S. National Science Foundation-Physics Division,
U.S. National Science Foundation-EPSCoR,
U.S. National Science Foundation-Office of Advanced Cyberinfrastructure,
Wisconsin Alumni Research Foundation,
Center for High Throughput Computing (CHTC) at the University of Wisconsin{\textendash}Madison,
Open Science Grid (OSG),
Partnership to Advance Throughput Computing (PATh),
Advanced Cyberinfrastructure Coordination Ecosystem: Services {\&} Support (ACCESS),
Frontera computing project at the Texas Advanced Computing Center,
U.S. Department of Energy-National Energy Research Scientific Computing Center,
Particle astrophysics research computing center at the University of Maryland,
Institute for Cyber-Enabled Research at Michigan State University,
Astroparticle physics computational facility at Marquette University,
NVIDIA Corporation,
and Google Cloud Platform;
Belgium {\textendash} Funds for Scientific Research (FRS-FNRS and FWO),
FWO Odysseus and Big Science programmes,
and Belgian Federal Science Policy Office (Belspo);
Germany {\textendash} Bundesministerium f{\"u}r Bildung und Forschung (BMBF),
Deutsche Forschungsgemeinschaft (DFG),
Helmholtz Alliance for Astroparticle Physics (HAP),
Initiative and Networking Fund of the Helmholtz Association,
Deutsches Elektronen Synchrotron (DESY),
and High Performance Computing cluster of the RWTH Aachen;
Sweden {\textendash} Swedish Research Council,
Swedish Polar Research Secretariat,
Swedish National Infrastructure for Computing (SNIC),
and Knut and Alice Wallenberg Foundation;
European Union {\textendash} EGI Advanced Computing for research;
Australia {\textendash} Australian Research Council;
Canada {\textendash} Natural Sciences and Engineering Research Council of Canada,
Calcul Qu{\'e}bec, Compute Ontario, Canada Foundation for Innovation, WestGrid, and Digital Research Alliance of Canada;
Denmark {\textendash} Villum Fonden, Carlsberg Foundation, and European Commission;
New Zealand {\textendash} Marsden Fund;
Japan {\textendash} Japan Society for Promotion of Science (JSPS)
and Institute for Global Prominent Research (IGPR) of Chiba University;
Korea {\textendash} National Research Foundation of Korea (NRF);
Switzerland {\textendash} Swiss National Science Foundation (SNSF).

This work has made use of data from the Asteroid Terrestrial-impact Last Alert
System (ATLAS) project. The Asteroid Terrestrial-impact Last Alert System
(ATLAS) project is primarily funded to search for near earth asteroids through
NASA grants NN12AR55G, 80NSSC18K0284, and 80NSSC18K1575; byproducts of
the NEO search include images and catalogs from the survey area. This work was
partially funded by Kepler/K2 grant J1944/80NSSC19K0112 and HST GO-15889,
and STFC grants ST/T000198/1 and ST/S006109/1. The ATLAS science products
have been made possible through the contributions of the University of Hawaii
Institute for Astronomy, the Queen’s University Belfast, the Space Telescope
Science Institute, the South African Astronomical Observatory, and The
Millennium Institute of Astrophysics (MAS), Chile.
Part of this work is based on archival data, software or online services provided by the Space Science Data Center - ASI.

E.S-S's work at Barnard College was supported by grants: NASA 80NSSC23K0239 and NSF PHY 120096.

\end{acknowledgments}




\bibliography{bib_all}{}

\begin{thebibliography}{}
\expandafter\ifx\csname natexlab\endcsname\relax\def\natexlab#1{#1}\fi
\providecommand{\url}[1]{\href{#1}{#1}}
\providecommand{\dodoi}[1]{doi:~\href{http://doi.org/#1}{\nolinkurl{#1}}}
\providecommand{\doeprint}[1]{\href{http://ascl.net/#1}{\nolinkurl{http://ascl.net/#1}}}
\providecommand{\doarXiv}[1]{\href{https://arxiv.org/abs/#1}{\nolinkurl{https://arxiv.org/abs/#1}}}

\bibitem[{{Aartsen} {et~al.}(2013){Aartsen}, {Abbasi}, {Abdou}, {Ackermann}, {Adams}, {et~al.}}]{IceCube13}
{Aartsen}, M.~G., {Abbasi}, R., {Abdou}, Y., {et~al.} 2013, Science, 342, 1242856, \dodoi{10.1126/science.1242856}

\bibitem[{Aartsen {et~al.}(2014)Aartsen, Abbasi, Ackermann, Adams, Aguilar, Ahlers, Altmann, Arguelles, Auffenberg, Bai, Baker, Barwick, Baum, Bay, Beatty, Tjus, Becker, BenZvi, Berghaus, Berley, Bernardini, Bernhard, Besson, Binder, Bindig, Bissok, Blaufuss, Blumenthal, Boersma, Bohm, Bose, Böser, Botner, Brayeur, Bretz, Brown, Bruijn, Casey, Casier, Chirkin, Christov, Christy, Clark, Classen, Clevermann, Coenders, Cohen, Cowen, Silva, Danninger, Daughhetee, Davis, Day, Clercq, Ridder, Desiati, de~Vries, de~With, DeYoung, D{\'{\i} }az-V{\'{e}}lez, Dunkman, Eagan, Eberhardt, Eichmann, Eisch, Euler, Evenson, Fadiran, Fazely, Fedynitch, Feintzeig, Feusels, Filimonov, Finley, Fischer-Wasels, Flis, Franckowiak, Frantzen, Fuchs, Gaisser, Gallagher, Gerhardt, Gladstone, Glüsenkamp, Goldschmidt, Golup, Gonzalez, Goodman, G{\'{o}}ra, Grandmont, Grant, Gretskov, Groh, Gro{\ss}, Ha, Ismail, Hallen, Hallgren, Halzen, Hanson, Hebecker, Heereman, Heinen, Helbing, Hellauer, Hickford, Hill, Hoffman, Hoffmann, Homeier,
  Hoshina, Huang, Huelsnitz, Hulth, Hultqvist, Hussain, Ishihara, Jackson, Jacobi, Jacobsen, Jagielski, Japaridze, Jero, Jlelati, Kaminsky, Kappes, Karg, Karle, Kauer, Kelley, Kiryluk, Kläs, Klein, Köhne, Kohnen, Kolanoski, Köpke, Kopper, Kopper, Koskinen, Kowalski, Krasberg, Kriesten, Krings, Kroll, Kunnen, Kurahashi, Kuwabara, Labare, Landsman, Larson, Lesiak-Bzdak, Leuermann, Leute, Lünemann, Mac{\'{\i}}as, Madsen, Maggi, Maruyama, Mase, Matis, McNally, Meagher, Merck, Meures, Miarecki, Middell, Milke, Miller, Mohrmann, Montaruli, Morse, Nahnhauer, Naumann, Niederhausen, Nowicki, Nygren, Obertacke, Odrowski, Olivas, Omairat, O{\textquotesingle}Murchadha, Paul, Pepper, de~los Heros, Pfendner, Pieloth, Pinat, Posselt, Price, Przybylski, Quinnan, Rädel, Rameez, Rawlins, Redl, Reimann, Resconi, Rhode, Ribordy, Richman, Riedel, Robertson, Rodrigues, Rott, Ruhe, Ruzybayev, Ryckbosch, Saba, Sander, Santander, Sarkar, Schatto, Scheriau, Schmidt, Schmitz, Schoenen, Schöneberg, Schönwald, Schukraft, Schulte,
  Schulz, Seckel, Sestayo, Seunarine, Shanidze, Sheremata, Smith, Soldin, Spiczak, Spiering, Stamatikos, Stanev, Stanisha, Stasik, Stezelberger, Stokstad, Stö{\ss}l, Strahler, Ström, Strotjohann, Sullivan, Taavola, Taboada, Tamburro, Tepe, Ter-Antonyan, Te$\lbrace${\v{s}}$\rbrace$i{\'{c}}, Tilav, Toale, Tobin, Toscano, Tselengidou, Unger, Usner, Vallecorsa, van Eijndhoven, Overloop, van Santen, Vehring, Voge, Vraeghe, Walck, Waldenmaier, Wallraff, Weaver, Wellons, Wendt, Westerhoff, Whelan, Whitehorn, Wiebe, Wiebusch, Williams, Wissing, Wolf, Wood, Woschnagg, Xu, Xu, Yanez, Yodh, Yoshida, Zarzhitsky, Ziemann, Zierke, \& Zoll}]{Aartsen_2014}
Aartsen, M.~G., Abbasi, R., Ackermann, M., {et~al.} 2014, Journal of Instrumentation, 9, P03009, \dodoi{10.1088/1748-0221/9/03/p03009}

\bibitem[{{Aartsen} {et~al.}(2017{\natexlab{a}}){Aartsen}, {Ackermann}, {Adams}, {Aguilar}, {Ahlers}, {Ahrens}, {Altmann}, {Andeen}, {Anderson}, {Ansseau}, {Anton}, {Archinger}, {Arg{\"u}elles}, {Auer}, {Auffenberg}, {Axani}, {Baccus}, {Bai}, {Barnet}, {Barwick}, {Baum}, {Bay}, {Beattie}, {Beatty}, {Becker Tjus}, {Becker}, {Bendfelt}, {BenZvi}, {Berley}, {Bernardini}, {Bernhard}, {Besson}, {Binder}, {Bindig}, {Bissok}, {Blaufuss}, {Blot}, {Boersma}, {Bohm}, {B{\"o}rner}, {Bos}, {Bose}, {B{\"o}ser}, {Botner}, {Bouchta}, {Braun}, {Brayeur}, {Bretz}, {Bron}, {Burgman}, {Burreson}, {Carver}, {Casier}, {Cheung}, {Chirkin}, {Christov}, {Clark}, {Classen}, {Coenders}, {Collin}, {Conrad}, {Cowen}, {Cross}, {Day}, {Day}, {de Andr{\'e}}, {De Clercq}, {del Pino Rosendo}, {Dembinski}, {De Ridder}, {Descamps}, {Desiati}, {de Vries}, {de Wasseige}, {de With}, {DeYoung}, {D{\'\i}az-V{\'e}lez}, {di Lorenzo}, {Dujmovic}, {Dumm}, {Dunkman}, {Eberhardt}, {Edwards}, {Ehrhardt}, {Eichmann}, {Eller}, {Euler}, {Evenson}, {Fahey},
  {Fazely}, {Feintzeig}, {Felde}, {Filimonov}, {Finley}, {Flis}, {F{\"o}sig}, {Franckowiak}, {Fr{\`e}re}, {Friedman}, {Fuchs}, {Gaisser}, {Gallagher}, {Gerhardt}, {Ghorbani}, {Giang}, {Gladstone}, {Glauch}, {Glowacki}, {Gl{\"u}senkamp}, {Goldschmidt}, {Gonzalez}, {Grant}, {Griffith}, {Gustafsson}, {Haack}, {Hallgren}, {Halzen}, {Hansen}, {Hansmann}, {Hanson}, {Haugen}, {Hebecker}, {Heereman}, {Helbing}, {Hellauer}, {Heller}, {Hickford}, {Hignight}, {Hill}, {Hoffman}, {Hoffmann}, {Hoshina}, {Huang}, {Huber}, {Hulth}, {Hultqvist}, {In}, {Inaba}, {Ishihara}, {Jacobi}, {Jacobsen}, {Japaridze}, {Jeong}, {Jero}, {Jones}, {Jones}, {Joseph}, {Kang}, {Kappes}, {Karg}, {Karle}, {Katz}, {Kauer}, {Keivani}, {Kelley}, {Kemp}, {Kheirandish}, {Kim}, {Kim}, {Kintscher}, {Kiryluk}, {Kitamura}, {Kittler}, {Klein}, {Kleinfelder}, {Kleist}, {Kohnen}, {Koirala}, {Kolanoski}, {Konietz}, {K{\"o}pke}, {Kopper}, {Kopper}, {Koskinen}, {Kowalski}, {Krasberg}, {Krings}, {Kroll}, {Kr{\"u}ckl}, {Kr{\"u}ger}, {Kunnen}, {Kunwar},
  {Kurahashi}, {Kuwabara}, {Labare}, {Laihem}, {Landsman}, {Lanfranchi}, {Larson}, {Lauber}, {Laundrie}, {Lennarz}, {Leich}, {Lesiak-Bzdak}, {Leuermann}, {Lu}, {Ludwig}, {L{\"u}nemann}, {Mackenzie}, {Madsen}, {Maggi}, {Mahn}, {Mancina}, {Mandelartz}, {Maruyama}, {Mase}, {Matis}, {Maunu}, {McNally}, {McParland}, {Meade}, {Meagher}, {Medici}, {Meier}, {Meli}, {Menne}, {Merino}, {Meures}, {Miarecki}, {Minor}, {Montaruli}, {Moulai}, {Murray}, {Nahnhauer}, {Naumann}, {Neer}, {Newcomb}, {Niederhausen}, {Nowicki}, {Nygren}, {Obertacke Pollmann}, {Olivas}, {O'Murchadha}, {Palczewski}, {Pandya}, {Pankova}, {Patton}, {Peiffer}, {Penek}, {Pepper}, {P{\'e}rez de los Heros}, {Pettersen}, {Pieloth}, {Pinat}, {Price}, {Przybylski}, {Quinnan}, {Raab}, {R{\"a}del}, {Rameez}, {Rawlins}, {Reimann}, {Relethford}, {Relich}, {Resconi}, {Rhode}, {Richman}, {Riedel}, {Robertson}, {Rongen}, {Roucelle}, {Rott}, {Ruhe}, {Ryckbosch}, {Rysewyk}, {Sabbatini}, {Sanchez Herrera}, {Sandrock}, {Sandroos}, {Sandstrom}, {Sarkar}, {Satalecka},
  {Schlunder}, {Schmidt}, {Schoenen}, {Sch{\"o}neberg}, {Schukraft}, {Schumacher}, {Seckel}, {Seunarine}, {Solarz}, {Soldin}, {Song}, {Spiczak}, {Spiering}, {Stanev}, {Stasik}, {Stettner}, {Steuer}, {Stezelberger}, {Stokstad}, {St{\"o}{\ss}l}, {Str{\"o}m}, {Strotjohann}, {Sulanke}, {Sullivan}, {Sutherland}, {Taavola}, {Taboada}, {Tatar}, {Tenholt}, {Ter-Antonyan}, {Terliuk}, {Te{\v{s}}i{\'c}}, {Thollander}, {Tilav}, {Toale}, {Tobin}, {Toscano}, {Tosi}, {Tselengidou}, {Turcati}, {Unger}, {Usner}, {Vandenbroucke}, {van Eijndhoven}, {Vanheule}, {van Rossem}, {van Santen}, {Vehring}, {Voge}, {Vogel}, {Vraeghe}, {Wahl}, {Walck}, {Wallace}, {Wallraff}, {Wandkowsky}, {Weaver}, {Weiss}, {Wendt}, {Westerhoff}, {Wharton}, {Whelan}, {Wickmann}, {Wiebe}, {Wiebusch}, {Wille}, {Williams}, {Wills}, {Wisniewski}, {Wolf}, {Wood}, {Woolsey}, {Woschnagg}, {Xu}, {Xu}, {Xu}, {Yanez}, {Yodh}, {Yoshida}, \& {Zoll}}]{icecube}
{Aartsen}, M.~G., {Ackermann}, M., {Adams}, J., {et~al.} 2017{\natexlab{a}}, Journal of Instrumentation, 12, P03012, \dodoi{10.1088/1748-0221/12/03/P03012}

\bibitem[{{Aartsen} {et~al.}(2017{\natexlab{b}}){Aartsen}, {Ackermann}, {Adams}, {Aguilar}, {Ahlers}, {Ahrens}, {Altmann}, {Andeen}, {Anderson}, {Ansseau}, {Anton}, {Archinger}, {Arg{\"u}elles}, {Auffenberg}, {Axani}, {Bai}, {Barwick}, {Baum}, {Bay}, {Beatty}, {Becker Tjus}, {Becker}, {BenZvi}, {Berley}, {Bernardini}, {Bernhard}, {Besson}, {Binder}, {Bindig}, {Bissok}, {Blaufuss}, {Blot}, {Bohm}, {B{\"o}rner}, {Bos}, {Bose}, {B{\"o}ser}, {Botner}, {Braun}, {Brayeur}, {Bretz}, {Bron}, {Burgman}, {Carver}, {Casier}, {Cheung}, {Chirkin}, {Christov}, {Clark}, {Classen}, {Coenders}, {Collin}, {Conrad}, {Cowen}, {Cross}, {Day}, {de Andr{\'e}}, {De Clercq}, {del Pino Rosendo}, {Dembinski}, {De Ridder}, {Desiati}, {de Vries}, {de Wasseige}, {de With}, {DeYoung}, {D{\'\i}az-V{\'e}lez}, {di Lorenzo}, {Dujmovic}, {Dumm}, {Dunkman}, {Eberhardt}, {Ehrhardt}, {Eichmann}, {Eller}, {Euler}, {Evenson}, {Fahey}, {Fazely}, {Feintzeig}, {Felde}, {Filimonov}, {Finley}, {Flis}, {F{\"o}sig}, {Franckowiak}, {Friedman}, {Fuchs},
  {Gaisser}, {Gallagher}, {Gerhardt}, {Ghorbani}, {Giang}, {Gladstone}, {Glauch}, {Gl{\"u}senkamp}, {Goldschmidt}, {Gonzalez}, {Grant}, {Griffith}, {Haack}, {Hallgren}, {Halzen}, {Hansen}, {Hansmann}, {Hanson}, {Hebecker}, {Heereman}, {Helbing}, {Hellauer}, {Hickford}, {Hignight}, {Hill}, {Hoffman}, {Hoffmann}, {Hoshina}, {Huang}, {Huber}, {Hultqvist}, {In}, {Ishihara}, {Jacobi}, {Japaridze}, {Jeong}, {Jero}, {Jones}, {Kang}, {Kappes}, {Karg}, {Karle}, {Katz}, {Kauer}, {Keivani}, {Kelley}, {Kheirandish}, {Kim}, {Kim}, {Kintscher}, {Kiryluk}, {Kittler}, {Klein}, {Kohnen}, {Koirala}, {Kolanoski}, {Konietz}, {K{\"o}pke}, {Kopper}, {Kopper}, {Koskinen}, {Kowalski}, {Krings}, {Kroll}, {Kr{\"u}ckl}, {Kr{\"u}ger}, {Kunnen}, {Kunwar}, {Kurahashi}, {Kuwabara}, {Labare}, {Lanfranchi}, {Larson}, {Lauber}, {Lennarz}, {Lesiak-Bzdak}, {Leuermann}, {Lu}, {L{\"u}nemann}, {Madsen}, {Maggi}, {Mahn}, {Mancina}, {Mandelartz}, {Maruyama}, {Mase}, {Maunu}, {McNally}, {Meagher}, {Medici}, {Meier}, {Meli}, {Menne}, {Merino},
  {Meures}, {Miarecki}, {Montaruli}, {Moulai}, {Nahnhauer}, {Naumann}, {Neer}, {Niederhausen}, {Nowicki}, {Nygren}, {Obertacke Pollmann}, {Olivas}, {O'Murchadha}, {Palczewski}, {Pandya}, {Pankova}, {Peiffer}, {Penek}, {Pepper}, {P{\'e}rez de los Heros}, {Pieloth}, {Pinat}, {Price}, {Przybylski}, {Quinnan}, {Raab}, {R{\"a}del}, {Rameez}, {Rawlins}, {Reimann}, {Relethford}, {Relich}, {Resconi}, {Rhode}, {Richman}, {Riedel}, {Robertson}, {Rongen}, {Rott}, {Ruhe}, {Ryckbosch}, {Rysewyk}, {Sabbatini}, {Sanchez Herrera}, {Sandrock}, {Sandroos}, {Sarkar}, {Satalecka}, {Schlunder}, {Schmidt}, {Schoenen}, {Sch{\"o}neberg}, {Schumacher}, {Seckel}, {Seunarine}, {Soldin}, {Song}, {Spiczak}, {Spiering}, {Stanev}, {Stasik}, {Stettner}, {Steuer}, {Stezelberger}, {Stokstad}, {St{\"o}{\ss}l}, {Str{\"o}m}, {Strotjohann}, {Sullivan}, {Sutherland}, {Taavola}, {Taboada}, {Tatar}, {Tenholt}, {Ter-Antonyan}, {Terliuk}, {Te{\v{s}}i{\'c}}, {Tilav}, {Toale}, {Tobin}, {Toscano}, {Tosi}, {Tselengidou}, {Turcati}, {Unger}, {Usner},
  {Vandenbroucke}, {van Eijndhoven}, {Vanheule}, {van Rossem}, {van Santen}, {Vehring}, {Voge}, {Vogel}, {Vraeghe}, {Walck}, {Wallace}, {Wallraff}, {Wandkowsky}, {Weaver}, {Weiss}, {Wendt}, {Westerhoff}, {Whelan}, {Wickmann}, {Wiebe}, {Wiebusch}, {Wille}, {Williams}, {Wills}, {Wolf}, {Wood}, {Woolsey}, {Woschnagg}, {Xu}, {Xu}, {Xu}, {Yanez}, {Yodh}, {Yoshida}, \& {Zoll}}]{2017APh....92...30A}
---. 2017{\natexlab{b}}, Astroparticle Physics, 92, 30, \dodoi{10.1016/j.astropartphys.2017.05.002}

\bibitem[{{Abbasi} {et~al.}(2023{\natexlab{a}}){Abbasi}, {Ackermann}, {Adams}, {Agarwalla}, {Aguilar}, {Ahlers}, {Alameddine}, {Amin}, {Andeen}, {Anton}, {Arg{\"u}elles}, {Ashida}, {Athanasiadou}, {Axani}, {Bai}, {Balagopal}, {Baricevic}, {Barwick}, {Basu}, {Bay}, {Beatty}, {Becker}, {Becker Tjus}, {Beise}, {Bellenghi}, {Benning}, {BenZvi}, {Berley}, {Bernardini}, {Besson}, {Binder}, {Blaufuss}, {Blot}, {Bontempo}, {Book}, {Meneguolo}, {B{\"o}ser}, {Botner}, {B{\"o}ttcher}, {Bourbeau}, {Braun}, {Brinson}, {Brostean-Kaiser}, {Burley}, {Busse}, {Butterfield}, {Campana}, {Carloni}, {Carnie-Bronca}, {Chattopadhyay}, {Chau}, {Chen}, {Chen}, {Chirkin}, {Choi}, {Clark}, {Classen}, {Coleman}, {Collin}, {Connolly}, {Conrad}, {Coppin}, {Correa}, {Countryman}, {Cowen}, {Dave}, {De Clercq}, {DeLaunay}, {Delgado}, {Dembinski}, {Deng}, {Deoskar}, {Desai}, {Desiati}, {de Vries}, {de Wasseige}, {DeYoung}, {Diaz}, {D{\'\i}az-V{\'e}lez}, {Dittmer}, {Domi}, {Dujmovic}, {DuVernois}, {Ehrhardt}, {Eller}, {El Mentawi}, {Engel},
  {Erpenbeck}, {Evans}, {Evenson}, {Fan}, {Fang}, {Farrag}, {Fazely}, {Fedynitch}, {Feigl}, {Fiedlschuster}, {Finley}, {Fischer}, {Fox}, {Franckowiak}, {Friedman}, {Fritz}, {F{\"u}rst}, {Gaisser}, {Gallagher}, {Ganster}, {Garcia}, {Gerhardt}, {Ghadimi}, {Glaser}, {Glauch}, {Gl{\"u}senkamp}, {Goehlke}, {Gonzalez}, {Goswami}, {Grant}, {Gray}, {Gries}, {Griffin}, {Griswold}, {G{\"u}nther}, {Gutjahr}, {Haack}, {Hallgren}, {Halliday}, {Halve}, {Halzen}, {Hamdaoui}, {Minh}, {Hanson}, {Hardin}, {Harnisch}, {Hatch}, {Haungs}, {Helbing}, {Hellrung}, {Henningsen}, {Heuermann}, {Heyer}, {Hickford}, {Hidvegi}, {Hill}, {Hill}, {Hoffman}, {Hori}, {Hoshina}, {Hou}, {Huber}, {Hultqvist}, {H{\"u}nnefeld}, {Hussain}, {Hymon}, {In}, {Ishihara}, {Jacquart}, {Janik}, {Jansson}, {Japaridze}, {Jayakumar}, {Jeong}, {Jin}, {Jones}, {Kang}, {Kang}, {Kang}, {Kappes}, {Kappesser}, {Kardum}, {Karg}, {Karl}, {Karle}, {Katz}, {Kauer}, {Kelley}, {Zathul}, {Kheirandish}, {Kiryluk}, {Klein}, {Kochocki}, {Koirala}, {Kolanoski}, {Kontrimas},
  {K{\"o}pke}, {Kopper}, {Koskinen}, {Koundal}, {Kovacevich}, {Kowalski}, {Kozynets}, {Kruiswijk}, {Krupczak}, {Kumar}, {Kun}, {Kurahashi}, {Lad}, {Lagunas Gualda}, {Lamoureux}, {Larson}, {Latseva}, {Lauber}, {Lazar}, {Lee}, {Leonard DeHolton}, {Leszczy{\'n}ska}, {Lincetto}, {Liu}, {Liubarska}, {Lohfink}, {Love}, {Mariscal}, {Lu}, {Lucarelli}, {Ludwig}, {Luszczak}, {Lyu}, {Madsen}, {Mahn}, {Makino}, {Manao}, {Mancina}, {Sainte}, {Mari{\c{s}}}, {Marka}, {Marka}, {Marsee}, {Martinez-Soler}, {Maruyama}, {Mayhew}, {McElroy}, {McNally}, {Mead}, {Meagher}, {Mechbal}, {Medina}, {Meier}, {Merckx}, {Merten}, {Micallef}, {Montaruli}, {Moore}, {Morii}, {Morse}, {Moulai}, {Mukherjee}, {Naab}, {Nagai}, {Nakos}, {Naumann}, {Necker}, {Neumann}, {Niederhausen}, {Nisa}, {Noell}, {Nowicki}, {Obertacke Pollmann}, {O'Dell}, {Oehler}, {Oeyen}, {Olivas}, {Orsoe}, {Osborn}, {O'Sullivan}, {Pandya}, {Park}, {Parker}, {Paudel}, {Paul}, {P{\'e}rez de los Heros}, {Peterson}, {Philippen}, {Pieper}, {Pizzuto}, {Plum}, {Pont{\'e}n},
  {Popovych}, {Prado Rodriguez}, {Pries}, {Procter-Murphy}, {Przybylski}, {Rack-Helleis}, {Rawlins}, {Rechav}, {Rehman}, {Reichherzer}, {Renzi}, {Resconi}, {Reusch}, {Rhode}, {Richman}, {Riedel}, {Rifaie}, {Roberts}, {Robertson}, {Rodan}, {Roellinghoff}, {Rongen}, {Rott}, {Ruhe}, {Ruohan}, {Ryckbosch}, {Safa}, {Saffer}, {Salazar-Gallegos}, {Sampathkumar}, {Sanchez Herrera}, {Sandrock}, {Santander}, {Sarkar}, {Sarkar}, {Savelberg}, {Savina}, {Schaufel}, {Schieler}, {Schindler}, {Schlickmann}, {Schl{\"u}ter}, {Schl{\"u}ter}, {Schmidt}, {Schneider}, {Schr{\"o}der}, {Schumacher}, {Schwefer}, {Sclafani}, {Seckel}, {Seikh}, {Seunarine}, {Shah}, {Sharma}, {Shefali}, {Shimizu}, {Silva}, {Skrzypek}, {Smithers}, {Snihur}, {Soedingrekso}, {S{\o}gaard}, {Soldin}, {Soldin}, {Sommani}, {Spannfellner}, {Spiczak}, {Spiering}, {Stamatikos}, {Stanev}, {Stezelberger}, {St{\"u}rwald}, {Stuttard}, {Sullivan}, {Taboada}, {Ter-Antonyan}, {Thiesmeyer}, {Thompson}, {Thwaites}, {Tilav}, {Tollefson}, {T{\"o}nnis}, {Toscano}, {Tosi},
  {Trettin}, {Tung}, {Turcotte}, {Twagirayezu}, {Ty}, {Unland Elorrieta}, {Upadhyay}, {Upshaw}, {Valtonen-Mattila}, {Vandenbroucke}, {van Eijndhoven}, {Vannerom}, {van Santen}, {Vara}, {Veitch-Michaelis}, {Venugopal}, {Vereecken}, {Verpoest}, {Veske}, {Walck}, {Watson}, {Weaver}, {Weigel}, {Weindl}, {Weldert}, {Wendt}, {Werthebach}, {Weyrauch}, {Whitehorn}, {Wiebusch}, {Willey}, {Williams}, {Wolf}, {Wolf}, {Wrede}, {Xu}, {Yanez}, {Yildizci}, {Yoshida}, {Young}, {Yu}, {Yu}, {Yuan}, {Zhang}, {Zhelnin}, \& {IceCube Collaboration}}]{2023ApJ...954...75A}
{Abbasi}, R., {Ackermann}, M., {Adams}, J., {et~al.} 2023{\natexlab{a}}, \apj, 954, 75, \dodoi{10.3847/1538-4357/acdfcb}

\bibitem[{{Abbasi} {et~al.}(2023{\natexlab{b}}){Abbasi}, {Ackermann}, {Adams}, {Agarwalla}, {Aguilar}, {Ahlers}, {Alameddine}, {Amin}, {Andeen}, {Anton}, {Arg{\"u}elles}, {Ashida}, {Athanasiadou}, {Axani}, {Bai}, {Balagopal}, {Baricevic}, {Barwick}, {Basu}, {Bay}, {Beatty}, {Becker}, {Becker Tjus}, {Beise}, {Bellenghi}, {BenZvi}, {Berley}, {Bernardini}, {Besson}, {Binder}, {Bindig}, {Blaufuss}, {Blot}, {Bontempo}, {Book}, {Boscolo Meneguolo}, {B{\"o}ser}, {Botner}, {B{\"o}ttcher}, {Bourbeau}, {Braun}, {Brinson}, {Brostean-Kaiser}, {Burley}, {Busse}, {Butterfield}, {Campana}, {Carloni}, {Carnie-Bronca}, {Chattopadhyay}, {Chau}, {Chen}, {Chen}, {Chirkin}, {Choi}, {Clark}, {Classen}, {Coleman}, {Collin}, {Connolly}, {Conrad}, {Coppin}, {Correa}, {Countryman}, {Cowen}, {Dave}, {De Clercq}, {DeLaunay}, {Delgado}, {Dembinski}, {Deng}, {Deoskar}, {Desai}, {Desiati}, {de Vries}, {de Wasseige}, {DeYoung}, {Diaz}, {D{\'\i}az-V{\'e}lez}, {Dittmer}, {Domi}, {Dujmovic}, {DuVernois}, {Ehrhardt}, {Eller}, {Engel},
  {Erpenbeck}, {Evans}, {Evenson}, {Fan}, {Fang}, {Farrag}, {Fazely}, {Fedynitch}, {Feigl}, {Fiedlschuster}, {Finley}, {Fischer}, {Fox}, {Franckowiak}, {Friedman}, {Fritz}, {F{\"u}rst}, {Gaisser}, {Gallagher}, {Ganster}, {Garcia}, {Gerhardt}, {Ghadimi}, {Glaser}, {Glauch}, {Gl{\"u}senkamp}, {Goehlke}, {Gonzalez}, {Goswami}, {Grant}, {Gray}, {Griffin}, {Griswold}, {G{\"u}nther}, {Gutjahr}, {Haack}, {Hallgren}, {Halliday}, {Halve}, {Halzen}, {Hamdaoui}, {Ha Minh}, {Hanson}, {Hardin}, {Harnisch}, {Hatch}, {Haungs}, {Helbing}, {Hellrung}, {Henningsen}, {Heuermann}, {Heyer}, {Hickford}, {Hidvegi}, {Hill}, {Hill}, {Hoffman}, {Hoshina}, {Hou}, {Huber}, {Hultqvist}, {H{\"u}nnefeld}, {Hussain}, {Hymon}, {In}, {Ishihara}, {Jacquart}, {Janik}, {Jansson}, {Japaridze}, {Jayakumar}, {Jeong}, {Jin}, {Jones}, {Kang}, {Kang}, {Kang}, {Kappes}, {Kappesser}, {Kardum}, {Karg}, {Karl}, {Karle}, {Katz}, {Kauer}, {Kelley}, {Zathul}, {Kheirandish}, {Kiryluk}, {Klein}, {Kochocki}, {Koirala}, {Kolanoski}, {Kontrimas}, {K{\"o}pke},
  {Kopper}, {Koskinen}, {Koundal}, {Kovacevich}, {Kowalski}, {Kozynets}, {Kruiswijk}, {Krupczak}, {Kumar}, {Kun}, {Kurahashi}, {Lad}, {Lagunas Gualda}, {Lamoureux}, {Larson}, {Lauber}, {Lazar}, {Lee}, {Leonard DeHolton}, {Leszczy{\'n}ska}, {Lincetto}, {Liu}, {Liubarska}, {Lohfink}, {Love}, {Lozano Mariscal}, {Lu}, {Lucarelli}, {Ludwig}, {Luszczak}, {Lyu}, {Madsen}, {Mahn}, {Makino}, {Manao}, {Mancina}, {Marie Sainte}, {Mari{\c{s}}}, {Marka}, {Marka}, {Marsee}, {Martinez-Soler}, {Maruyama}, {Mayhew}, {McElroy}, {McNally}, {Mead}, {Meagher}, {Mechbal}, {Medina}, {Meier}, {Merckx}, {Merten}, {Micallef}, {Montaruli}, {Moore}, {Morii}, {Morse}, {Moulai}, {Mukherjee}, {Naab}, {Nagai}, {Nakos}, {Naumann}, {Necker}, {Neumann}, {Niederhausen}, {Nisa}, {Noell}, {Nowicki}, {Obertacke Pollmann}, {O'Dell}, {Oehler}, {Oeyen}, {Olivas}, {Orsoe}, {Osborn}, {O'Sullivan}, {Pandya}, {Park}, {Parker}, {Paudel}, {Paul}, {P{\'e}rez de los Heros}, {Peterson}, {Philippen}, {Pieper}, {Pizzuto}, {Plum}, {Pont{\'e}n}, {Popovych},
  {Prado Rodriguez}, {Pries}, {Procter-Murphy}, {Przybylski}, {Rack-Helleis}, {Rawlins}, {Rechav}, {Rehman}, {Reichherzer}, {Renzi}, {Resconi}, {Reusch}, {Rhode}, {Richman}, {Riedel}, {Roberts}, {Robertson}, {Rodan}, {Roellinghoff}, {Rongen}, {Rott}, {Ruhe}, {Ruohan}, {Ryckbosch}, {Safa}, {Saffer}, {Salazar-Gallegos}, {Sampathkumar}, {Sanchez Herrera}, {Sandrock}, {Santander}, {Sarkar}, {Sarkar}, {Savelberg}, {Savina}, {Schaufel}, {Schieler}, {Schindler}, {Schl{\"u}ter}, {Schl{\"u}ter}, {Schmidt}, {Schneider}, {Schr{\"o}der}, {Schumacher}, {Schwefer}, {Sclafani}, {Seckel}, {Seunarine}, {Shah}, {Sharma}, {Shefali}, {Shimizu}, {Silva}, {Skrzypek}, {Smithers}, {Snihur}, {Soedingrekso}, {S{\o}gaard}, {Soldin}, {Sommani}, {Spannfellner}, {Spiczak}, {Spiering}, {Stamatikos}, {Stanev}, {Stezelberger}, {St{\"u}rwald}, {Stuttard}, {Sullivan}, {Taboada}, {Ter-Antonyan}, {Thiesmeyer}, {Thompson}, {Thwaites}, {Tilav}, {Tollefson}, {T{\"o}nnis}, {Toscano}, {Tosi}, {Trettin}, {Tung}, {Turcotte}, {Twagirayezu}, {Ty},
  {Unland Elorrieta}, {Upadhyay}, {Upshaw}, {Valtonen-Mattila}, {Vandenbroucke}, {van Eijndhoven}, {Vannerom}, {van Santen}, {Vara}, {Veitch-Michaelis}, {Venugopal}, {Verpoest}, {Veske}, {Walck}, {Watson}, {Weaver}, {Weigel}, {Weindl}, {Weldert}, {Wendt}, {Werthebach}, {Weyrauch}, {Whitehorn}, {Wiebusch}, {Willey}, {Williams}, {Wolf}, {Wolf}, {Wrede}, {Xu}, {Yanez}, {Yildizci}, {Yoshida}, {Yu}, {Yu}, {Yuan}, {Zhang}, \& {Zhelnin}}]{2023ApJS..269...25A}
---. 2023{\natexlab{b}}, \apjs, 269, 25, \dodoi{10.3847/1538-4365/acfa95}

\bibitem[{{Abbasi} {et~al.}(2024){Abbasi}, {Ackermann}, {Adams}, {Agarwalla}, {Aguilar}, {Ahlers}, {Alameddine}, {Amin}, {Andeen}, {Arg{\"u}elles}, {Ashida}, {Athanasiadou}, {Ausborm}, {Axani}, {Bai}, {Balagopal V.}, {Baricevic}, {Barwick}, {Bash}, {Basu}, {Bay}, {Beatty}, {Becker Tjus}, {Beise}, {Bellenghi}, {Benning}, {BenZvi}, {Berley}, {Bernardini}, {Besson}, {Blaufuss}, {Bloom}, {Blot}, {Bontempo}, {Book Motzkin}, {Boscolo Meneguolo}, {B{\"o}ser}, {Botner}, {B{\"o}ttcher}, {Braun}, {Brinson}, {Brostean-Kaiser}, {Brusa}, {Burley}, {Butterfield}, {Campana}, {Caracas}, {Carloni}, {Carpio}, {Chattopadhyay}, {Chau}, {Chen}, {Chirkin}, {Choi}, {Clark}, {Coleman}, {Collin}, {Connolly}, {Conrad}, {Coppin}, {Corley}, {Correa}, {Cowen}, {Dave}, {De Clercq}, {DeLaunay}, {Delgado}, {Deng}, {Desai}, {Desiati}, {de Vries}, {de Wasseige}, {DeYoung}, {Diaz}, {D{\'\i}az-V{\'e}lez}, {Dierichs}, {Dittmer}, {Domi}, {Draper}, {Dujmovic}, {Dutta}, {DuVernois}, {Ehrhardt}, {Eidenschink}, {Eimer}, {Eller}, {Ellinger}, {El
  Mentawi}, {Els{\"a}sser}, {Engel}, {Erpenbeck}, {Evans}, {Evenson}, {Fan}, {Fang}, {Farrag}, {Fazely}, {Fedynitch}, {Feigl}, {Fiedlschuster}, {Finley}, {Fischer}, {Fox}, {Franckowiak}, {Fukami}, {F{\"u}rst}, {Gallagher}, {Ganster}, {Garcia}, {Garcia}, {Garg}, {Genton}, {Gerhardt}, {Ghadimi}, {Girard-Carillo}, {Glaser}, {Gl{\"u}senkamp}, {Gonzalez}, {Goswami}, {Granados}, {Grant}, {Gray}, {Gries}, {Griffin}, {Griswold}, {Groth}, {G{\"u}nther}, {Gutjahr}, {Ha}, {Haack}, {Hallgren}, {Halve}, {Halzen}, {Hamdaoui}, {Minh}, {Handt}, {Hanson}, {Hardin}, {Harnisch}, {Hatch}, {Haungs}, {H{\"a}u{\ss}ler}, {Helbing}, {Hellrung}, {Hermannsgabner}, {Heuermann}, {Heyer}, {Hickford}, {Hidvegi}, {Hill}, {Hill}, {Hoffman}, {Hori}, {Hoshina}, {Hostert}, {Hou}, {Huber}, {Hultqvist}, {H{\"u}nnefeld}, {Hussain}, {Hymon}, {Ishihara}, {Iwakiri}, {Jacquart}, {Jain}, {Janik}, {Jansson}, {Japaridze}, {Jeong}, {Jin}, {Jones}, {Kamp}, {Kang}, {Kang}, {Kang}, {Kappes}, {Kappesser}, {Kardum}, {Karg}, {Karl}, {Karle}, {Katil}, {Katz},
  {Kauer}, {Kelley}, {Khanal}, {Khatee Zathul}, {Kheirandish}, {Kiryluk}, {Klein}, {Kochocki}, {Koirala}, {Kolanoski}, {Kontrimas}, {K{\"o}pke}, {Kopper}, {Koskinen}, {Koundal}, {Kovacevich}, {Kowalski}, {Kozynets}, {Krishnamoorthi}, {Kruiswijk}, {Krupczak}, {Kumar}, {Kun}, {Kurahashi}, {Lad}, {Lagunas Gualda}, {Lamoureux}, {Larson}, {Latseva}, {Lauber}, {Lazar}, {Lee}, {DeHolton}, {Leszczy{\'n}ska}, {Liao}, {Lincetto}, {Liu}, {Liubarska}, {Love}, {Lozano Mariscal}, {Lu}, {Lucarelli}, {Luszczak}, {Lyu}, {Madsen}, {Magnus}, {Mahn}, {Makino}, {Manao}, {Mancina}, {Sainte}, {Mari{\c{s}}}, {Marka}, {Marka}, {Marsee}, {Martinez-Soler}, {Maruyama}, {Mayhew}, {McNally}, {Mead}, {Meagher}, {Mechbal}, {Medina}, {Meier}, {Merckx}, {Merten}, {Micallef}, {Mitchell}, {Montaruli}, {Moore}, {Morii}, {Morse}, {Moulai}, {Mukherjee}, {Naab}, {Nagai}, {Nakos}, {Naumann}, {Necker}, {Negi}, {Neste}, {Neumann}, {Niederhausen}, {Nisa}, {Noda}, {Noell}, {Novikov}, {Obertacke Pollmann}, {O'Dell}, {Oeyen}, {Olivas}, {Orsoe}, {Osborn},
  {O'Sullivan}, {Palusova}, {Pandya}, {Park}, {Parker}, {Paudel}, {Paul}, {P{\'e}rez de los Heros}, {Pernice}, {Peterson}, {Philippen}, {Pizzuto}, {Plum}, {Pont{\'e}n}, {Popovych}, {Prado Rodriguez}, {Pries}, {Privon}, {Procter-Murphy}, {Przybylski}, {Raab}, {Rack-Helleis}, {Ravn}, {Rawlins}, {Rechav}, {Rehman}, {Reichherzer}, {Resconi}, {Reusch}, {Rhode}, {Riedel}, {Rifaie}, {Roberts}, {Robertson}, {Rodan}, {Roellinghoff}, {Rongen}, {Rosted}, {Rott}, {Ruhe}, {Ruohan}, {Ryckbosch}, {Safa}, {Saffer}, {Salazar-Gallegos}, {Sampathkumar}, {Sandrock}, {Santander}, {Sarkar}, {Sarkar}, {Savelberg}, {Savina}, {Schaile}, {Schaufel}, {Schieler}, {Schindler}, {Schlickmann}, {Schl{\"u}ter}, {Schl{\"u}ter}, {Schmeisser}, {Schmidt}, {Schneider}, {Schr{\"o}der}, {Schumacher}, {Sclafani}, {Seckel}, {Seikh}, {Seo}, {Seunarine}, {Sevle Myhr}, {Shah}, {Shefali}, {Shimizu}, {Silva}, {Skrzypek}, {Smithers}, {Snihur}, {Soedingrekso}, {S{\o}gaard}, {Soldin}, {Soldin}, {Sommani}, {Spannfellner}, {Spiczak}, {Spiering}, {Stamatikos},
  {Stanev}, {Stezelberger}, {St{\"u}rwald}, {Stuttard}, {Sullivan}, {Taboada}, {Ter-Antonyan}, {Terliuk}, {Thiesmeyer}, {Thompson}, {Thwaites}, {Tilav}, {Tollefson}, {T{\"o}nnis}, {Toscano}, {Tosi}, {Trettin}, {Turcotte}, {Twagirayezu}, {Unland Elorrieta}, {Upadhyay}, {Upshaw}, {Vaidyanathan}, {Valtonen-Mattila}, {Vandenbroucke}, {van Eijndhoven}, {Vannerom}, {van Santen}, {Vara}, {Varsi}, {Veitch-Michaelis}, {Venugopal}, {Vereecken}, {Verpoest}, {Veske}, {Vijai}, {Walck}, {Wang}, {Weaver}, {Weigel}, {Weindl}, {Weldert}, {Wen}, {Wendt}, {Werthebach}, {Weyrauch}, {Whitehorn}, {Wiebusch}, {Williams}, {Witthaus}, {Wolf}, {Wolf}, {Wrede}, {Xu}, {Yanez}, {Yildizci}, {Yoshida}, {Young}, {Yu}, {Yuan}, {Zhang}, {Zhelnin}, {Zilberman}, \& {Zimmerman}}]{icecube_hard_x_ray}
---. 2024, arXiv e-prints, arXiv:2406.06684, \dodoi{10.48550/arXiv.2406.06684}

\bibitem[{{Abdollahi} {et~al.}(2022){Abdollahi}, {Acero}, {Baldini}, {Ballet}, {Bastieri}, {Bellazzini}, {Berenji}, {Berretta}, {Bissaldi}, {Blandford}, {Bloom}, {Bonino}, {Brill}, {Britto}, {Bruel}, {Burnett}, {Buson}, {Cameron}, {Caputo}, {Caraveo}, {Castro}, {Chaty}, {Cheung}, {Chiaro}, {Cibrario}, {Ciprini}, {Coronado-Bl{\'a}zquez}, {Crnogorcevic}, {Cutini}, {D'Ammando}, {De Gaetano}, {Digel}, {Di Lalla}, {Dirirsa}, {Di Venere}, {Dom{\'\i}nguez}, {Fallah Ramazani}, {Fegan}, {Ferrara}, {Fiori}, {Fleischhack}, {Franckowiak}, {Fukazawa}, {Funk}, {Fusco}, {Galanti}, {Gammaldi}, {Gargano}, {Garrappa}, {Gasparrini}, {Giacchino}, {Giglietto}, {Giordano}, {Giroletti}, {Glanzman}, {Green}, {Grenier}, {Grondin}, {Guillemot}, {Guiriec}, {Gustafsson}, {Harding}, {Hays}, {Hewitt}, {Horan}, {Hou}, {J{\'o}hannesson}, {Karwin}, {Kayanoki}, {Kerr}, {Kuss}, {Landriu}, {Larsson}, {Latronico}, {Lemoine-Goumard}, {Li}, {Liodakis}, {Longo}, {Loparco}, {Lott}, {Lubrano}, {Maldera}, {Malyshev}, {Manfreda}, {Mart{\'\i}-Devesa},
  {Mazziotta}, {Mereu}, {Meyer}, {Michelson}, {Mirabal}, {Mitthumsiri}, {Mizuno}, {Moiseev}, {Monzani}, {Morselli}, {Moskalenko}, {Negro}, {Nuss}, {Omodei}, {Orienti}, {Orlando}, {Paneque}, {Pei}, {Perkins}, {Persic}, {Pesce-Rollins}, {Petrosian}, {Pillera}, {Poon}, {Porter}, {Principe}, {Rain{\`o}}, {Rando}, {Rani}, {Razzano}, {Razzaque}, {Reimer}, {Reimer}, {Reposeur}, {S{\'a}nchez-Conde}, {Saz Parkinson}, {Scotton}, {Serini}, {Sgr{\`o}}, {Siskind}, {Smith}, {Spandre}, {Spinelli}, {Sueoka}, {Suson}, {Tajima}, {Tak}, {Thayer}, {Thompson}, {Torres}, {Troja}, {Valverde}, {Wood}, \& {Zaharijas}}]{4fgl_dr3}
{Abdollahi}, S., {Acero}, F., {Baldini}, L., {et~al.} 2022, \apjs, 260, 53, \dodoi{10.3847/1538-4365/ac6751}

\bibitem[{{Abeysekara} {et~al.}(2018){Abeysekara}, {Archer}, {Benbow}, {Bird}, {Brill}, {Brose}, {Buckley}, {Christiansen}, {Chromey}, {Daniel}, {Falcone}, {Feng}, {Finley}, {Fortson}, {Furniss}, {Gillanders}, {Gueta}, {Hanna}, {Hervet}, {Holder}, {Hughes}, {Humensky}, {Johnson}, {Kaaret}, {Kar}, {Kelley-Hoskins}, {Kertzman}, {Kieda}, {Krause}, {Krennrich}, {Lang}, {Moriarty}, {Mukherjee}, {O'Brien}, {Ong}, {Otte}, {Park}, {Petrashyk}, {Pohl}, {Pueschel}, {Quinn}, {Ragan}, {Reynolds}, {Richards}, {Roache}, {Rulten}, {Sadeh}, {Santander}, {Scott}, {Sembroski}, {Shahinyan}, {Tyler}, {Wakely}, {Weinstein}, {Wells}, {Wilcox}, {Wilhelm}, {Williams}, {Williamson}, {Zitzer}, {VERITAS Collaboration}, \& {Kaur}}]{2018ApJ...861L..20A}
{Abeysekara}, A.~U., {Archer}, A., {Benbow}, W., {et~al.} 2018, \apjl, 861, L20, \dodoi{10.3847/2041-8213/aad053}

\bibitem[{Acciari {et~al.}(2021)Acciari, Ansoldi, Antonelli, Arbet~Engels, Artero, Asano, Baack, Babic, Baquero, Barres~de Almeida, Barrio, Batković, Becerra~Gonzalez, Bednarek, Bellizzi, Bernardini, Bernardos, Berti, Besenrieder, Bhattacharyya, Bigongiari, Biland, Blanch, Bökenkamp, Bonnoli, Bosnjak, Busetto, Carosi, Ceribella, Cerruti, Chai, Chilingarian, Cikota, Colak, Colombo, Contreras, Cortina, Covino, D'Amico, D'Elia, Da~Vela, Dazzi, De~Angelis, De~Lotto, Delfino, Delgado, Delgado~Mendez, Depaoli, Di~Pierro, Di~Venere, Do~Souto~Espiñeira, Dominis~Prester, Donini, Dorner, Doro, Elsaesser, Fallah~Ramazani, Fattorini, Fonseca, Font, Fruck, Fukami, Fukazawa, García~López, Garczarczyk, Gasparyan, Gaug, Giglietto, Giordano, Gliwny, Godinovic, Green, Green, Hadasch, Hahn, Heckmann, Herrera, Hoang, Hrupec, Hütten, Inada, Ishio, Iwamura, Jiménez~Martínez, Jormanainen, Jouvin, Karjalainen, Kerszberg, Kobayashi, Kubo, Kushida, Lamastra, LELAS, Leone, Lindfors, Linhoff, Lombardi, Longo, Lopez-Coto,
  López-Moya, López-Oramas, Loporchio, Machado~de Oliveira~Fraga, Maggio, Majumdar, MAKARIEV, Mallamaci, Maneva, Manganaro, Mannheim, Maraschi, Mariotti, Martinez, Mazin, Menchiari, Mender, Mićanović, Miceli, Miener, Miranda, Mirzoyan, Molina, Moralejo, Morcuende, Moreno, Moretti, Nakamori, Nava, Neustroev, Nigro, Nilsson, Nishijima, Noda, Nozaki, Ohtani, Oka, Otero-Santos, Paiano, Palatiello, Paneque, Paoletti, Paredes, Pavletić, Peñil, Persic, Pihet, Prada~Moroni, Prandini, Priyadarshi, Puljak, Rhode, Ribó, Rico, Righi, Rugliancich, Sahakyan, Saito, Sakurai, Satalecka, Saturni, Schleicher, Schmidt, Schweizer, Sitarek, Šnidarić, Sobczyńska, Spolon, Stamerra, Strišković, Strom, Strzys, Suda, Surić, Takahashi, Takeishi, Tavecchio, Temnikov, Terzic, Teshima, Tosti, Truzzi, Tutone, Ubach, van Scherpenberg, Vanzo, VAZQUEZ~ACOSTA, Ventura, VERGUILOV, Vigorito, Vitale, Vovk, Will, Wunderlich, Yamamoto, Zarić, Balbo, Bretz, Buss, Eisenberger, Hildebrand, Iotov, Kalenski, Neise, Noethe, Paravac,
  Sliusar, Walter, Abbasi, Ackermann, Adams, Aguilar, Ahlers, Ahrens, Alispach, Alves~Junior, Amin, An, Andeen, Anderson, Anton, Arguelles, Ashida, Axani, Bai, Balagopal~V., Barbano, Barwick, Bastian, Basu, Baur, Bay, Beatty, Becker, Becker~Tjus, Bellenghi, BenZvi, Berley, Besson, Binder, Bindig, Blaufuss, Blot, Boddenberg, Bontempo, Borowka, Boser, Botner, Bottcher, Bourbeau, Bradascio, Braun, Bron, Brostean-Kaiser, Browne, Burgman, Burley, Busse, Campana, Carnie-Bronca, Chen, Chirkin, Choi, Clark, Clark, Classen, Coleman, Collin, Conrad, Coppin, Correa, Cowen, Cross, Dappen, Dave, DE~CLERCQ, DeLaunay, Dembinski, Deoskar, De~Ridder, Desai, Desiati, de~Vries, de~Wasseige, De~With, DeYoung, Dharani, Diaz, Diaz-Velez, Dittmer, Dujmovic, Dunkman, DuVernois, Dvorak, Ehrhardt, Eller, Engel, Erpenbeck, Evans, Evenson, Fan, Fazely, Fiedlschuster, Fienberg, Filimonov, Finley, Fischer, Fox, Franckowiak, Friedman, Fritz, Furst, Gaisser, Gallagher, Ganster, Garcia, Garrappa, Gerhardt, Ghadimi, Glaser, Glauch,
  Glusenkamp, Goldschmidt, Gonzalez, Goswami, Grant, Grégoire, Griswold, Gunduz, Günther, Haack, Hallgren, Halliday, Halve, Halzen, Ha~Minh, Hanson, Hardin, Harnisch, Haungs, Hauser, Hebecker, Helbing, Henningsen, Hettinger, Hickford, Hignight, Hill, Hill, Hoffman, Hoffmann, Hoinka, Hokanson-Fasig, Hoshina, Huang, Huber, Huber, Hultqvist, Hunnefeld, Hussain, In, Iovine, Ishihara, Jansson, Japaridze, Jeong, Jones, Kang, Kang, Kang, Kappes, Kappesser, Karg, Karl, Karle, Katz, Kauer, Kellermann, Kelley, Kheirandish, Kin, Kintscher, Kiryluk, Klein, Koirala, Kolanoski, Kontrimas, Kopke, Kopper, Kopper, Koskinen, Koundal, Kovacevich, Kowalski, Kozynets, Kun, Kurahashi, Lad, Lagunas~Gualda, Lanfranchi, Larson, Lauber, Lazar, Lee, Leonard, Leszczyńska, Li, Lincetto, Liu, Liubarska, Lohfink, Lozano~Mariscal, Lu, Lucarelli, Ludwig, Luszczak, Lyu, Ma, Madsen, Mahn, Makino, Mancina, Maris, Maruyama, Mase, McElroy, McNally, Mead, Meagher, Medina, Meier, Meighen-Berger, Micallef, Mockler, Montaruli, Moore, Morse,
  Moulai, Naab, Nagai, Naumann, Necker, Nguyen, Niederhausen, Nisa, Nowicki, Nygren, Obertacke~Pollmann, Oehler, Olivas, O'Sullivan, Pandya, Pankova, Park, Parker, Paudel, Paul, Perez de~los Heros, Peters, Philippen, Pieloth, Pieper, Pittermann, Pizzuto, Plum, Popovych, Porcelli, Prado~Rodriguez, Price, Pries, Przybylski, Raab, Raissi, Rameez, Rawlins, Rea, Rehman, Reimann, Renzi, Resconi, Reusch, Richman, Riedel, Roberts, Robertson, Roellinghoff, Rongen, Rott, Ruhe, Ryckbosch, Rysewyk~Cantu, Safa, Saffer, Sanchez~Herrera, Sandrock, Sandroos, Santander, Sarkar, Sarkar, Scharf, Schaufel, Schieler, Schindler, Schlunder, Schmidt, Schneider, Schneider, Schröder, Schumacher, Schwefer, Sclafani, Seckel, Seunarine, Sharma, Shefali, Silva, Skrzypek, Smithers, Snihur, Soedingrekso, Soldin, Spannfellner, Spiczak, Spiering, Stachurska, Stamatikos, Stanev, Stein, Stettner, Steuer, Stezelberger, Sturwald, Stuttard, Sullivan, Taboada, Tenholt, Ter-Antonyan, Tilav, Tischbein, Tollefson, Tönnis, Toscano, Tosi, Trettin,
  Tselengidou, Tung, Turcati, Turcotte, Turley, Twagirayezu, Ty, Unland~Elorrieta, Valtonen-Mattila, Vandenbroucke, van Eijndhoven, Vannerom, van Santen, Verpoest, Vraeghe, Walck, Watson, Weaver, Weigel, Weindl, Weiss, Weldert, Wendt, Werthebach, Weyrauch, Whitehorn, Wiebusch, Williams, Wolf, Woschnagg, Wrede, Wulff, Xu, Xu, Yanez, Yoshida, Yu, Yuan, Zhang, Jin, Abdalla, Aharonian, Ait-Benkhali, Anguener, Arcaro, Armand, Armstrong, Ashkar, Backes, Baghmanyan, Barbosa~Martins, Barnacka, Barnard, Batzofin, Becherini, Berge, Bernlöhr, Bi, Böttcher, Boisson, Bolmont, de~Bony, Breuhaus, Brose, Brun, Bulik, Bylund, Cangemi, Caroff, Casanova, Catalano, Chambery, Chand, Chen, Cotter, Curlo, Damascene~Mbarubucyeye, Davids, Davies, Devin, Djannati-Ataï, Dmytriev, Donath, Doroshenko, Dreyer, Du~Plessis, Duffy, Egberts, Einecke, Emery, ERNENWEIN, Fegan, Feijen, Fiasson, Fichet~de Clairfontaine, Fontaine, Frans, Fuessling, Funk, Gabici, Gallant, Ghafourizade, Giavitto, Giunti, Glawion, Glicenstein, Grondin, Hattingh,
  Haupt, HERMANN, Hinton, Hofmann, Hoischen, Holch, Holler, Horns, Huang, Huber, Hörbe, Jamrozy, Jankowsky, Joshi, JUNG, Kasai, Katarzynski, Katz, Khangulyan, Khelifi, Klepser, Kluzniak, Komin, Konno, Kosack, Kostunin, Kreter, Kukec~Mezek, Kundu, Lamanna, Le~Stum, Lemiere, Lemoine-Goumard, Lenain, Leuschner, Levy, Lohse, Luashvili, Lypova, Mackey, Majumdar, Malyshev, Malyshev, Marandon, Marchegiani, Marcowith, Mares, Martí-Devesa, Marx, Maurin, Meintjes, Meyer, Mitchell, Moderski, Mohrmann, Montanari, Moore, Morris, Moulin, Muller, Murach, Nakashima, Naurois~(de), Nayerhoda, Davids, Niemiec, Noel, O'Brien, Oberholzer, Ohm, Olivera-Nieto, Ona-Wilhelmi~(de), Ostrowski, Panny, Panter, Parsons, Peron, Pita, Poireau, Prokhorov, Prokoph, PUEHLHOFER, Punch, Quirrenbach, Reichherzer, Reimer, Reimer, Remy, Renaud, Reville, Rieger, Romoli, Rowell, Rudak, Rueda~Ricarte, Ruiz~Velasco, Sahakian, Sailer, Salzmann, Sanchez, Santangelo, Sasaki, Schaefer, Schutte, Schwanke, Schüssler, Senniappan, Seyffert, Shapopi,
  Shiningayamwe, Simoni, Sinha, Sol, Spackman, Specovius, Spencer, Spir-Jacob, Stawarz, Steenkamp, Stegmann, Steinmassl, Steppa, Sun, Takahashi, Tanaka, Tavernier, Taylor, Terrier, Thiersen, Thorpe-Morgan, Tluczykont, Tomankova, Tsirou, Tsuji, Tuffs, Uchiyama, van~der Walt, van Eldik, van Rensburg, van Soelen, Vasileiadis, Veh, Venter, Vincent, Vink, Völk, Wagner, Watson, Werner, White, Wierzcholska, Wong, Yassin, Yusafzai, Zacharias, Zanin, Zargaryan, Zdziarski, Zech, Zhu, Zmija, Zouari, \& Żywucka}]{2022icrc.confE.960T}
Acciari, V.~A., Ansoldi, S., Antonelli, L.~A., {et~al.} 2021, PoS, ICRC2021, 960, \dodoi{10.22323/1.395.0960}

\bibitem[{{Acciari} {et~al.}(2022){Acciari}, {Aniello}, {Ansoldi}, {Antonelli}, {Arbet Engels}, {Artero}, {Asano}, {Baack}, {Babi{\'c}}, {Baquero}, {Barres de Almeida}, {Barrio}, {Batkovi{\'c}}, {Becerra Gonz{\'a}lez}, {Bednarek}, {Bernardini}, {Bernardos}, {Berti}, {Besenrieder}, {Bhattacharyya}, {Bigongiari}, {Biland}, {Blanch}, {B{\"o}kenkamp}, {Bonnoli}, {Bo{\v{s}}njak}, {Busetto}, {Carosi}, {Ceribella}, {Cerruti}, {Chai}, {Chilingarian}, {Cikota}, {Colombo}, {Contreras}, {Cortina}, {Covino}, {D'Amico}, {D'Elia}, {Vela}, {Dazzi}, {De Angelis}, {De Lotto}, {Del Popolo}, {Delfino}, {Delgado}, {Mendez}, {Depaoli}, {Di Pierro}, {Di Venere}, {Do Souto Espi{\~n}eira}, {Dominis Prester}, {Donini}, {Dorner}, {Doro}, {Elsaesser}, {Fallah Ramazani}, {Fari{\~n}a}, {Fattorini}, {Font}, {Fruck}, {Fukami}, {Fukazawa}, {Garc{\'\i}a L{\'o}pez}, {Garczarczyk}, {Gasparyan}, {Gaug}, {Giglietto}, {Giordano}, {Gliwny}, {Godinovi{\'c}}, {Green}, {Green}, {Hadasch}, {Hahn}, {Hassan}, {Heckmann}, {Herrera}, {Hoang}, {Hrupec},
  {H{\"u}tten}, {Inada}, {Iotov}, {Ishio}, {Iwamura}, {Jim{\'e}nez Mart{\'\i}nez}, {Jormanainen}, {Jouvin}, {Kerszberg}, {Kobayashi}, {Kubo}, {Kushida}, {Lamastra}, {Lelas}, {Leone}, {Lindfors}, {Linhoff}, {Lombardi}, {Longo}, {L{\'o}pez-Coto}, {L{\'o}pez-Moya}, {L{\'o}pez-Oramas}, {Loporchio}, {Machado de Oliveira Fraga}, {Maggio}, {Majumdar}, {Makariev}, {Mallamaci}, {Maneva}, {Manganaro}, {Mannheim}, {Mariotti}, {Mart{\'\i}nez}, {Mas Aguilar}, {Mazin}, {Menchiari}, {Mender}, {Mi{\'c}anovi{\'c}}, {Miceli}, {Miener}, {Miranda}, {Mirzoyan}, {Molina}, {Moralejo}, {Morcuende}, {Moreno}, {Moretti}, {Nakamori}, {Nava}, {Neustroev}, {Nievas Rosillo}, {Nigro}, {Nilsson}, {Nishijima}, {Noda}, {Nozaki}, {Ohtani}, {Oka}, {Otero-Santos}, {Paiano}, {Palatiello}, {Paneque}, {Paoletti}, {Paredes}, {Pavleti{\'c}}, {Pe{\~n}il}, {Persic}, {Pihet}, {Prada Moroni}, {Prandini}, {Priyadarshi}, {Puljak}, {Rhode}, {Rib{\'o}}, {Rico}, {Righi}, {Rugliancich}, {Sahakyan}, {Saito}, {Sakurai}, {Satalecka}, {Saturni}, {Schleicher},
  {Schmidt}, {Schmuckermaier}, {Schweizer}, {Sitarek}, {{\v{S}}nidari{\'c}}, {Sobczynska}, {Spolon}, {Stamerra}, {Stri{\v{s}}kovi{\'c}}, {Strom}, {Strzys}, {Suda}, {Suri{\'c}}, {Takahashi}, {Takeishi}, {Tavecchio}, {Temnikov}, {Terzi{\'c}}, {Teshima}, {Tosti}, {Truzzi}, {Tutone}, {Ubach}, {van Scherpenberg}, {Vanzo}, {Vazquez Acosta}, {Ventura}, {Verguilov}, {Viale}, {Vigorito}, {Vitale}, {Vovk}, {Will}, {Wunderlich}, {Yamamoto}, {Zari{\'c}}, {Hodges}, {Hovatta}, {Kiehlmann}, {Liodakis}, {Max-Moerbeck}, {Pearson}, {Readhead}, {Reeves}, {L{\"a}hteenm{\"a}ki}, {Tornikoski}, {Tammi}, {D'Ammando}, \& {Marchini}}]{2022ApJ...927..197A}
{Acciari}, V.~A., {Aniello}, T., {Ansoldi}, S., {et~al.} 2022, \apj, 927, 197, \dodoi{10.3847/1538-4357/ac531d}

\bibitem[{Acero {et~al.}(2015)Acero, Ackermann, Ajello, Albert, Atwood, Axelsson, Baldini, Ballet, Barbiellini, Bastieri, Belfiore, Bellazzini, Bissaldi, Blandford, Bloom, Bogart, Bonino, Bottacini, Bregeon, Britto, Bruel, Buehler, Burnett, Buson, Caliandro, Cameron, Caputo, Caragiulo, Caraveo, Casandjian, Cavazzuti, Charles, Chaves, Chekhtman, Cheung, Chiang, Chiaro, Ciprini, Claus, Cohen-Tanugi, Cominsky, Conrad, Cutini, D’Ammando, de~Angelis, DeKlotz, de~Palma, Desiante, Digel, Venere, Drell, Dubois, Dumora, Favuzzi, Fegan, Ferrara, Finke, Franckowiak, Fukazawa, Funk, Fusco, Gargano, Gasparrini, Giebels, Giglietto, Giommi, Giordano, Giroletti, Glanzman, Godfrey, Grenier, Grondin, Grove, Guillemot, Guiriec, Hadasch, Harding, Hays, Hewitt, Hill, Horan, Iafrate, Jogler, Jóhannesson, Johnson, Johnson, Johnson, Johnson, Kamae, Kataoka, Katsuta, Kuss, Mura, Landriu, Larsson, Latronico, Lemoine-Goumard, Li, Li, Longo, Loparco, Lott, Lovellette, Lubrano, Madejski, Massaro, Mayer, Mazziotta, McEnery, Michelson,
  Mirabal, Mizuno, Moiseev, Mongelli, Monzani, Morselli, Moskalenko, Murgia, Nuss, Ohno, Ohsugi, Omodei, Orienti, Orlando, Ormes, Paneque, Panetta, Perkins, Pesce-Rollins, Piron, Pivato, Porter, Racusin, Rando, Razzano, Razzaque, Reimer, Reimer, Reposeur, Rochester, Romani, Salvetti, Sánchez-Conde, Parkinson, Schulz, Siskind, Smith, Spada, Spandre, Spinelli, Stephens, Strong, Suson, Takahashi, Takahashi, Tanaka, Thayer, Thayer, Thompson, Tibaldo, Tibolla, Torres, Torresi, Tosti, Troja, Klaveren, Vianello, Winer, Wood, Wood, \& Zimmer}]{Acero_2015}
Acero, F., Ackermann, M., Ajello, M., {et~al.} 2015, The Astrophysical Journal Supplement Series, 218, 23, \dodoi{10.1088/0067-0049/218/2/23}

\bibitem[{{Acharyya} \& {Santander}(2023)}]{2023arXiv230906164A}
{Acharyya}, A., \& {Santander}, M. 2023, arXiv e-prints, arXiv:2309.06164, \dodoi{10.48550/arXiv.2309.06164}

\bibitem[{{Acharyya} {et~al.}(2023){Acharyya}, {Adams}, {Archer}, {Bangale}, {Bartkoske}, {Batista}, {Benbow}, {Brill}, {Buckley}, {Christiansen}, {Chromey}, {Errando}, {Falcone}, {Feng}, {Foote}, {Fortson}, {Furniss}, {Gallagher}, {Hanlon}, {Hanna}, {Hervet}, {Hinrichs}, {Hoang}, {Holder}, {Humensky}, {Jin}, {Kaaret}, {Kertzman}, {Kherlakian}, {Kieda}, {Kleiner}, {Korzoun}, {Kumar}, {Lang}, {Lundy}, {Maier}, {McGrath}, {Millard}, {Millis}, {Mooney}, {Moriarty}, {Mukherjee}, {O'Brien}, {Ong}, {Pohl}, {Pueschel}, {Quinn}, {Ragan}, {Reynolds}, {Ribeiro}, {Roache}, {Sadeh}, {Sadun}, {Saha}, {Santander}, {Sembroski}, {Shang}, {Splettstoesser}, {Talluri}, {Tucci}, {Vassiliev}, {Weinstein}, {Williams}, {Wong}, {Woo}, {Aharonian}, {Aschersleben}, {Backes}, {Martins}, {Batzofin}, {Becherini}, {Berge}, {Bernl{\"o}hr}, {Bi}, {B{\"o}ttcher}, {Boisson}, {Bolmont}, {de Bony de Lavergne}, {Borowska}, {Bouyahiaoui}, {Bradascio}, {Breuhaus}, {Brose}, {Brun}, {Bruno}, {Bulik}, {Burger-Scheidlin}, {Caroff}, {Casanova},
  {Cecil}, {Celic}, {Cerruti}, {Chand}, {Chandra}, {Chen}, {Chibueze}, {Chibueze}, {Cotter}, {Dai}, {Mbarubucyeye}, {Djannati-Ata{\"\i}}, {Dmytriiev}, {Doroshenko}, {Einecke}, {Ernenwein}, {de Clairfontaine}, {Filipovic}, {Fontaine}, {F{\"u}{\ss}ling}, {Funk}, {Gabici}, {Ghafourizadeh}, {Giavitto}, {Glawion}, {Glicenstein}, {Goswami}, {Grolleron}, {Haerer}, {Hinton}, {Holch}, {Holler}, {Horns}, {Jamrozy}, {Jankowsky}, {Joshi}, {Jung-Richardt}, {Kasai}, {Katarzy{\'n}ski}, {Khatoon}, {Kh{\'e}lifi}, {Klepser}, {Klu{\'z}niak}, {Kosack}, {Kostunin}, {Lang}, {Le Stum}, {Lemi{\`e}re}, {Lenain}, {Leuschner}, {Lohse}, {Luashvili}, {Lypova}, {Mackey}, {Malyshev}, {Marandon}, {Marchegiani}, {Marcowith}, {Mart{\'\i}-Devesa}, {Marx}, {Mitchell}, {Moderski}, {Mohrmann}, {Montanari}, {Moulin}, {Murach}, {Nakashima}, {Niemiec}, {Noel}, {O'Brien}, {Olivera-Nieto}, {de Ona Wilhelmi}, {Ostrowski}, {Panny}, {Panter}, {Peron}, {Prokhorov}, {P{\"u}hlhofer}, {Punch}, {Quirrenbach}, {Reichherzer}, {Reimer}, {Reimer}, {Ren},
  {Renaud}, {Rieger}, {Rudak}, {Ruiz-Velasco}, {Sahakian}, {Santangelo}, {Sasaki}, {Sch{\"a}fer}, {Sch{\"u}ssler}, {Schutte}, {Schwanke}, {Shapopi}, {Specovius}, {Spencer}, {Stawarz}, {Steenkamp}, {Steinmassl}, {Sushch}, {Suzuki}, {Takahashi}, {Tanaka}, {Terrier}, {van Eldik}, {Vecchi}, {Veh}, {Venter}, {Vink}, {White}, {Wierzcholska}, {Wong}, {Zacharias}, {Zargaryan}, {Zdziarski}, {Zech}, {Zouari}, {{\.Z}ywucka}, {Mori}, \& {H.~E.~S.~S. Collaboration}}]{2023_Qi}
{Acharyya}, A., {Adams}, C.~B., {Archer}, A., {et~al.} 2023, \apj, 954, 70, \dodoi{10.3847/1538-4357/ace327}

\bibitem[{Adams {et~al.}(2022)Adams, Batshoun, Benbow, Brill, Buckley, Capasso, Cavins, Christiansen, Coppi, Errando, Farrell, Feng, Finley, Foote, Fortson, Furniss, Gent, Giuri, Hanna, Hassan, Hervet, Holder, Houck, Humensky, Jin, Kaaret, Kertzman, Kieda, Krennrich, Kumar, Lundy, Maier, McGrath, Moriarty, Mukherjee, Nieto, Nievas-Rosillo, O’Brien, Ong, Oppenheimer, Otte, Patel, Pfrang, Pohl, Prado, Pueschel, Quinn, Ragan, Reynolds, Rhatigan, Ribeiro, Roache, Ryan, Santander, Sembroski, Williams, Williamson, Collaboration), Valverde, Horan, Buson, Cheung, Ciprini, Gasparrini, Ojha, van Zyl, Collaboration), \& Sironi}]{Adams_2022}
Adams, C.~B., Batshoun, J., Benbow, W., {et~al.} 2022, The Astrophysical Journal, 924, 95, \dodoi{10.3847/1538-4357/ac32bd}

\bibitem[{{Aiello} {et~al.}(2024){Aiello}, {Albert}, {Alshamsi}, {Alves Garre}, {Aly}, {Ambrosone}, {Ameli}, {Andre}, {Androutsou}, {Anguita}, {Aphecetche}, {Ardid}, {Ardid}, {Atmani}, {Aublin}, {Badaracco}, {Bailly-Salins}, {Bardacov{\'a}}, {Baret}, {Bariego-Quintana}, {Baruzzi}, {Basegmez du Pree}, {Becherini}, {Bendahman}, {Benfenati}, {Benhassi}, {Benoit}, {Berbee}, {Bertin}, {Biagi}, {Boettcher}, {Bonanno}, {Boumaaza}, {Bouta}, {Bouwhuis}, {Bozza}, {Bozza}, {Br{\^a}nzas}, {Bretaudeau}, {Breuhaus}, {Bruijn}, {Brunner}, {Bruno}, {Buis}, {Buompane}, {Busto}, {Caiffi}, {Calvo}, {Campion}, {Capone}, {Carenini}, {Carretero}, {Cartraud}, {Castaldi}, {Cecchini}, {Celli}, {Cerisy}, {Chabab}, {Chadolias}, {Chen}, {Cherubini}, {Chiarusi}, {Circella}, {Cocimano}, {Coelho}, {Coleiro}, {Coniglione}, {Coyle}, {Creusot}, {Cuttone}, {Dallier}, {Darras}, {De Benedittis}, {De Martino}, {Decoene}, {Del Burgo}, {Del Rosso}, {Di Mauro}, {Di Palma}, {D{\'\i}az}, {Diaz}, {Diego-Tortosa}, {Distefano}, {Domi}, {Donzaud},
  {Dornic}, {D{\"o}rr}, {Drakopoulou}, {Drouhin}, {Ducoin}, {Dvornick{\'y}}, {Eberl}, {Eckerov{\'a}}, {Eddymaoui}, {van Eeden}, {Eff}, {van Eijk}, {El Bojaddaini}, {El Hedri}, {Enzenh{\"o}fer}, {Ferrara}, {Filipovic}, {Filippini}, {Franciotti}, {Fusco}, {Gabriel}, {Gagliardini}, {Gal}, {Garc{\'\i}a M{\'e}ndez}, {Garcia Soto}, {Gatius Oliver}, {Gei{\ss}elbrecht}, {Ghaddari}, {Gialanella}, {Gibson}, {Giorgio}, {Goos}, {Goswami}, {Goupilliere}, {Gozzini}, {Gracia}, {Graf}, {Guidi}, {Guillon}, {Guti{\'e}rrez}, {van Haren}, {Heijboer}, {Hekalo}, {Hennig}, {Hern{\'a}ndez-Rey}, {Idrissi Ibnsalih}, {Illuminati}, {de Jong}, {de Jong}, {Jung}, {Kalaczynski}, {Kalekin}, {Katz}, {Kistauri}, {Kopper}, {Kouchner}, {Kueviakoe}, {Kulikovskiy}, {Kvatadze}, {Labalme}, {Lahmann}, {Larosa}, {Lastoria}, {Lazo}, {Le Stum}, {Lehaut}, {Leonora}, {Lessing}, {Levi}, {Clark}, {Longhitano}, {Magnani}, {Majumdar}, {Malerba}, {Mamedov}, {Manczak}, {Manfreda}, {Marconi}, {Margiotta}, {Marinelli}, {Markou}, {Martin}, {Mart{\'\i}nez-Mora},
  {Marzaioli}, {Mastrodicasa}, {Mastroianni}, {Miccich{\`e}}, {Miele}, {Migliozzi}, {Migneco}, {Mitsou}, {Mollo}, {Morales-Gallegos}, {Morga}, {Moussa}, {Mozun Mateo}, {Muller}, {Musone}, {Musumeci}, {Navas}, {Nayerhoda}, {Nicolau}, {Nkosi}, {Fearraigh}, {Oliviero}, {Orlando}, {Oukacha}, {Paesani}, {Palacios Gonz{\'a}lez}, {Papalashvili}, {Parisi}, {Pastor Gomez}, {Paun}, {Pavalas}, {Pelegris}, {Pe{\~n}a Mart{\'\i}nez}, {Perrin-Terrin}, {Perronnel}, {Pestel}, {Pestes}, {Piattelli}, {Poir{\`e}}, {Popa}, {Pradier}, {Prado}, {Pulvirenti}, {Quiroz-Rangel}, {Rahaman}, {Randazzo}, {Randriatoamanana}, {Razzaque}, {Rea}, {Real}, {Riccobene}, {Robinson}, {Romanov}, {{\v{S}}aina}, {Salesa Greus}, {Samtleben}, {S{\'a}nchez Losa}, {Sanfilippo}, {Sanguineti}, {Santonastaso}, {Santonocito}, {Sapienza}, {Schnabel}, {Schumann}, {Schutte}, {Seneca}, {Sennan}, {Setter}, {Sgura}, {Shanidze}, {Sharma}, {Shitov}, {{\v{S}}imkovic}, {Simonelli}, {Sinopoulou}, {Smirnov}, {Spisso}, {Spurio}, {Stavropoulos}, {{\v{S}}tekl}, {Taiuti},
  {Tayalati}, {Thiersen}, {Tosta e Melo}, {Tragia}, {Trocm{\'e}}, {Tsourapis}, {Tudorache}, {Tzamariudaki}, {Vacheret}, {Valer Melchor}, {Valsecchi}, {Van Elewyck}, {Vannoye}, {Vasileiadis}, {Vazquez de Sola}, {Verilhac}, {Veutro}, {Viola}, {Vivolo}, {Wilms}, {de Wolf}, {Yepes-Ramirez}, {Zarpapis}, {Zavatarelli}, {Zegarelli}, {Zito}, {Zornoza}, {Z{\'u}{\~n}iga}, \& {Zywucka}}]{2024arXiv240208363A}
{Aiello}, S., {Albert}, A., {Alshamsi}, M., {et~al.} 2024, arXiv e-prints, arXiv:2402.08363, \dodoi{10.48550/arXiv.2402.08363}

\bibitem[{Ajello {et~al.}(2017)Ajello, Atwood, Baldini, Ballet, Barbiellini, Bastieri, Bellazzini, Bissaldi, Blandford, Bloom, Bonino, Bregeon, Britto, Bruel, Buehler, Buson, Cameron, Caputo, Caragiulo, Caraveo, Cavazzuti, Cecchi, Charles, Chekhtman, Cheung, Chiaro, Ciprini, Cohen, Costantin, Costanza, Cuoco, Cutini, D’Ammando, de~Palma, Desiante, Digel, Lalla, Mauro, Venere, Domínguez, Drell, Dumora, Favuzzi, Fegan, Ferrara, Fortin, Franckowiak, Fukazawa, Funk, Fusco, Gargano, Gasparrini, Giglietto, Giommi, Giordano, Giroletti, Glanzman, Green, Grenier, Grondin, Grove, Guillemot, Guiriec, Harding, Hays, Hewitt, Horan, Jóhannesson, Kensei, Kuss, Mura, Larsson, Latronico, Lemoine-Goumard, Li, Longo, Loparco, Lott, Lubrano, Magill, Maldera, Manfreda, Mazziotta, McEnery, Meyer, Michelson, Mirabal, Mitthumsiri, Mizuno, Moiseev, Monzani, Morselli, Moskalenko, Negro, Nuss, Ohsugi, Omodei, Orienti, Orlando, Palatiello, Paliya, Paneque, Perkins, Persic, Pesce-Rollins, Piron, Porter, Principe, Rainò, Rando,
  Razzano, Razzaque, Reimer, Reimer, Reposeur, Parkinson, Sgrò, Simone, Siskind, Spada, Spandre, Spinelli, Stawarz, Suson, Takahashi, Tak, Thayer, Thayer, Thompson, Torres, Torresi, Troja, Vianello, Wood, \& Wood}]{Ajello_2017}
Ajello, M., Atwood, W.~B., Baldini, L., {et~al.} 2017, The Astrophysical Journal Supplement Series, 232, 18, \dodoi{10.3847/1538-4365/aa8221}

\bibitem[{{Aleksi{\'c}} {et~al.}(2012){Aleksi{\'c}}, {Alvarez}, {Antonelli}, {Antoranz}, {Asensio}, {Backes}, {Barrio}, {Bastieri}, {Becerra Gonz{\'a}lez}, {Bednarek}, {Berdyugin}, {Berger}, {Bernardini}, {Biland}, {Blanch}, {Bock}, {Boller}, {Bonnoli}, {Borla Tridon}, {Braun}, {Bretz}, {Ca{\~n}ellas}, {Carmona}, {Carosi}, {Colin}, {Colombo}, {Contreras}, {Cortina}, {Cossio}, {Covino}, {Dazzi}, {de Angelis}, {de Caneva}, {de Cea Del Pozo}, {de Lotto}, {Delgado Mendez}, {Diago Ortega}, {Doert}, {Dom{\'\i}nguez}, {Dominis Prester}, {Dorner}, {Doro}, {Elsaesser}, {Ferenc}, {Fonseca}, {Font}, {Fruck}, {Garc{\'\i}a L{\'o}pez}, {Garczarczyk}, {Garrido}, {Giavitto}, {Godinovi{\'c}}, {Hadasch}, {H{\"a}fner}, {Herrero}, {Hildebrand}, {H{\"o}hne-M{\"o}nch}, {Hose}, {Hrupec}, {Huber}, {Jogler}, {Kellermann}, {Klepser}, {Kr{\"a}henb{\"u}hl}, {Krause}, {La Barbera}, {Lelas}, {Leonardo}, {Lindfors}, {Lombardi}, {L{\'o}pez}, {L{\'o}pez-Oramas}, {Lorenz}, {Makariev}, {Maneva}, {Mankuzhiyil}, {Mannheim}, {Maraschi},
  {Mariotti}, {Mart{\'\i}nez}, {Mazin}, {Meucci}, {Miranda}, {Mirzoyan}, {Miyamoto}, {Mold{\'o}n}, {Moralejo}, {Munar-Adrover}, {Nieto}, {Nilsson}, {Orito}, {Oya}, {Paneque}, {Paoletti}, {Pardo}, {Paredes}, {Partini}, {Pasanen}, {Pauss}, {Perez-Torres}, {Persic}, {Peruzzo}, {Pilia}, {Pochon}, {Prada}, {Prada Moroni}, {Prandini}, {Puljak}, {Reichardt}, {Reinthal}, {Rhode}, {Rib{\'o}}, {Rico}, {R{\"u}gamer}, {Saggion}, {Saito}, {Saito}, {Salvati}, {Satalecka}, {Scalzotto}, {Scapin}, {Schultz}, {Schweizer}, {Shayduk}, {Shore}, {Sillanp{\"a}{\"a}}, {Sitarek}, {Snidaric}, {Sobczynska}, {Spanier}, {Spiro}, {Stamatescu}, {Stamerra}, {Steinke}, {Storz}, {Strah}, {Suri{\'c}}, {Takalo}, {Takami}, {Tavecchio}, {Temnikov}, {Terzi{\'c}}, {Tescaro}, {Teshima}, {Tibolla}, {Torres}, {Treves}, {Uellenbeck}, {Vankov}, {Vogler}, {Wagner}, {Weitzel}, {Zabalza}, {Zandanel}, {Zanin}, {Kadenius}, {Weidinger}, \& {Buson}}]{2012A&A...539A.118A}
{Aleksi{\'c}}, J., {Alvarez}, E.~A., {Antonelli}, L.~A., {et~al.} 2012, \aap, 539, A118, \dodoi{10.1051/0004-6361/201117967}

\bibitem[{Archer {et~al.}(2018)Archer, Benbow, Bird, Brose, Buchovecky, Bugaev, Cui, Daniel, Falcone, Feng, Finley, Flinders, Fortson, Furniss, Gillanders, Hütten, Hanna, Hervet, Holder, Hughes, Humensky, Johnson, Kaaret, Kar, Kelley-Hoskins, Kieda, Krause, Krennrich, Kumar, Lang, Lin, McArthur, Moriarty, Mukherjee, Nieto, O’Brien, Ong, Otte, Park, Petrashyk, Pohl, Popkow, Pueschel, Quinn, Ragan, Reynolds, Richards, Roache, Rulten, Sadeh, Sembroski, Shahinyan, Tyler, Wakely, Weiner, Weinstein, Wells, Wilcox, Wilhelm, Williams, Collaboration, Brisken, \& Pontrelli}]{Archer_2018}
Archer, A., Benbow, W., Bird, R., {et~al.} 2018, The Astrophysical Journal, 862, 41, \dodoi{10.3847/1538-4357/aacbd0}

\bibitem[{{Arnaud}(1996)}]{xspec}
{Arnaud}, K.~A. 1996, in Astronomical Society of the Pacific Conference Series, Vol. 101, Astronomical Data Analysis Software and Systems V, ed. G.~H. {Jacoby} \& J.~{Barnes}, 17

\bibitem[{{Atwood} {et~al.}(2013){Atwood}, {Albert}, {Baldini}, {Tinivella}, {Bregeon}, {Pesce-Rollins}, {Sgr{\`o}}, {Bruel}, {Charles}, {Drlica-Wagner}, {Franckowiak}, {Jogler}, {Rochester}, {Usher}, {Wood}, {Cohen-Tanugi}, \& {Zimmer}}]{atwood2013pass}
{Atwood}, W., {Albert}, A., {Baldini}, L., {et~al.} 2013, arXiv e-prints, arXiv:1303.3514.
\newblock \doarXiv{1303.3514}

\bibitem[{{Atwood} {et~al.}(2009){Atwood}, {Abdo}, {Ackermann}, {Althouse}, {Anderson}, {Axelsson}, {Baldini}, {Ballet}, {Band}, {Barbiellini}, {Bartelt}, {Bastieri}, {Baughman}, {Bechtol}, {B{\'e}d{\'e}r{\`e}de}, {Bellardi}, {Bellazzini}, {Berenji}, {Bignami}, {Bisello}, {Bissaldi}, {Blandford}, {Bloom}, {Bogart}, {Bonamente}, {Bonnell}, {Borgland }, {Bouvier}, {Bregeon}, {Brez}, {Brigida}, {Bruel}, {Burnett}, {Busetto}, {Caliandro}, {Cameron}, {Caraveo}, {Carius}, {Carlson}, {Casandjian}, {Cavazzuti}, {Ceccanti}, {Cecchi}, {Charles}, {Chekhtman}, {Cheung}, {Chiang}, {Chipaux}, {Cillis}, {Ciprini}, {Claus}, {Cohen-Tanugi}, {Condamoor}, {Conrad}, {Corbet}, {Corucci}, {Costamante}, {Cutini}, {Davis}, {Decotigny}, {DeKlotz}, {Dermer}, {de Angelis}, {Digel}, {do Couto e Silva}, {Drell}, {Dubois}, {Dumora}, {Edmonds}, {Fabiani}, {Farnier}, {Favuzzi}, {Flath}, {Fleury}, {Focke}, {Funk}, {Fusco}, {Gargano}, {Gasparrini}, {Gehrels}, {Gentit}, {Germani}, {Giebels}, {Giglietto}, {Giommi}, {Giordano}, {Glanzman},
  {Godfrey}, {Grenier}, {Grondin}, {Grove}, {Guillemot}, {Guiriec}, {Haller}, {Harding}, {Hart}, {Hays}, {Healey}, {Hirayama}, {Hjalmarsdotter}, {Horn}, {Hughes}, {J{\'o}hannesson}, {Johansson}, {Johnson}, {Johnson}, {Johnson}, {Johnson}, {Kamae}, {Katagiri}, {Kataoka}, {Kavelaars}, {Kawai}, {Kelly}, {Kerr}, {Klamra}, {Kn{\"o}dlseder}, {Kocian}, {Komin}, {Kuehn}, {Kuss}, {Landriu}, {Latronico}, {Lee}, {Lee}, {Lemoine-Goumard}, {Lionetto}, {Longo}, {Loparco}, {Lott}, {Lovellette}, {Lubrano}, {Madejski}, {Makeev}, {Marangelli}, {Massai}, {Mazziotta}, {McEnery}, {Menon}, {Meurer}, {Michelson}, {Minuti}, {Mirizzi}, {Mitthumsiri}, {Mizuno}, {Moiseev}, {Monte}, {Monzani}, {Moretti}, {Morselli}, {Moskalenko}, {Murgia}, {Nakamori}, {Nishino}, {Nolan}, {Norris}, {Nuss}, {Ohno}, {Ohsugi}, {Omodei}, {Orlando}, {Ormes}, {Paccagnella}, {Paneque}, {Panetta}, {Parent}, {Pearce}, {Pepe}, {Perazzo}, {Pesce-Rollins}, {Picozza}, {Pieri}, {Pinchera}, {Piron}, {Porter}, {Poupard}, {Rain{\`o}}, {Rando}, {Rapposelli}, {Razzano},
  {Reimer}, {Reimer}, {Reposeur}, {Reyes}, {Ritz}, {Rochester}, {Rodriguez}, {Romani}, {Roth}, {Russell}, {Ryde}, {Sabatini}, {Sadrozinski}, {Sanchez}, {Sand er}, {Sapozhnikov}, {Parkinson}, {Scargle}, {Schalk}, {Scolieri}, {Sgr{\`o}}, {Share}, {Shaw}, {Shimokawabe}, {Shrader}, {Sierpowska-Bartosik}, {Siskind}, {Smith}, {Smith}, {Spandre}, {Spinelli}, {Starck}, {Stephens}, {Strickman}, {Strong}, {Suson}, {Tajima}, {Takahashi}, {Takahashi}, {Tanaka}, {Tenze}, {Tether}, {Thayer}, {Thayer}, {Thompson}, {Tibaldo}, {Tibolla}, {Torres}, {Tosti}, {Tramacere}, {Turri}, {Usher}, {Vilchez}, {Vitale}, {Wang}, {Watters}, {Winer}, {Wood}, {Ylinen}, \& {Ziegler}}]{Fermi_LAT}
{Atwood}, W.~B., {Abdo}, A.~A., {Ackermann}, M., {et~al.} 2009, \apj, 697, 1071, \dodoi{10.1088/0004-637X/697/2/1071}

\bibitem[{Bennett {et~al.}(2014)Bennett, Larson, Weiland, \& Hinshaw}]{Bennett_2014}
Bennett, C.~L., Larson, D., Weiland, J.~L., \& Hinshaw, G. 2014, The Astrophysical Journal, 794, 135, \dodoi{10.1088/0004-637X/794/2/135}

\bibitem[{{Berge} {et~al.}(2007){Berge}, {Funk}, \& {Hinton}}]{2007A&A...466.1219B}
{Berge}, D., {Funk}, S., \& {Hinton}, J. 2007, \aap, 466, 1219, \dodoi{10.1051/0004-6361:20066674}

\bibitem[{{Burrows} {et~al.}(2005){Burrows}, {Hill}, {Nousek}, {Kennea}, {Wells}, {Osborne}, {Abbey}, {Beardmore}, {Mukerjee}, {Short}, {Chincarini}, {Campana}, {Citterio}, {Moretti}, {Pagani}, {Tagliaferri}, {Giommi}, {Capalbi}, {Tamburelli}, {Angelini}, {Cusumano}, {Br{\"a}uninger}, {Burkert}, \& {Hartner}}]{2005_Burrows}
{Burrows}, D.~N., {Hill}, J.~E., {Nousek}, J.~A., {et~al.} 2005, \ssr, 120, 165, \dodoi{10.1007/s11214-005-5097-2}

\bibitem[{{Cerruti} {et~al.}(2019){Cerruti}, {Zech}, {Boisson}, {Emery}, {Inoue}, \& {Lenain}}]{2019MNRAS.483L..12C}
{Cerruti}, M., {Zech}, A., {Boisson}, C., {et~al.} 2019, \mnras, 483, L12, \dodoi{10.1093/mnrasl/sly210}

\bibitem[{{Cogan}(2008)}]{VEGAS}
{Cogan}, P. 2008, in International Cosmic Ray Conference, Vol.~3, International Cosmic Ray Conference, 1385--1388.
\newblock \doarXiv{0709.4233}

\bibitem[{{de Vaucouleurs} {et~al.}(1991){de Vaucouleurs}, {de Vaucouleurs}, {Corwin}, {Buta}, {Paturel}, \& {Fouque}}]{deVaucouleurs_1991}
{de Vaucouleurs}, G., {de Vaucouleurs}, A., {Corwin}, Herold~G., J., {et~al.} 1991, {Third Reference Catalogue of Bright Galaxies}

\bibitem[{{Evans} {et~al.}(2007){Evans}, {Beardmore}, {Page}, {Tyler}, {Osborne}, {Goad}, {O'Brien}, {Vetere}, {Racusin}, {Morris}, {Burrows}, {Capalbi}, {Perri}, {Gehrels}, \& {Romano}}]{Evans07}
{Evans}, P.~A., {Beardmore}, A.~P., {Page}, K.~L., {et~al.} 2007, \aap, 469, 379, \dodoi{10.1051/0004-6361:20077530}

\bibitem[{{Evans} {et~al.}(2009){Evans}, {Beardmore}, {Page}, {Osborne}, {O'Brien}, {Willingale}, {Starling}, {Burrows}, {Godet}, {Vetere}, {Racusin}, {Goad}, {Wiersema}, {Angelini}, {Capalbi}, {Chincarini}, {Gehrels}, {Kennea}, {Margutti}, {Morris}, {Mountford}, {Pagani}, {Perri}, {Romano}, \& {Tanvir}}]{Evans09}
---. 2009, \mnras, 397, 1177, \dodoi{10.1111/j.1365-2966.2009.14913.x}

\bibitem[{{Fact Collaboration} {et~al.}(2022){Fact Collaboration}, {The H.~E.~S.~S. Collaboration}, {The IceCube Collaboration}, {The MAGIC Collaboration}, {The VERITAS Collaboration}, {Acciari}, {Ansoldi}, {Antonelli}, {Arbet Engels}, {Artero}, {Asano}, {Baack}, {Babic}, {Baquero}, {Barres de Almeida}, {Barrio}, {Batkovi{\'c}}, {Becerra Gonzalez}, {Bednarek}, {Bellizzi}, {Bernardini}, {Bernardos}, {Berti}, {Besenrieder}, {Bhattacharyya}, {Bigongiari}, {Biland}, {Blanch}, {B{\"o}kenkamp}, {Bonnoli}, {Bosnjak}, {Busetto}, {Carosi}, {Ceribella}, {Cerruti}, {Chai}, {Chilingarian}, {Cikota}, {Colak}, {Colombo}, {Contreras}, {Cortina}, {Covino}, {D'Amico}, {D'Elia}, {da Vela}, {Dazzi}, {de Angelis}, {de Lotto}, {Delfino}, {Delgado}, {Delgado Mendez}, {Depaoli}, {di Pierro}, {di Venere}, {Do Souto Espi{\~n}eira}, {Dominis Prester}, {Donini}, {Dorner}, {Doro}, {Elsaesser}, {Fallah Ramazani}, {Fattorini}, {Fonseca}, {Font}, {Fruck}, {Fukami}, {Fukazawa}, {Garc{\'\i}a L{\'o}pez}, {Garczarczyk}, {Gasparyan}, {Gaug},
  {Giglietto}, {Giordano}, {Gliwny}, {Godinovic}, {Green}, {Green}, {Hadasch}, {Hahn}, {Heckmann}, {Herrera}, {Hoang}, {Hrupec}, {H{\"u}tten}, {Inada}, {Ishio}, {Iwamura}, {Jim{\'e}nez Mart{\'\i}nez}, {Jormanainen}, {Jouvin}, {Karjalainen}, {Kerszberg}, {Kobayashi}, {Kubo}, {Kushida}, {Lamastra}, {Lelas}, {Leone}, {Lindfors}, {Linhoff}, {Lombardi}, {Longo}, {Lopez-Coto}, {L{\'o}pez-Moya}, {L{\'o}pez-Oramas}, {Loporchio}, {Machado de Oliveira Fraga}, {Maggio}, {Majumdar}, {Makariev}, {Mallamaci}, {Maneva}, {Manganaro}, {Mannheim}, {Maraschi}, {Mariotti}, {Martinez}, {Mazin}, {Menchiari}, {Mender}, {Mi{\'c}anovi{\'c}}, {Miceli}, {Miener}, {Miranda}, {Mirzoyan}, {Molina}, {Moralejo}, {Morcuende}, {Moreno}, {Moretti}, {Nakamori}, {Nava}, {Neustroev}, {Nigro}, {Nilsson}, {Nishijima}, {Noda}, {Nozaki}, {Ohtani}, {Oka}, {Otero-Santos}, {Paiano}, {Palatiello}, {Paneque}, {Paoletti}, {Paredes}, {Pavleti{\'c}}, {Pe{\~n}il}, {Persic}, {Pihet}, {Prada Moroni}, {Prandini}, {Priyadarshi}, {Puljak}, {Rhode}, {Rib{\'o}},
  {Rico}, {Righi}, {Rugliancich}, {Sahakyan}, {Saito}, {Sakurai}, {Satalecka}, {Saturni}, {Schleicher}, {Schmidt}, {Schweizer}, {Sitarek}, {{\v{S}}nidari{\'c}}, {Sobczy{\'n}ska}, {Spolon}, {Stamerra}, {Stri{\v{s}}kovi{\'c}}, {Strom}, {Strzys}, {Suda}, {Suri{\'c}}, {Takahashi}, {Takeishi}, {Tavecchio}, {Temnikov}, {Terzic}, {Teshima}, {Tosti}, {Truzzi}, {Tutone}, {Ubach}, {van Scherpenberg}, {Vanzo}, {Vazquez Acosta}, {Ventura}, {Verguilov}, {Vigorito}, {Vitale}, {Vovk}, {Will}, {Wunderlich}, {Yamamoto}, {Zari{\'c}}, {Balbo}, {Bretz}, {Buss}, {Eisenberger}, {Hildebrand}, {Iotov}, {Kalenski}, {Neise}, {Noethe}, {Paravac}, {Sliusar}, {Walter}, {Abbasi}, {Ackermann}, {Adams}, {Aguilar}, {Ahlers}, {Ahrens}, {Alispach}, {Alves Junior}, {Amin}, {An}, {Andeen}, {Anderson}, {Anton}, {Arguelles}, {Ashida}, {Axani}, {Bai}, {Balagopal v.}, {Barbano}, {Barwick}, {Bastian}, {Basu}, {Baur}, {Bay}, {Beatty}, {Becker}, {Becker Tjus}, {Bellenghi}, {Benzvi}, {Berley}, {Besson}, {Binder}, {Bindig}, {Blaufuss}, {Blot},
  {Boddenberg}, {Bontempo}, {Borowka}, {Boser}, {Botner}, {Bottcher}, {Bourbeau}, {Bradascio}, {Braun}, {Bron}, {Brostean-Kaiser}, {Browne}, {Burgman}, {Burley}, {Busse}, {Campana}, {Carnie-Bronca}, {Chen}, {Chirkin}, {Choi}, {Clark}, {Clark}, {Classen}, {Coleman}, {Collin}, {Conrad}, {Coppin}, {Correa}, {Cowen}, {Cross}, {Dappen}, {Dave}, {de Clercq}, {Delaunay}, {Dembinski}, {Deoskar}, {De Ridder}, {Desai}, {Desiati}, {de Vries}, {de Wasseige}, {de With}, {Deyoung}, {Dharani}, {Diaz}, {Diaz-Velez}, {Dittmer}, {Dujmovic}, {Dunkman}, {Duvernois}, {Dvorak}, {Ehrhardt}, {Eller}, {Engel}, {Erpenbeck}, {Evans}, {Evenson}, {Fan}, {Fazely}, {Fiedlschuster}, {Fienberg}, {Filimonov}, {Finley}, {Fischer}, {Fox}, {Franckowiak}, {Friedman}, {Fritz}, {Furst}, {Gaisser}, {Gallagher}, {Ganster}, {Garcia}, {Garrappa}, {Gerhardt}, {Ghadimi}, {Glaser}, {Glauch}, {Glusenkamp}, {Goldschmidt}, {Gonzalez}, {Goswami}, {Grant}, {Gr{\'e}goire}, {Griswold}, {Gunduz}, {G{\"u}nther}, {Haack}, {Hallgren}, {Halliday}, {Halve}, {Halzen},
  {Minh}, {Hanson}, {Hardin}, {Harnisch}, {Haungs}, {Hauser}, {Hebecker}, {Helbing}, {Henningsen}, {Hettinger}, {Hickford}, {Hignight}, {Hill}, {Hill}, {Hoffman}, {Hoffmann}, {Hoinka}, {Hokanson-Fasig}, {Hoshina}, {Huang}, {Huber}, {Huber}, {Hultqvist}, {Hunnefeld}, {Hussain}, {in}, {Iovine}, {Ishihara}, {Jansson}, {Japaridze}, {Jeong}, {Jones}, {Kang}, {Kang}, {Kang}, {Kappes}, {Kappesser}, {Karg}, {Karl}, {Karle}, {Katz}, {Kauer}, {Kellermann}, {Kelley}, {Kheirandish}, {Kin}, {Kintscher}, {Kiryluk}, {Klein}, {Koirala}, {Kolanoski}, {Kontrimas}, {Kopke}, {Kopper}, {Kopper}, {Koskinen}, {Koundal}, {Kovacevich}, {Kowalski}, {Kozynets}, {Kun}, {Kurahashi}, {Lad}, {Lagunas Gualda}, {Lanfranchi}, {Larson}, {Lauber}, {Lazar}, {Lee}, {Leonard}, {Leszczy{\'n}ska}, {Li}, {Lincetto}, {Liu}, {Liubarska}, {Lohfink}, {Lozano Mariscal}, {Lu}, {Lucarelli}, {Ludwig}, {Luszczak}, {Lyu}, {Ma}, {Madsen}, {Mahn}, {Makino}, {Mancina}, {Maris}, {Maruyama}, {Mase}, {McElroy}, {McNally}, {Mead}, {Meagher}, {Medina}, {Meier},
  {Meighen-Berger}, {Micallef}, {Mockler}, {Montaruli}, {Moore}, {Morse}, {Moulai}, {Naab}, {Nagai}, {Naumann}, {Necker}, {Nguyen}, {Niederhausen}, {Nisa}, {Nowicki}, {Nygren}, {Obertacke Pollmann}, {Oehler}, {Olivas}, {O'Sullivan}, {Pandya}, {Pankova}, {Park}, {Parker}, {Paudel}, {Paul}, {Perez de Los Heros}, {Peters}, {Philippen}, {Pieloth}, {Pieper}, {Pittermann}, {Pizzuto}, {Plum}, {Popovych}, {Porcelli}, {Prado Rodriguez}, {Price}, {Pries}, {Przybylski}, {Raab}, {Raissi}, {Rameez}, {Rawlins}, {Rea}, {Rehman}, {Reimann}, {Renzi}, {Resconi}, {Reusch}, {Richman}, {Riedel}, {Roberts}, {Robertson}, {Roellinghoff}, {Rongen}, {Rott}, {Ruhe}, {Ryckbosch}, {Rysewyk Cantu}, {Safa}, {Saffer}, {Sanchez Herrera}, {Sandrock}, {Sandroos}, {Santander}, {Sarkar}, {Sarkar}, {Scharf}, {Schaufel}, {Schieler}, {Schindler}, {Schlunder}, {Schmidt}, {Schneider}, {Schneider}, {Schr{\"o}der}, {Schumacher}, {Schwefer}, {Sclafani}, {Seckel}, {Seunarine}, {Sharma}, {Shefali}, {Silva}, {Skrzypek}, {Smithers}, {Snihur},
  {Soedingrekso}, {Soldin}, {Spannfellner}, {Spiczak}, {Spiering}, {Stachurska}, {Stamatikos}, {Stanev}, {Stein}, {Stettner}, {Steuer}, {Stezelberger}, {Sturwald}, {Stuttard}, {Sullivan}, {Taboada}, {Tenholt}, {Ter-Antonyan}, {Tilav}, {Tischbein}, {Tollefson}, {T{\"o}nnis}, {Toscano}, {Tosi}, {Trettin}, {Tselengidou}, {Tung}, {Turcati}, {Turcotte}, {Turley}, {Twagirayezu}, {Ty}, {Unland Elorrieta}, {Valtonen-Mattila}, {Vandenbroucke}, {van Eijndhoven}, {Vannerom}, {van Santen}, {Verpoest}, {Vraeghe}, {Walck}, {Watson}, {Weaver}, {Weigel}, {Weindl}, {Weiss}, {Weldert}, {Wendt}, {Werthebach}, {Weyrauch}, {Whitehorn}, {Wiebusch}, {Williams}, {Wolf}, {Woschnagg}, {Wrede}, {Wulff}, {Xu}, {Xu}, {Yanez}, {Yoshida}, {Yu}, {Yuan}, {Zhang}, {Jin}, {Abdalla}, {Aharonian}, {Ait-Benkhali}, {Anguener}, {Arcaro}, {Armand}, {Armstrong}, {Ashkar}, {Backes}, {Baghmanyan}, {Barbosa Martins}, {Barnacka}, {Barnard}, {Batzofin}, {Becherini}, {Berge}, {Bernl{\"o}hr}, {Bi}, {B{\"o}ttcher}, {Boisson}, {Bolmont}, {de Bony},
  {Breuhaus}, {Brose}, {Brun}, {Bulik}, {Bylund}, {Cangemi}, {Caroff}, {Casanova}, {Catalano}, {Chambery}, {Chand}, {Chen}, {Cotter}, {Curlo}, {Damascene Mbarubucyeye}, {Davids}, {Davies}, {Devin}, {Djannati-Ata{\"\i}}, {Dmytriiev}, {Donath}, {Doroshenko}, {Dreyer}, {Du Plessis}, {Duffy}, {Egberts}, {Einecke}, {Emery}, {Ernenwein}, {Fegan}, {Feijen}, {Fiasson}, {Fichet de Clairfontaine}, {Fontaine}, {Frans}, {Fuessling}, {Funk}, {Gabici}, {Gallant}, {Ghafourizade}, {Giavitto}, {Giunti}, {Glawion}, {Glicenstein}, {Grondin}, {Hattingh}, {Haupt}, {Hermann}, {Hinton}, {Hofmann}, {Hoischen}, {Holch}, {Holler}, {Horns}, {Huang}, {Huber}, {H{\"o}rbe}, {Jamrozy}, {Jankowsky}, {Joshi}, {Jung}, {Kasai}, {Katarzynski}, {Katz}, {Khangulyan}, {Khelifi}, {Klepser}, {Kluzniak}, {Komin}, {Konno}, {Kosack}, {Kostunin}, {Kreter}, {Kukec Mezek}, {Kundu}, {Lamanna}, {Le Stum}, {Lemiere}, {Lemoine-Goumard}, {Lenain}, {Leuschner}, {Levy}, {Lohse}, {Luashvili}, {Lypova}, {Mackey}, {Majumdar}, {Malyshev}, {Malyshev}, {Marandon},
  {Marchegiani}, {Marcowith}, {Mares}, {Marti'I-Devesa}, {Marx}, {Maurin}, {Meintjes}, {Meyer}, {Mitchell}, {Moderski}, {Mohrmann}, {Montanari}, {Moore}, {Morris}, {Moulin}, {Muller}, {Murach}, {Nakashima}, {Naurois (de)}, {Nayerhoda}, {Davids}, {Niemiec}, {Noel}, {O'Brien}, {Oberholzer}, {Ohm}, {Olivera-Nieto}, {Ona-Wilhelmi (de)}, {Ostrowski}, {Panny}, {Panter}, {Parsons}, {Peron}, {Pita}, {Poireau}, {Prokhorov}, {Prokoph}, {Puehlhofer}, {Punch}, {Quirrenbach}, {Reichherzer}, {Reimer}, {Reimer}, {Remy}, {Renaud}, {Reville}, {Rieger}, {Romoli}, {Rowell}, {Rudak}, {Rueda Ricarte}, {Ruiz Velasco}, {Sahakian}, {Sailer}, {Salzmann}, {Sanchez}, {Santangelo}, {Sasaki}, {Schaefer}, {Schutte}, {Schwanke}, {Sch{\"u}ssler}, {Senniappan}, {Seyffert}, {Shapopi}, {Shiningayamwe}, {Simoni}, {Sinha}, {Sol}, {Spackman}, {Specovius}, {Spencer}, {Spir-Jacob}, {Stawarz}, {Steenkamp}, {Stegmann}, {Steinmassl}, {Steppa}, {Sun}, {Takahashi}, {Tanaka}, {Tavernier}, {Taylor}, {Terrier}, {Thiersen}, {Thorpe-Morgan}, {Tluczykont},
  {Tomankova}, {Tsirou}, {Tsuji}, {Tuffs}, {Uchiyama}, {van der Walt}, {van Eldik}, {van Rensburg}, {van Soelen}, {Vasileiadis}, {Veh}, {Venter}, {Vincent}, {Vink}, {V{\"o}lk}, {Wagner}, {Watson}, {Werner}, {White}, {Wierzcholska}, {Wong}, {Yassin}, {Yusafzai}, {Zacharias}, {Zanin}, {Zargaryan}, {Zdziarski}, {Zech}, {Zhu}, {Zmija}, {Zouari}, \& {{\.Z}ywucka}}]{2022_gfu}
{Fact Collaboration}, {The H.~E.~S.~S. Collaboration}, {The IceCube Collaboration}, {et~al.} 2022, in 37th International Cosmic Ray Conference, 960, \dodoi{10.22323/1.395.0960}

\bibitem[{{Falco} {et~al.}(1998){Falco}, {Kochanek}, \& {Mu{\~n}oz}}]{1998ApJ...494...47F}
{Falco}, E.~E., {Kochanek}, C.~S., \& {Mu{\~n}oz}, J.~A. 1998, \apj, 494, 47, \dodoi{10.1086/305207}

\bibitem[{{Fioc} \& {Rocca-Volmerange}(1999)}]{Fioc_1999}
{Fioc}, M., \& {Rocca-Volmerange}, B. 1999, arXiv e-prints, astro.
\newblock \doarXiv{astro-ph/9912179}

\bibitem[{Fomin {et~al.}(1994)Fomin, Fennell, Lamb, Lewis, Punch, \& Weekes}]{FOMIN1994151}
Fomin, V., Fennell, S., Lamb, R., {et~al.} 1994, Astroparticle Physics, 2, 151 , \dodoi{http://dx.doi.org/10.1016/0927-6505(94)90037-X}

\bibitem[{{Franceschini} \& {Rodighiero}(2017)}]{2017A&A...603A..34F}
{Franceschini}, A., \& {Rodighiero}, G. 2017, \aap, 603, A34, \dodoi{10.1051/0004-6361/201629684}

\bibitem[{{Franckowiak} {et~al.}(2020){Franckowiak}, {Garrappa}, {Paliya}, {Shappee}, {Stein}, {Strotjohann}, {Kowalski}, {Buson}, {Kiehlmann}, {Max-Moerbeck}, \& {Angioni}}]{2020ApJ...893..162F}
{Franckowiak}, A., {Garrappa}, S., {Paliya}, V., {et~al.} 2020, \apj, 893, 162, \dodoi{10.3847/1538-4357/ab8307}

\bibitem[{{Gaisser} {et~al.}(1995){Gaisser}, {Halzen}, \& {Stanev}}]{1995PhR...258..173G}
{Gaisser}, T.~K., {Halzen}, F., \& {Stanev}, T. 1995, \physrep, 258, 173, \dodoi{10.1016/0370-1573(95)00003-Y}

\bibitem[{{Gao} {et~al.}(2019){Gao}, {Fedynitch}, {Winter}, \& {Pohl}}]{2019_Gao}
{Gao}, S., {Fedynitch}, A., {Winter}, W., \& {Pohl}, M. 2019, Nature Astronomy, 3, 88, \dodoi{10.1038/s41550-018-0610-1}

\bibitem[{{Gao} {et~al.}(2017){Gao}, {Pohl}, \& {Winter}}]{2017ApJ...843..109G}
{Gao}, S., {Pohl}, M., \& {Winter}, W. 2017, \apj, 843, 109, \dodoi{10.3847/1538-4357/aa7754}

\bibitem[{{Garrappa} {et~al.}(2024){Garrappa}, {Buson}, {Sinapius}, {Franckowiak}, {Liodakis}, {Bartolini}, {Giroletti}, {Nanci}, {Principe}, \& {Venters}}]{2024A&A...687A..59G}
{Garrappa}, S., {Buson}, S., {Sinapius}, J., {et~al.} 2024, \aap, 687, A59, \dodoi{10.1051/0004-6361/202449221}

\bibitem[{{Gehrels}(2004)}]{2004AIPC..727..637G}
{Gehrels}, N. 2004, in American Institute of Physics Conference Series, Vol. 727, Gamma-Ray Bursts: 30 Years of Discovery, ed. E.~{Fenimore} \& M.~{Galassi}, 637--641, \dodoi{10.1063/1.1810924}

\bibitem[{{Giommi} {et~al.}(2020){Giommi}, {Padovani}, {Oikonomou}, {Glauch}, {Paiano}, \& {Resconi}}]{2020A&A...640L...4G}
{Giommi}, P., {Padovani}, P., {Oikonomou}, F., {et~al.} 2020, \aap, 640, L4, \dodoi{10.1051/0004-6361/202038423}

\bibitem[{{Harrison} {et~al.}(2013){Harrison}, {Craig}, {Christensen}, {Hailey}, {Zhang}, {Boggs}, {Stern}, {Cook}, {Forster}, {Giommi}, {Grefenstette}, {Kim}, {Kitaguchi}, {Koglin}, {Madsen}, {Mao}, {Miyasaka}, {Mori}, {Perri}, {Pivovaroff}, {Puccetti}, {Rana}, {Westergaard}, {Willis}, {Zoglauer}, {An}, {Bachetti}, {Barri{\`e}re}, {Bellm}, {Bhalerao}, {Brejnholt}, {Fuerst}, {Liebe}, {Markwardt}, {Nynka}, {Vogel}, {Walton}, {Wik}, {Alexander}, {Cominsky}, {Hornschemeier}, {Hornstrup}, {Kaspi}, {Madejski}, {Matt}, {Molendi}, {Smith}, {Tomsick}, {Ajello}, {Ballantyne}, {Balokovi{\'c}}, {Barret}, {Bauer}, {Blandford}, {Brandt}, {Brenneman}, {Chiang}, {Chakrabarty}, {Chenevez}, {Comastri}, {Dufour}, {Elvis}, {Fabian}, {Farrah}, {Fryer}, {Gotthelf}, {Grindlay}, {Helfand}, {Krivonos}, {Meier}, {Miller}, {Natalucci}, {Ogle}, {Ofek}, {Ptak}, {Reynolds}, {Rigby}, {Tagliaferri}, {Thorsett}, {Treister}, \& {Urry}}]{2013_NuSTAR}
{Harrison}, F.~A., {Craig}, W.~W., {Christensen}, F.~E., {et~al.} 2013, \apj, 770, 103, \dodoi{10.1088/0004-637X/770/2/103}

\bibitem[{Hervet {et~al.}(2024)Hervet, Johnson, \& Youngquist}]{2023_Bjet}
Hervet, O., Johnson, C.~A., \& Youngquist, A. 2024, The Astrophysical Journal, 962, 140, \dodoi{10.3847/1538-4357/ad09c0}

\bibitem[{{Hervet, O.} {et~al.}(2015){Hervet, O.}, {Boisson, C.}, \& {Sol, H.}}]{refId0}
{Hervet, O.}, {Boisson, C.}, \& {Sol, H.} 2015, A\&A, 578, A69, \dodoi{10.1051/0004-6361/201425330}

\bibitem[{{Holder}(2011)}]{vts_paper}
{Holder}, J. 2011, in International Cosmic Ray Conference, Vol.~11, International Cosmic Ray Conference, 137, \dodoi{10.7529/ICRC2011/V12/H11}

\bibitem[{{IceCube Collaboration}(2021)}]{2021GCN.31191....1I}
{IceCube Collaboration}. 2021, GRB Coordinates Network, 31191, 1

\bibitem[{{IceCube Collaboration} {et~al.}(2016){IceCube Collaboration}, {Aartsen}, {Abraham}, {Ackermann}, {Adams}, {Aguilar}, {Ahlers}, {Ahrens}, {Altmann}, {Andeen}, {Anderson}, {Ansseau}, {Anton}, {Archinger}, {Arg{\"u}elles}, {Auffenberg}, {Axani}, {Bai}, {Barwick}, {Baum}, {Bay}, {Beatty}, {Becker Tjus}, {Becker}, {BenZvi}, {Berley}, {Bernardini}, {Bernhard}, {Besson}, {Binder}, {Bindig}, {Bissok}, {Blaufuss}, {Blot}, {Bohm}, {B{\"o}rner}, {Bos}, {Bose}, {B{\"o}ser}, {Botner}, {Braun}, {Brayeur}, {Bretz}, {Bron}, {Burgman}, {Carver}, {Casier}, {Cheung}, {Chirkin}, {Christov}, {Clark}, {Classen}, {Coenders}, {Collin}, {Conrad}, {Cowen}, {Cross}, {Day}, {de Andr{\'e}}, {De Clercq}, {del Pino Rosendo}, {Dembinski}, {De Ridder}, {Desiati}, {de Vries}, {de Wasseige}, {de With}, {DeYoung}, {D{\'\i}az-V{\'e}lez}, {di Lorenzo}, {Dujmovic}, {Dumm}, {Dunkman}, {Eberhardt}, {Ehrhardt}, {Eichmann}, {Eller}, {Euler}, {Evenson}, {Fahey}, {Fazely}, {Feintzeig}, {Felde}, {Filimonov}, {Finley}, {Flis}, {F{\"o}sig},
  {Franckowiak}, {Franke}, {Friedman}, {Fuchs}, {Gaisser}, {Gallagher}, {Gerhardt}, {Ghorbani}, {Giang}, {Gladstone}, {Glauch}, {Gl{\"u}senkamp}, {Goldschmidt}, {Golup}, {Gonzalez}, {Grant}, {Griffith}, {Haack}, {Haj Ismail}, {Hallgren}, {Halzen}, {Hansen}, {Hansmann}, {Hanson}, {Hebecker}, {Heereman}, {Helbing}, {Hellauer}, {Hickford}, {Hignight}, {Hill}, {Hoffman}, {Hoffmann}, {Holzapfel}, {Hoshina}, {Huang}, {Huber}, {Hultqvist}, {In}, {Ishihara}, {Jacobi}, {Japaridze}, {Jeong}, {Jero}, {Jones}, {Jurkovic}, {Kappes}, {Karg}, {Karle}, {Katz}, {Kauer}, {Keivani}, {Kelley}, {Kheirandish}, {Kim}, {Kintscher}, {Kiryluk}, {Kittler}, {Klein}, {Kohnen}, {Koirala}, {Kolanoski}, {Konietz}, {K{\"o}pke}, {Kopper}, {Kopper}, {Koskinen}, {Kowalski}, {Krings}, {Kroll}, {Kr{\"u}ckl}, {Kr{\"u}ger}, {Kunnen}, {Kunwar}, {Kurahashi}, {Kuwabara}, {Labare}, {Lanfranchi}, {Larson}, {Lauber}, {Lennarz}, {Lesiak-Bzdak}, {Leuermann}, {Lu}, {L{\"u}nemann}, {Madsen}, {Maggi}, {Mahn}, {Mancina}, {Mandelartz}, {Maruyama}, {Mase},
  {Maunu}, {McNally}, {Meagher}, {Medici}, {Meier}, {Meli}, {Menne}, {Merino}, {Meures}, {Miarecki}, {Mohrmann}, {Montaruli}, {Moulai}, {Nahnhauer}, {Naumann}, {Neer}, {Niederhausen}, {Nowicki}, {Nygren}, {Obertacke Pollmann}, {Olivas}, {O'Murchadha}, {Palczewski}, {Pandya}, {Pankova}, {Peiffer}, {Penek}, {Pepper}, {P{\'e}rez de los Heros}, {Pieloth}, {Pinat}, {Price}, {Przybylski}, {Quinnan}, {Raab}, {R{\"a}del}, {Rameez}, {Rawlins}, {Reimann}, {Relethford}, {Relich}, {Resconi}, {Rhode}, {Richman}, {Riedel}, {Robertson}, {Rongen}, {Rott}, {Ruhe}, {Ryckbosch}, {Rysewyk}, {Sabbatini}, {Sanchez Herrera}, {Sandrock}, {Sandroos}, {Sarkar}, {Satalecka}, {Schlunder}, {Schmidt}, {Schoenen}, {Sch{\"o}neberg}, {Schumacher}, {Seckel}, {Seunarine}, {Soldin}, {Song}, {Spiczak}, {Spiering}, {Stanev}, {Stasik}, {Stettner}, {Steuer}, {Stezelberger}, {Stokstad}, {St{\"o}{\ss}l}, {Str{\"o}m}, {Strotjohann}, {Sullivan}, {Sutherland}, {Taavola}, {Taboada}, {Tatar}, {Tenholt}, {Ter-Antonyan}, {Terliuk}, {Te{\v{s}}i{\'c}},
  {Tilav}, {Toale}, {Tobin}, {Toscano}, {Tosi}, {Tselengidou}, {Turcati}, {Unger}, {Usner}, {Vandenbroucke}, {van Eijndhoven}, {Vanheule}, {van Rossem}, {van Santen}, {Veenkamp}, {Vehring}, {Voge}, {Vogel}, {Vraeghe}, {Walck}, {Wallace}, {Wallraff}, {Wandkowsky}, {Weaver}, {Weiss}, {Wendt}, {Westerhoff}, {Whelan}, {Wickmann}, {Wiebe}, {Wiebusch}, {Wille}, {Williams}, {Wills}, {Wolf}, {Wood}, {Woolsey}, {Woschnagg}, {Xu}, {Xu}, {Xu}, {Yanez}, {Yodh}, {Yoshida}, {Zoll}, {MAGIC Collaboration}, {Ahnen}, {Ansoldi}, {Antonelli}, {Antoranz}, {Babic}, {Banerjee}, {Bangale}, {Barres de Almeida}, {Barrio}, {Becerra Gonz{\'a}lez}, {Bednarek}, {Bernardini}, {Berti}, {Biasuzzi}, {Biland}, {Blanch}, {Bonnefoy}, {Bonnoli}, {Borracci}, {Bretz}, {Buson}, {Carosi}, {Chatterjee}, {Clavero}, {Colin}, {Colombo}, {Contreras}, {Cortina}, {Covino}, {Da Vela}, {Dazzi}, {De Angelis}, {De Lotto}, {de O{\~n}a Wilhelmi}, {Di Pierro}, {Doert}, {Dom{\'\i}nguez}, {Dominis Prester}, {Dorner}, {Doro}, {Einecke}, {Eisenacher Glawion},
  {Elsaesser}, {Engelkemeier}, {Fallah Ramazani}, {Fern{\'a}ndez-Barral}, {Fidalgo}, {Fonseca}, {Font}, {Frantzen}, {Fruck}, {Galindo}, {Garc{\'\i}a L{\'o}pez}, {Garczarczyk}, {Garrido Terrats}, {Gaug}, {Giammaria}, {Godinovi{\'c}}, {Gonz{\'a}lez Mu{\~n}oz}, {G{\'o}ra}, {Guberman}, {Hadasch}, {Hahn}, {Hanabata}, {Hayashida}, {Herrera}, {Hose}, {Hrupec}, {Hughes}, {Idec}, {Kodani}, {Konno}, {Kubo}, {Kushida}, {La Barbera}, {Lelas}, {Lindfors}, {Lombardi}, {Longo}, {L{\'o}pez}, {L{\'o}pez-Coto}, {Majumdar}, {Makariev}, {Mallot}, {Maneva}, {Manganaro}, {Mannheim}, {Maraschi}, {Marcote}, {Mariotti}, {Mart{\'\i}nez}, {Mazin}, {Menzel}, {Miranda}, {Mirzoyan}, {Moralejo}, {Moretti}, {Nakajima}, {Neustroev}, {Niedzwiecki}, {Nievas Rosillo}, {Nilsson}, {Nishijima}, {Noda}, {Nogu{\'e}s}, {Overkemping}, {Paiano}, {Palacio}, {Palatiello}, {Paneque}, {Paoletti}, {Paredes}, {Paredes-Fortuny}, {Pedaletti}, {Peresano}, {Perri}, {Persic}, {Poutanen}, {Prada Moroni}, {Prandini}, {Puljak}, {Reichardt}, {Rhode}, {Rib{\'o}},
  {Rico}, {Rodriguez Garcia}, {Saito}, {Satalecka}, {Schroeder}, {Schultz}, {Schweizer}, {Sillanp{\"a}{\"a}}, {Sitarek}, {Snidaric}, {Sobczynska}, {Stamerra}, {Steinbring}, {Strzys}, {Suri{\'c}}, {Takalo}, {Tavecchio}, {Temnikov}, {Terzi{\'c}}, {Tescaro}, {Teshima}, {Thaele}, {Torres}, {Toyama}, {Treves}, {Vanzo}, {Verguilov}, {Vovk}, {Ward}, {Will}, {Wu}, {Zanin}, {VERITAS Collaboration}, {Abeysekara}, {Archambault}, {Archer}, {Benbow}, {Bird}, {Bourbeau}, {Buchovecky}, {Bugaev}, {Byrum}, {Cardenzana}, {Cerruti}, {Ciupik}, {Connolly}, {Cui}, {Dickinson}, {Dumm}, {Eisch}, {Errando}, {Falcone}, {Feng}, {Finley}, {Fleischhack}, {Flinders}, {Fortson}, {Furniss}, {Gillanders}, {Griffin}, {H{\"u}tten}, {H{\r{a}}kansson}, {Hervet}, {Holder}, {Humensky}, {Johnson}, {Kaaret}, {Kar}, {Kelley-Hoskins}, {Kertzman}, {Kieda}, {Krause}, {Krennrich}, {Kumar}, {Lang}, {Maier}, {McArthur}, {McCann}, {Moriarty}, {Mukherjee}, {Nguyen}, {Nieto}, {O'Brien}, {Ong}, {Otte}, {Park}, {Pohl}, {Popkow}, {Pueschel}, {Quinn}, {Ragan},
  {Reynolds}, {Richards}, {Roache}, {Rulten}, {Sadeh}, {Santander}, {Sembroski}, {Shahinyan}, {Staszak}, {Telezhinsky}, {Tucci}, {Tyler}, {Wakely}, {Weinstein}, {Wilcox}, {Wilhelm}, {Williams}, \& {Zitzer}}]{2016JInst..1111009I}
{IceCube Collaboration}, {Aartsen}, M.~G., {Abraham}, K., {et~al.} 2016, Journal of Instrumentation, 11, P11009, \dodoi{10.1088/1748-0221/11/11/P11009}

\bibitem[{{IceCube Collaboration} {et~al.}(2022){IceCube Collaboration}, {Abbasi}, {Ackermann}, {Adams}, {Aguilar}, {Ahlers}, {Ahrens}, {Alameddine}, {Alispach}, {Alves}, {Amin}, {Andeen}, {Anderson}, {Anton}, {Arg{\"u}elles}, {Ashida}, {Axani}, {Bai}, {Balagopal}, {Barbano}, {Barwick}, {Bastian}, {Basu}, {Baur}, {Bay}, {Beatty}, {Becker}, {Becker Tjus}, {Bellenghi}, {Benzvi}, {Berley}, {Bernardini}, {Besson}, {Binder}, {Bindig}, {Blaufuss}, {Blot}, {Boddenberg}, {Bontempo}, {Borowka}, {B{\"o}ser}, {Botner}, {B{\"o}ttcher}, {Bourbeau}, {Bradascio}, {Braun}, {Brinson}, {Bron}, {Brostean-Kaiser}, {Browne}, {Burgman}, {Burley}, {Busse}, {Campana}, {Carnie-Bronca}, {Chen}, {Chen}, {Chirkin}, {Choi}, {Clark}, {Clark}, {Classen}, {Coleman}, {Collin}, {Conrad}, {Coppin}, {Correa}, {Cowen}, {Cross}, {Dappen}, {Dave}, {de Clercq}, {Delaunay}, {Delgado L{\'o}pez}, {Dembinski}, {Deoskar}, {Desai}, {Desiati}, {de Vries}, {de Wasseige}, {de With}, {Deyoung}, {Diaz}, {D{\'\i}az-V{\'e}lez}, {Dittmer}, {Dujmovic}, {Dunkman},
  {Duvernois}, {Dvorak}, {Ehrhardt}, {Eller}, {Engel}, {Erpenbeck}, {Evans}, {Evenson}, {Fan}, {Fazely}, {Fedynitch}, {Feigl}, {Fiedlschuster}, {Fienberg}, {Filimonov}, {Finley}, {Fischer}, {Fox}, {Franckowiak}, {Friedman}, {Fritz}, {F{\"u}rst}, {Gaisser}, {Gallagher}, {Ganster}, {Garcia}, {Garrappa}, {Gerhardt}, {Ghadimi}, {Glaser}, {Glauch}, {Gl{\"u}senkamp}, {Goldschmidt}, {Gonzalez}, {Goswami}, {Grant}, {Gr{\'e}goire}, {Griswold}, {G{\"u}nther}, {Gutjahr}, {Haack}, {Hallgren}, {Halliday}, {Halve}, {Halzen}, {Hanson}, {Hardin}, {Harnisch}, {Haungs}, {Hebecker}, {Helbing}, {Henningsen}, {Hettinger}, {Hickford}, {Hignight}, {Hill}, {Hill}, {Hoffman}, {Hoffmann}, {Hokanson-Fasig}, {Hoshina}, {Huang}, {Huber}, {Huber}, {Hultqvist}, {H{\"u}nnefeld}, {Hussain}, {Hymon}, {in}, {Iovine}, {Ishihara}, {Jansson}, {Japaridze}, {Jeong}, {Jin}, {Jones}, {Kang}, {Kang}, {Kang}, {Kappes}, {Kappesser}, {Kardum}, {Karg}, {Karl}, {Karle}, {Katz}, {Kauer}, {Kellermann}, {Kelley}, {Kheirandish}, {Kin}, {Kintscher}, {Kiryluk},
  {Klein}, {Koirala}, {Kolanoski}, {Kontrimas}, {K{\"o}pke}, {Kopper}, {Kopper}, {Koskinen}, {Koundal}, {Kovacevich}, {Kowalski}, {Kozynets}, {Kun}, {Kurahashi}, {Lad}, {Lagunas Gualda}, {Lanfranchi}, {Larson}, {Lauber}, {Lazar}, {Lee}, {Leonard}, {Leszczy{\'n}ska}, {Li}, {Lincetto}, {Liu}, {Liubarska}, {Lohfink}, {Lozano Mariscal}, {Lu}, {Lucarelli}, {Ludwig}, {Luszczak}, {Lyu}, {Ma}, {Madsen}, {Mahn}, {Makino}, {Mancina}, {Mari{\c{s}}}, {Martinez-Soler}, {Maruyama}, {Mase}, {McElroy}, {McNally}, {Mead}, {Meagher}, {Mechbal}, {Medina}, {Meier}, {Meighen-Berger}, {Micallef}, {Mockler}, {Montaruli}, {Moore}, {Morse}, {Moulai}, {Naab}, {Nagai}, {Nahnhauer}, {Naumann}, {Necker}, {Nguyen}, {Niederhausen}, {Nisa}, {Nowicki}, {Nygren}, {Obertack}, {Pollmann}, {Oehler}, {Oeyen}, {Olivas}, {O'Sullivan}, {Pandya}, {Pankova}, {Park}, {Parker}, {Paudel}, {Paul}, {P{\'e}rez de Los Heros}, {Peters}, {Peterson}, {Philippen}, {Pieper}, {Pittermann}, {Pizzuto}, {Plum}, {Popovych}, {Porcelli}, {Prado Rodriguez}, {Price},
  {Pries}, {Przybylski}, {Rack-Helleis}, {Raissi}, {Rameez}, {Rawlins}, {Rea}, {Rehman}, {Reichherzer}, {Reimann}, {Renzi}, {Resconi}, {Reusch}, {Rhode}, {Richman}, {Riedel}, {Roberts}, {Robertson}, {Roellinghoff}, {Rongen}, {Rott}, {Ruhe}, {Ryckbosch}, {Rysewyk Cantu}, {Safa}, {Saffer}, {Sanchez Herrera}, {Sandrock}, {Sandroos}, {Santander}, {Sarkar}, {Sarkar}, {Satalecka}, {Schaufel}, {Schieler}, {Schindler}, {Schmidt}, {Schneider}, {Schneider}, {Schr{\"o}der}, {Schumacher}, {Schwefer}, {Sclafani}, {Seckel}, {Seunarine}, {Sharma}, {Shefali}, {Silva}, {Skrzypek}, {Smithers}, {Snihur}, {Soedingrekso}, {Soldin}, {Spannfellner}, {Spiczak}, {Spiering}, {Stachurska}, {Stamatikos}, {Stanev}, {Stein}, {Stettner}, {Steuer}, {Stezelberger}, {Stokstad}, {St{\"u}rwald}, {Stuttard}, {Sullivan}, {Taboada}, {Ter-Antonyan}, {Tilav}, {Tischbein}, {Tollefson}, {T{\"o}nnis}, {Toscano}, {Tosi}, {Trettin}, {Tselengidou}, {Tung}, {Turcati}, {Turcotte}, {Turley}, {Twagirayezu}, {Ty}, {Unland Elorrieta}, {Valtonen-Mattila},
  {Vandenbroucke}, {van Eijndhoven}, {Vannerom}, {van Santen}, {Verpoest}, {Walck}, {Watson}, {Weaver}, {Weigel}, {Weindl}, {Weiss}, {Weldert}, {Wendt}, {Werthebach}, {Weyrauch}, {Whitehorn}, {Wiebusch}, {Williams}, {Wolf}, {Woschnagg}, {Wrede}, {Wulff}, {Xu}, {Yanez}, {Yoshida}, {Yu}, {Yuan}, {Zhangan}, \& {Zhelnin}}]{2022Sci...378..538I}
{IceCube Collaboration}, {Abbasi}, R., {Ackermann}, M., {et~al.} 2022, Science, 378, 538, \dodoi{10.1126/science.abg3395}

\bibitem[{{Ishihara}(2023)}]{2023arXiv230809427I}
{Ishihara}, A. 2023, arXiv e-prints, arXiv:2308.09427, \dodoi{10.48550/arXiv.2308.09427}

\bibitem[{{Jones} {et~al.}(1974){Jones}, {O'Dell}, \& {Stein}}]{1974_Jones}
{Jones}, T.~W., {O'Dell}, S.~L., \& {Stein}, W.~A. 1974, \apj, 188, 353, \dodoi{10.1086/152724}

\bibitem[{{Kadler} {et~al.}(2016){Kadler}, {Krau{\ss}}, {Mannheim}, {Ojha}, {M{\"u}ller}, {Schulz}, {Anton}, {Baumgartner}, {Beuchert}, {Buson}, {Carpenter}, {Eberl}, {Edwards}, {Eisenacher Glawion}, {Els{\"a}sser}, {Gehrels}, {Gr{\"a}fe}, {Gulyaev}, {Hase}, {Horiuchi}, {James}, {Kappes}, {Kappes}, {Katz}, {Kreikenbohm}, {Kreter}, {Kreykenbohm}, {Langejahn}, {Leiter}, {Litzinger}, {Longo}, {Lovell}, {McEnery}, {Natusch}, {Phillips}, {Pl{\"o}tz}, {Quick}, {Ros}, {Stecker}, {Steinbring}, {Stevens}, {Thompson}, {Tr{\"u}stedt}, {Tzioumis}, {Weston}, {Wilms}, \& {Zensus}}]{2016_Kadler}
{Kadler}, M., {Krau{\ss}}, F., {Mannheim}, K., {et~al.} 2016, Nature Physics, 12, 807, \dodoi{10.1038/nphys3715}

\bibitem[{{Keivani} {et~al.}(2018){Keivani}, {Murase}, {Petropoulou}, {Fox}, {Cenko}, {Chaty}, {Coleiro}, {DeLaunay}, {Dimitrakoudis}, {Evans}, {Kennea}, {Marshall}, {Mastichiadis}, {Osborne}, {Santander}, {Tohuvavohu}, \& {Turley}}]{2018_Keivani}
{Keivani}, A., {Murase}, K., {Petropoulou}, M., {et~al.} 2018, \apj, 864, 84, \dodoi{10.3847/1538-4357/aad59a}

\bibitem[{{Kochanek} {et~al.}(2017){Kochanek}, {Shappee}, {Stanek}, {Holoien}, {Thompson}, {Prieto}, {Dong}, {Shields}, {Will}, {Britt}, {Perzanowski}, \& {Pojma{\'n}ski}}]{2017Kochanek}
{Kochanek}, C.~S., {Shappee}, B.~J., {Stanek}, K.~Z., {et~al.} 2017, Publications of the Astronomical Society of the Pacific, 129, 104502, \dodoi{10.1088/1538-3873/aa80d9}

\bibitem[{{Kopper} \& {Blaufuss}(2017)}]{GCN21916}
{Kopper}, C., \& {Blaufuss}, E. 2017, GRB Coordinates Network, 21916, 1.
\newblock \url{https://ui.adsabs.harvard.edu/abs/2017GCN.21916....1K}

\bibitem[{{Krause} {et~al.}(2017){Krause}, {Pueschel}, \& {Maier}}]{2017APh....89....1K}
{Krause}, M., {Pueschel}, E., \& {Maier}, G. 2017, Astroparticle Physics, 89, 1, \dodoi{10.1016/j.astropartphys.2017.01.004}

\bibitem[{{Li} \& {Ma}(1983)}]{1983ApJ...272..317L}
{Li}, T.-P., \& {Ma}, Y.-Q. 1983, \apj, 272, 317, \dodoi{10.1086/161295}

\bibitem[{{Liao} {et~al.}(2022){Liao}, {Sheng}, {Jiang}, {Chang}, {Wang}, {Xu}, {Shu}, {Fan}, \& {Wang}}]{2022ApJ...932L..25L}
{Liao}, N.-H., {Sheng}, Z.-F., {Jiang}, N., {et~al.} 2022, \apjl, 932, L25, \dodoi{10.3847/2041-8213/ac756f}

\bibitem[{{Maier} \& {Holder}(2017)}]{Maier17}
{Maier}, G., \& {Holder}, J. 2017, in International Cosmic Ray Conference, Vol. 301, 35th International Cosmic Ray Conference (ICRC2017), 747.
\newblock \doarXiv{1708.04048}

\bibitem[{{Mattox} {et~al.}(1996){Mattox}, {Bertsch}, {Chiang}, {Dingus}, {Digel}, {Esposito}, {Fierro}, {Hartman}, {Hunter}, {Kanbach}, {Kniffen}, {Lin}, {Macomb}, {Mayer-Hasselwander}, {Michelson}, {von Montigny}, {Mukherjee}, {Nolan}, {Ramanamurthy}, {Schneid}, {Sreekumar}, {Thompson}, \& {Willis}}]{RN7}
{Mattox}, J.~R., {Bertsch}, D.~L., {Chiang}, J., {et~al.} 1996, \apj, 461, 396, \dodoi{10.1086/177068}

\bibitem[{{McBride} {et~al.}(2022){McBride}, {Fox}, {Cowen}, {Ayala Solares}, {Gregoire}, {Kennea}, {Coleiro}, {Evans}, \& {AMON}}]{XRT_IC220303A}
{McBride}, F., {Fox}, D.~B., {Cowen}, D., {et~al.} 2022, GRB Coordinates Network, 31687, 1

\bibitem[{{M{\'e}sz{\'a}ros}(2017)}]{2017ARNPS..67...45M}
{M{\'e}sz{\'a}ros}, P. 2017, Annual Review of Nuclear and Particle Science, 67, 45, \dodoi{10.1146/annurev-nucl-101916-123304}

\bibitem[{{Murase} \& {Stecker}(2023)}]{2023ecnp.book..483M}
{Murase}, K., \& {Stecker}, F.~W. 2023, in The Encyclopedia of Cosmology. Set 2: Frontiers in Cosmology. Volume 2: Neutrino Physics and Astrophysics, ed. F.~W. {Stecker}, 483--540, \dodoi{10.1142/9789811282645_0010}

\bibitem[{{Nilsson} {et~al.}(2003){Nilsson}, {Pursimo}, {Heidt}, {Takalo}, {Sillanp{\"a}{\"a}}, \& {Brinkmann}}]{2003A&A...400...95N}
{Nilsson}, K., {Pursimo}, T., {Heidt}, J., {et~al.} 2003, \aap, 400, 95, \dodoi{10.1051/0004-6361:20021861}

\bibitem[{{Nolan} {et~al.}(2012){Nolan}, {Abdo}, {Ackermann}, {Ajello}, {Allafort}, {Antolini}, {Atwood}, {Axelsson}, {Baldini}, {Ballet}, {Barbiellini}, {Bastieri}, {Bechtol}, {Belfiore}, {Bellazzini}, {Berenji}, {Bignami}, {Blandford}, {Bloom}, {Bonamente}, {Bonnell}, {Borgland}, {Bottacini}, {Bouvier}, {Brandt}, {Bregeon}, {Brigida}, {Bruel}, {Buehler}, {Burnett}, {Buson}, {Caliandro}, {Cameron}, {Campana}, {Ca{\~n}adas}, {Cannon}, {Caraveo}, {Casandjian}, {Cavazzuti}, {Ceccanti}, {Cecchi}, {{\c{C}}elik}, {Charles}, {Chekhtman}, {Cheung}, {Chiang}, {Chipaux}, {Ciprini}, {Claus}, {Cohen-Tanugi}, {Cominsky}, {Conrad}, {Corbet}, {Cutini}, {D'Ammando}, {Davis}, {de Angelis}, {DeCesar}, {DeKlotz}, {De Luca}, {den Hartog}, {de Palma}, {Dermer}, {Digel}, {Silva}, {Drell}, {Drlica-Wagner}, {Dubois}, {Dumora}, {Enoto}, {Escande}, {Fabiani}, {Falletti}, {Favuzzi}, {Fegan}, {Ferrara}, {Focke}, {Fortin}, {Frailis}, {Fukazawa}, {Funk}, {Fusco}, {Gargano}, {Gasparrini}, {Gehrels}, {Germani}, {Giebels}, {Giglietto},
  {Giommi}, {Giordano}, {Giroletti}, {Glanzman}, {Godfrey}, {Grenier}, {Grondin}, {Grove}, {Guillemot}, {Guiriec}, {Gustafsson}, {Hadasch}, {Hanabata}, {Harding}, {Hayashida}, {Hays}, {Hill}, {Horan}, {Hou}, {Hughes}, {Iafrate}, {Itoh}, {J{\'o}hannesson}, {Johnson}, {Johnson}, {Johnson}, {Johnson}, {Kamae}, {Katagiri}, {Kataoka}, {Katsuta}, {Kawai}, {Kerr}, {Kn{\"o}dlseder}, {Kocevski}, {Kuss}, {Lande}, {Landriu}, {Latronico}, {Lemoine-Goumard}, {Lionetto}, {Llena Garde}, {Longo}, {Loparco}, {Lott}, {Lovellette}, {Lubrano}, {Madejski}, {Marelli}, {Massaro}, {Mazziotta}, {McConville}, {McEnery}, {Mehault}, {Michelson}, {Minuti}, {Mitthumsiri}, {Mizuno}, {Moiseev}, {Mongelli}, {Monte}, {Monzani}, {Morselli}, {Moskalenko}, {Murgia}, {Nakamori}, {Naumann-Godo}, {Norris}, {Nuss}, {Nymark}, {Ohno}, {Ohsugi}, {Okumura}, {Omodei}, {Orlando}, {Ormes}, {Ozaki}, {Paneque}, {Panetta}, {Parent}, {Perkins}, {Pesce-Rollins}, {Pierbattista}, {Pinchera}, {Piron}, {Pivato}, {Porter}, {Racusin}, {Rain{\`o}}, {Rando}, {Razzano},
  {Razzaque}, {Reimer}, {Reimer}, {Reposeur}, {Ritz}, {Rochester}, {Romani}, {Roth}, {Rousseau}, {Ryde}, {Sadrozinski}, {Salvetti}, {Sanchez}, {Saz Parkinson}, {Sbarra}, {Scargle}, {Schalk}, {Sgr{\`o}}, {Shaw}, {Shrader}, {Siskind}, {Smith}, {Spandre}, {Spinelli}, {Stephens}, {Strickman}, {Suson}, {Tajima}, {Takahashi}, {Takahashi}, {Tanaka}, {Thayer}, {Thayer}, {Thompson}, {Tibaldo}, {Tibolla}, {Tinebra}, {Tinivella}, {Torres}, {Tosti}, {Troja}, {Uchiyama}, {Vandenbroucke}, {Van Etten}, {Van Klaveren}, {Vasileiou}, {Vianello}, {Vitale}, {Waite}, {Wallace}, {Wang}, {Werner}, {Winer}, {Wood}, {Wood}, {Wood}, {Yang}, \& {Zimmer}}]{2fgl}
{Nolan}, P.~L., {Abdo}, A.~A., {Ackermann}, M., {et~al.} 2012, \apjs, 199, 31, \dodoi{10.1088/0067-0049/199/2/31}

\bibitem[{{Park} \& {the VERITAS Collaboration}(2015)}]{Park15}
{Park}, N., \& {the VERITAS Collaboration}. 2015, in International Cosmic Ray Conference, Vol.~34, 34th International Cosmic Ray Conference (ICRC2015), 771.
\newblock \doarXiv{1508.07070}

\bibitem[{{Pohl} \& {Schlickeiser}(2000)}]{2000A&A...354..395P}
{Pohl}, M., \& {Schlickeiser}, R. 2000, \aap, 354, 395.
\newblock \doarXiv{astro-ph/9911452}

\bibitem[{{Poole} {et~al.}(2008){Poole}, {Breeveld}, {Page}, {Landsman}, {Holland}, {Roming}, {Kuin}, {Brown}, {Gronwall}, {Hunsberger}, {Koch}, {Mason}, {Schady}, {vanden Berk}, {Blustin}, {Boyd}, {Broos}, {Carter}, {Chester}, {Cucchiara}, {Hancock}, {Huckle}, {Immler}, {Ivanushkina}, {Kennedy}, {Marshall}, {Morgan}, {Pandey}, {de Pasquale}, {Smith}, \& {Still}}]{Poole08}
{Poole}, T.~S., {Breeveld}, A.~A., {Page}, M.~J., {et~al.} 2008, \mnras, 383, 627, \dodoi{10.1111/j.1365-2966.2007.12563.x}

\bibitem[{{Rolke} {et~al.}(2005){Rolke}, {L{\'o}pez}, \& {Conrad}}]{Rolke05}
{Rolke}, W.~A., {L{\'o}pez}, A.~M., \& {Conrad}, J. 2005, Nuclear Instruments and Methods in Physics Research A, 551, 493, \dodoi{10.1016/j.nima.2005.05.068}

\bibitem[{{Roming} {et~al.}(2005){Roming}, {Kennedy}, {Mason}, {Nousek}, {Ahr}, {Bingham}, {Broos}, {Carter}, {Hancock}, {Huckle}, {Hunsberger}, {Kawakami}, {Killough}, {Koch}, {McLelland}, {Smith}, {Smith}, {Soto}, {Boyd}, {Breeveld}, {Holland}, {Ivanushkina}, {Pryzby}, {Still}, \& {Stock}}]{2005SS_Roming}
{Roming}, P. W.~A., {Kennedy}, T.~E., {Mason}, K.~O., {et~al.} 2005, \ssr, 120, 95, \dodoi{10.1007/s11214-005-5095-4}

\bibitem[{{Roming} {et~al.}(2009){Roming}, {Koch}, {Oates}, {Porterfield}, {Vanden Berk}, {Boyd}, {Holland}, {Hoversten}, {Immler}, {Marshall}, {Page}, {Racusin}, {Schneider}, {Breeveld}, {Brown}, {Chester}, {Cucchiara}, {DePasquale}, {Gronwall}, {Hunsberger}, {Kuin}, {Landsman}, {Schady}, \& {Still}}]{2009_Roming}
{Roming}, P.~W.~A., {Koch}, T.~S., {Oates}, S.~R., {et~al.} 2009, \apj, 690, 163, \dodoi{10.1088/0004-637X/690/1/163}

\bibitem[{{Sahakyan} {et~al.}(2023){Sahakyan}, {Giommi}, {Padovani}, {Petropoulou}, {B{\'e}gu{\'e}}, {Boccardi}, \& {Gasparyan}}]{2023MNRAS.519.1396S}
{Sahakyan}, N., {Giommi}, P., {Padovani}, P., {et~al.} 2023, \mnras, 519, 1396, \dodoi{10.1093/mnras/stac3607}

\bibitem[{{Schlafly} \& {Finkbeiner}(2011)}]{2011_reddenning}
{Schlafly}, E.~F., \& {Finkbeiner}, D.~P. 2011, \apj, 737, 103, \dodoi{10.1088/0004-637X/737/2/103}

\bibitem[{Sch\"ussler {et~al.}(2023)}]{IceCube:2023kyd}
Sch\"ussler, F., {et~al.} 2023, PoS, ICRC2023, 1501, \dodoi{10.22323/1.444.1501}

\bibitem[{Shappee {et~al.}(2014)Shappee, Prieto, Grupe, Kochanek, Stanek, De~Rosa, Mathur, Zu, Peterson, Pogge, \& et~al.}]{2014Shappee}
Shappee, B.~J., Prieto, J.~L., Grupe, D., {et~al.} 2014, The Astrophysical Journal, 788, 48, \dodoi{10.1088/0004-637x/788/1/48}

\bibitem[{{Sikora} {et~al.}(1994){Sikora}, {Begelman}, \& {Rees}}]{1994_Sikora}
{Sikora}, M., {Begelman}, M.~C., \& {Rees}, M.~J. 1994, \apj, 421, 153, \dodoi{10.1086/173633}

\bibitem[{{Tavecchio} {et~al.}(2010){Tavecchio}, {Ghisellini}, {Ghirlanda}, {Foschini}, \& {Maraschi}}]{2010MNRAS.401.1570T}
{Tavecchio}, F., {Ghisellini}, G., {Ghirlanda}, G., {Foschini}, L., \& {Maraschi}, L. 2010, \mnras, 401, 1570, \dodoi{10.1111/j.1365-2966.2009.15784.x}

\bibitem[{{The IceCube Collaboration} {et~al.}(2018)}]{eaat1378}
{The IceCube Collaboration}, {et~al.} 2018, Science, 361, \dodoi{10.1126/science.aat1378}

\bibitem[{{Tonry} {et~al.}(2018){Tonry}, {Denneau}, {Heinze}, {Stalder}, {Smith}, {Smartt}, {Stubbs}, {Weiland}, \& {Rest}}]{ATLAS_main}
{Tonry}, J.~L., {Denneau}, L., {Heinze}, A.~N., {et~al.} 2018, \pasp, 130, 064505, \dodoi{10.1088/1538-3873/aabadf}

\bibitem[{Twagirayezu {et~al.}(2023)Twagirayezu, Niederhausen, Sclafani, Whitehorn, Nisa, Yu, \& Halliday}]{Twagirayezu:2023Sd}
Twagirayezu, J.~P., Niederhausen, H., Sclafani, S., {et~al.} 2023, PoS, ICRC2023, 1175, \dodoi{10.22323/1.444.1175}

\bibitem[{Valverde {et~al.}(2020)Valverde, Horan, Bernard, Fegan, Collaboration), Abeysekara, Archer, Benbow, Bird, Brill, Brose, Buchovecky, Buckley, Christiansen, Cui, Falcone, Feng, Finley, Fortson, Furniss, Gent, Gillanders, Giuri, Gueta, Hanna, Hassan, Hervet, Holder, Hughes, Humensky, Kaaret, Kelley-Hoskins, Kertzman, Kieda, Krause, Krennrich, Lang, Maier, Moriarty, Mukherjee, Nieto, Nievas-Rosillo, O’Brien, Ong, Otte, Park, Petrashyk, Pfrang, Pichel, Pohl, Prado, Pueschel, Quinn, Ragan, Reynolds, Ribeiro, Richards, Roache, Sadeh, Santander, Scott, Sembroski, Shahinyan, Shang, Sushch, Vassiliev, Weinstein, Wells, Wilcox, Wilhelm, Williams, Williamson, (VERITAS Collaboration), Noto, Edwards, Piner, Ramazani, Hovatta, Jormanainen, Lindfors, Nilsson, Takalo, Kovalev, Lister, Pushkarev, Savolainen, Kiehlmann, Max-Moerbeck, Readhead, Lähteenmäki, \& Tornikoski}]{Valverde_2020}
Valverde, J., Horan, D., Bernard, D., {et~al.} 2020, The Astrophysical Journal, 891, 170, \dodoi{10.3847/1538-4357/ab765d}

\bibitem[{{Willingale} {et~al.}(2013){Willingale}, {Starling}, {Beardmore}, {Tanvir}, \& {O'Brien}}]{column_density_paper}
{Willingale}, R., {Starling}, R.~L.~C., {Beardmore}, A.~P., {Tanvir}, N.~R., \& {O'Brien}, P.~T. 2013, \mnras, 431, 394, \dodoi{10.1093/mnras/stt175}

\bibitem[{{Wood} {et~al.}(2017){Wood}, {Caputo}, {Charles}, {Di Mauro}, {Magill}, {Perkins}, \& {Fermi-LAT Collaboration}}]{wood2017fermipy}
{Wood}, M., {Caputo}, R., {Charles}, E., {et~al.} 2017, in International Cosmic Ray Conference, Vol. 301, 35th International Cosmic Ray Conference (ICRC2017), 824.
\newblock \doarXiv{1707.09551}

\end{thebibliography}
\bibliographystyle{aasjournal}

\begin{appendix}
\section{Additional \texttt{Bjet\_MCMC} OUTPUTS FOR B3 2247+381}
\label{sec:appendix}
\begin{figure*}[h]
\centering
\includegraphics[width=0.45 \linewidth]{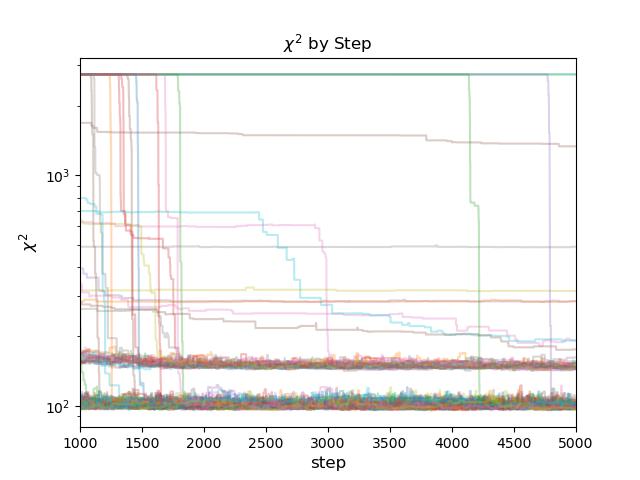}
\includegraphics[width=0.45 \linewidth]{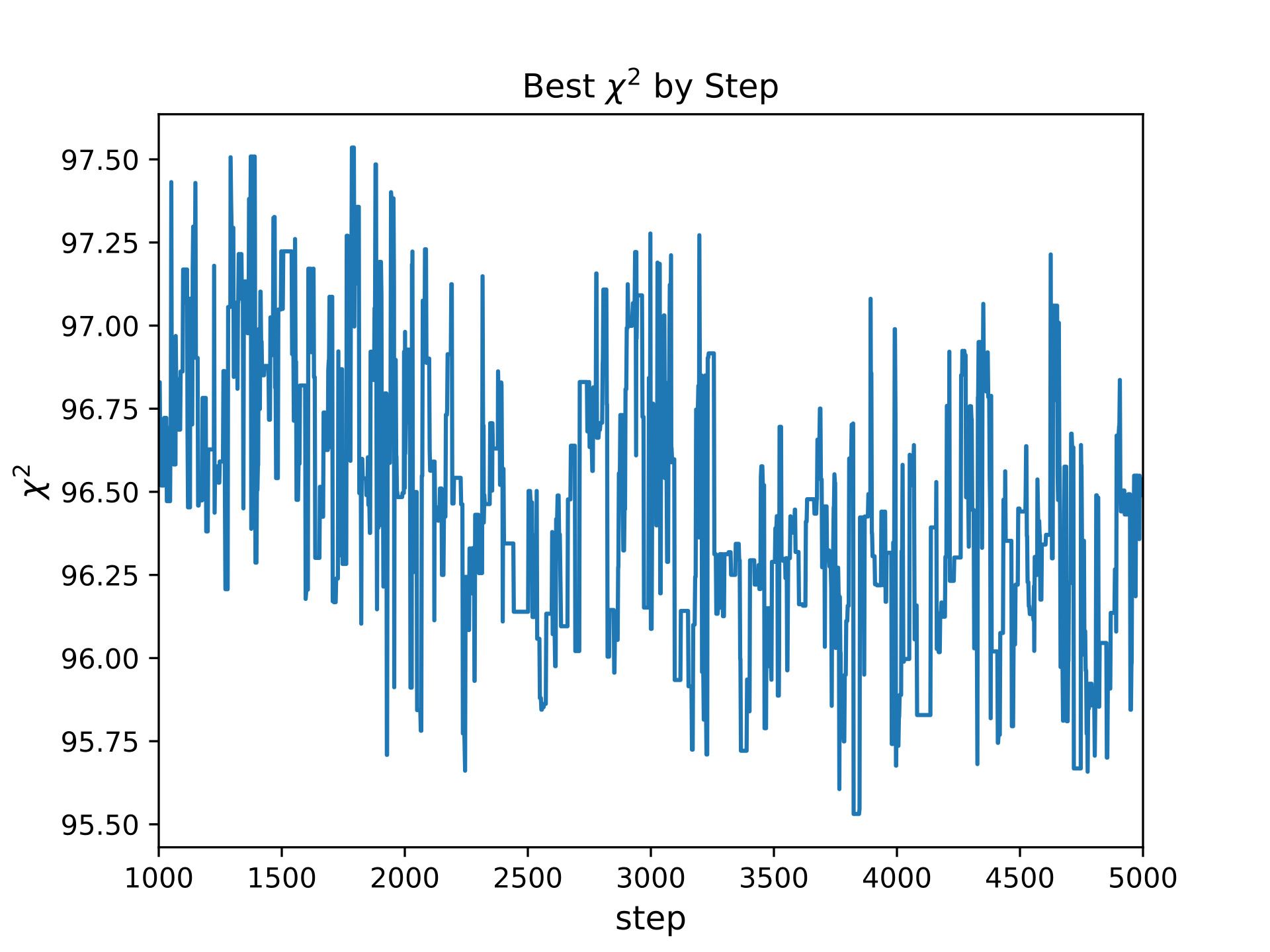}
\includegraphics[width=0.45 \linewidth]{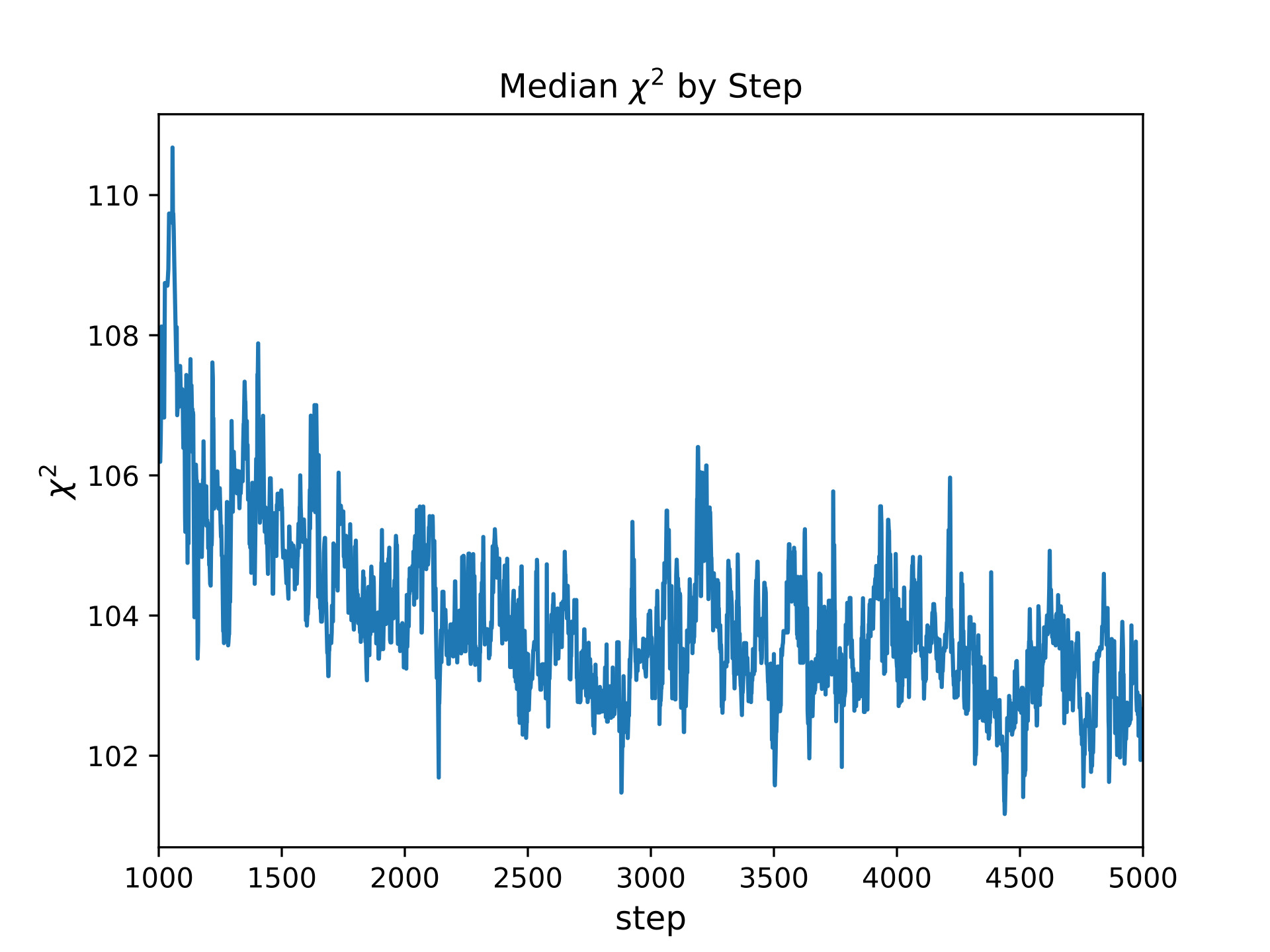}
\caption{The \texttt{Bjet\_MCMC} $\chi^{2}$ values of walkers at each step for the SED fit of B3 2247+381. Top left panel: all walkers, Top right panel: smallest $\chi^{2}$, Bottom panel: median $\chi^{2}$.}
\label{fig:walkers}
\end{figure*}

\begin{figure*}[h]
\centering
\includegraphics[width=0.8 \linewidth]{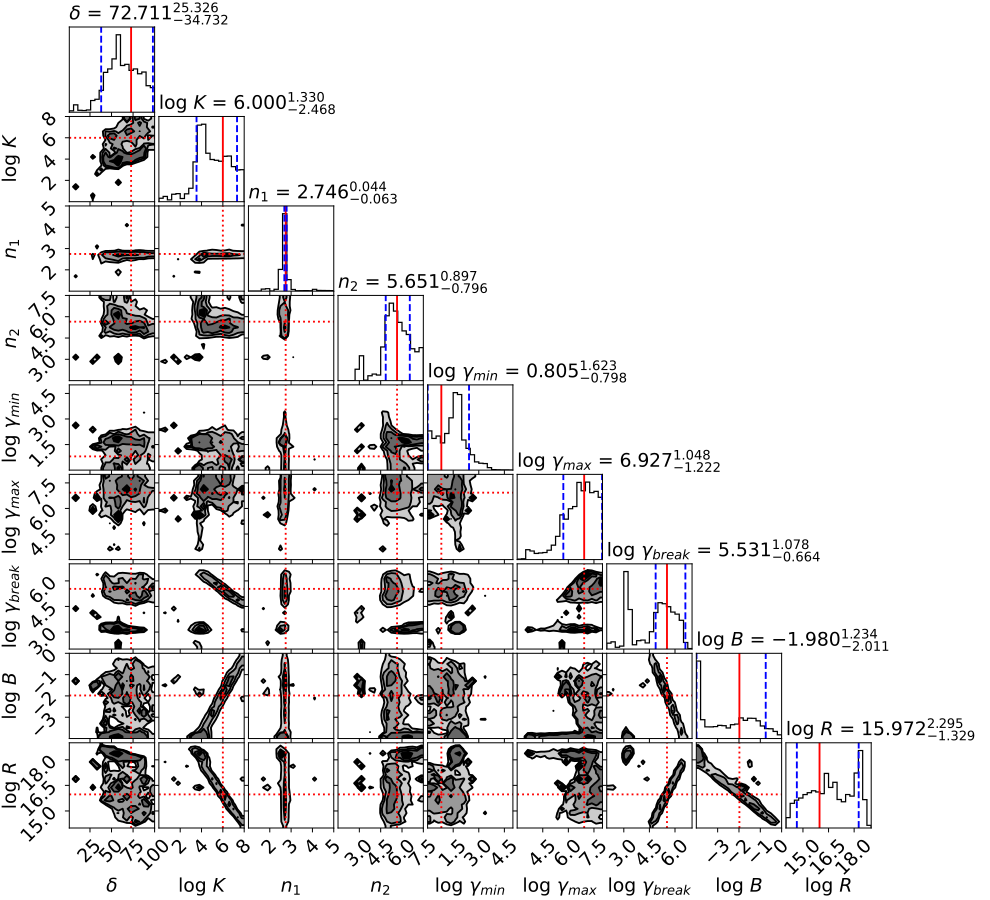}
\caption{The corner plot of the posterior probability distribution of the free parameters from the SED fit of B3 2247+381.}
\label{fig:corner}
\end{figure*}

Fig.~\ref{fig:walkers} shows the \texttt{Bjet\_MCMC} $\chi^{2}$ values of walkers at each step for the SED fit for all walkers, smallest $\chi^{2}$ and median $\chi^{2}$ respectively. Fig.~\ref{fig:corner} shows a corner plot of the posterior probability distribution of the free parameters from the SED fit.

\end{appendix}

\allauthors



\end{document}